\documentclass[aps,prd,a4paper,onecolumn,amsmath,showpacs,superscriptaddress,nofootinbib,preprintnumbers,notitlepage]{scrartcl}

\usepackage{amsmath,amssymb,latexsym,mathrsfs}
\usepackage{graphicx}
\usepackage{color}
\usepackage{hyperref}
\usepackage{enumitem,booktabs,cfr-lm}
\usepackage[referable]{threeparttablex}
\renewlist{tablenotes}{enumerate}{1}
\makeatletter
\setlist[tablenotes]{label=\tnote{\alph*},ref=\alph*,itemsep=\z@,topsep=\z@skip,partopsep=\z@skip,parsep=\z@,itemindent=\z@,labelindent=\tabcolsep,labelsep=.2em,leftmargin=*,align=left,before={\footnotesize}}
\makeatother

\newcommand{\zeq}{z_\mathrm{eq}}
\newcommand{\zrec}{z_\mathrm{rec}}
\newcommand{\zre}{z_\mathrm{re}}
\newcommand{\zdrag}{z_\mathrm{drag}}
\newcommand{\zde}{z_\Lambda}

\newcommand{\Neff}{N_\mathrm{eff}}
\newcommand{\Lya}{\mathrm{Ly}\alpha}
\newcommand{\der}{\mathrm{d}}

\newcommand{\eV}{\mathrm{eV}}
\newcommand{\MeV}{\mathrm{MeV}}
\newcommand{\kmsmpc}{\mathrm{km\,s^{-1}\,Mpc^{-1}}}
\newcommand{\GHz}{\mathrm{GHz}}
\newcommand{\uka}{\mu \mathrm{K\cdot arcmin}}
\newcommand{\sumnu}{\Sigma m_\nu}
\newcommand{\lcdm}{\Lambda\mathrm{CDM}}
\newcommand{\bb}{0\nu2\beta}
\newcommand{\dms}{\Delta m_{12}^2}
\newcommand{\dma}{\Delta m_{13}^2}
\newcommand{\mlight}{m_\mathrm{light}}
\newcommand{\mbb}{m_{\beta\beta}}

\begin{document}

\title{Status of neutrino properties and future prospects - Cosmological and astrophysical constraints}

\author{Martina Gerbino\thanks{The Oskar Klein Centre for Cosmoparticle Physics, Department of Physics, Stockholm University, SE-106 91 Stockholm, Sweden. Email: martina.gerbino@fysik.su.se} \and Massimiliano Lattanzi\thanks{Istituto Nazionale di Fisica Nucleare (INFN), Sezione di Ferrara, Via Giuseppe Saragat 1, I-44122 Ferrara, Italy. Email: lattanzi@fe.infn.it}}
%\author{Massimiliano Lattanzi}

\maketitle

\begin{abstract}
Cosmological observations are a powerful probe of neutrino properties, and in particular of their mass.
In this review, we first discuss the role of neutrinos in shaping the cosmological evolution at both the background and perturbation level,
and describe their effects on cosmological observables such as the cosmic microwave background and the distribution of matter at large scale.
We then present the state of the art concerning the constraints on neutrino masses from those observables, and also review
the prospects for future experiments. We also briefly discuss the prospects for determining the neutrino hierarchy from cosmology,
the complementarity with laboratory experiments, and the constraints on neutrino properties beyond their mass.
\end{abstract}

\tableofcontents

\section{Introduction}

Flavour oscillation experiments have by now firmly established that neutrinos have a mass. Current experiments
measure with great accuracy the three mixing angles, as well as the two mass-squared splittings between the three
active neutrinos. In the framework of the standard model (SM) of particle physics neutrinos are massless, and consequently do not mix,
since it is not possible to build a mass term for them using the particle content of the SM. Therefore, flavour oscillations
represent the only laboratory evidence for physics beyond the SM. Several unknowns in the neutrino sector still
remain, confirming these particles as being the most elusive within the SM. In particular,
the absolute scale of neutrino masses has yet to be determined. Moreover, the sign of the largest
mass squared splitting, the one governing atmospheric transitions, is still unknown. This leaves open 
two possibilities for the neutrino mass ordering, corresponding to the two signs of the atmospheric splitting: the normal hierarchy, in which
the atmospheric splitting is positive, and the inverted hierarchy, in which it is negative. 
Other unknowns are the value of a possible CP-violating phase in the neutrino mixing matrix,
and the Dirac or Majorana nature of neutrinos.

There are different ways of measuring the absolute neutrino mass scale. One is to use kinematic effects,
for example by measuring the energy spectrum of electrons produced in the $\beta$-decay of nuclei, looking
for the distortions due to the finite neutrino mass. This approach has the advantage of being very robust and providing 
model-independent results, as it basically relies only on energy conservation. Present constraints on the effective mass of the electron neutrino $m_\beta$ (an incoherent sum of the mass eigenvalues,
weighted with the elements of the mixing matrix) are $m_\beta < 2.05 \,\eV$ from the Troitsk~\cite{Aseev:2011dq} experiment, and $m_\beta < 2.3 \,\eV$ from the Mainz~\cite{Kraus:2004zw} experiment, at the $95\%$ CL.
The KATRIN spectrometer~\cite{katrin}, who will start its science run early in the next year, is expected to improve the sensitivity by an order of magnitude.
Another way to measure neutrino masses in the laboratory is to look for neutrinoless double $\beta$ decay ($0\nu2\beta$ in short) of nuclei,
a rare process that is allowed only if neutrinos are Majorana particles \cite{Schechter:1981bd}. The prospects for detection of neutrino mass with $0\nu2\beta$ searches
 are very promising: current constraints for the effective Majorana mass of the electron neutrino $\mbb$, 
 a \emph{coherent} sum of the mass eigenvalues, weighted with the elements of the mixing matrix, are in the $\mbb < 0.1\div 0.4\,\eV$ ballpark (see Sec~\ref{sec:lab} for more details).
 There are a few shortcomings, however. First of all, there is some amount of model dependence: one has
 to assume that neutrinos are Majorana particles to start, and even if this is, in some sense, a natural and very appealing scenario
 from the theoretical point of view - as it could explain the smallness of neutrino masses  \cite{Mohapatra:1979ia,Schechter:1980gr,Lazarides:1980nt,Chikashige:1980ui,Schechter:1981cv} - we have at the moment no indication that this is really the case. 
 Moreover, inferring the neutrino mass from a (non-)observation of $0\nu2\beta$ requires the implicit assumption that the mass mechanism
 is the only contribution to the amplitude of the process, i.e., that no other physics beyond the SM that violates lepton number is at play. Another issue is
 that the amplitude of the process also depends on nuclear matrix elements, that are known only with limited accuracy, introducing
 an additional layer of uncertainty in the interpretation of experimental results. Finally, given that $\mbb$ is a coherent superposition
 of the mass eigenvalues, it could be that the values of the Majorana phases arrange to make $\mbb$ vanishingly small.
 
 The third avenue to measure neutrino masses, and in fact the topic of this review, is to use cosmological observations.
 As we shall discuss in more detail in the following, the presence of a cosmic background of relic neutrinos (C$\nu$B) is a robust prediction
 of the standard cosmological model \cite{booknucosmo}. Even though a direct detection is extremely difficult and still lacking, 
(but experiments aiming at this are currently under development, like the PTOLEMY experiment  \cite{Betts:2013uya}),
nevertheless cosmological observations are in agreement with this prediction. The relic neutrinos affect the cosmological evolution, both at the background and perturbation level,
 so that cosmological observables can be used to constrain the neutrino properties, and in particular their mass (see e.g. Refs.  \cite{booknucosmo,Dolgov:2002wy, Lesgourgues:2006nd} for excellent reviews on this topic).  
 In fact, cosmology currently represent the most sensitive probe of neutrino masses. The observations of cosmic microwave background (CMB) anisotropies from the Planck satellite, without the addition of any external data, constrain the sum of neutrino masses already at the $0.6\,\eV$ level~\cite{PlanckXIII}, which is basically the same as the KATRIN sensitivity. Combinations of different datasets yield even stronger limits, at the same level or better than the ones
 from $0\nu2\beta$ searches, although a direct comparison is not immediate, due to the fact that different quantities are probed, and also due to the
theoretical assumptions involved in the interpretation of both kinds of data. Future-generation experiments will likely have the capability to detect neutrino masses, and to disentangle
 the hierarchy, provided that systematics effects can be kept under control - and that our theoretical understanding of the Universe is correct, of course!
Concerning this last point, the drawback of cosmological measurements of neutrino mass and other properties, is that they are somehow model dependent.
Inferences from cosmological observations are made in the framework of a model - the so-called $\Lambda$CDM model -, and of its simple extensions, that currently represents our best and simple description of the Universe that is compatible with observations. This model is based on General Relativity (with the assumption of a homogeneous and isotropic Universe at large scales) and on the SM of particle physics, complemented with a mechanism for the generation
of primordial perturbations, i.e., the inflationary paradigm. When cosmological data are interpreted in this framework, they 
point to the following picture: our Universe is spatially flat and is presently composed by baryons ($\sim 5\%$ of the total density), 
dark matter ($\sim25\%$), and an even more elusive component called dark energy ($\sim 70\%$), that behaves like a cosmological constant,
and is responsible for the present accelerated expansion, plus photons (a few parts in $10^5$) and light neutrinos. The constraints from Planck cited above imply that, in the framework of the $\lcdm$ model, neutrinos can contribute by $\sim 1\%$ of the present energy density at most. The structures that we observe today have evolved from adiabatic, nearly scale-invariant initial conditions. Even though this model is very successful, barring some intriguing but for the moment still mild (at the $\sim 2 \sigma$ level) discrepancies between observational probes, this dependence should be borne in mind. On the other hand,
such a healthy approach should not, in our opinion, be substituted with its contrary, i.e., a complete distrust towards cosmological constraints. 
A pragmatic approach to this problem is to test the robustness of our inferences concerning neutrino properties against different assumptions,
by exploring extensions of the $\Lambda$CDM model. This has been in fact done quite extensively in the literature, and we will take care, towards the end of the review, to report results obtained in extended models.
 
Another advantage of cosmological observations is that they are able to probe neutrino properties beyond their mass. A well-known
example is the effective number of neutrinos, basically a measure of the energy density in relativistic species in the early Universe, that is a powerful probe
of a wide range of beyond the SM model physics (in fact, not necessarily related to neutrinos). For example, it could 
probe the existence of an additional, light sterile mass eigenstate, as well as the physics of neutrino decoupling,
or the presence of lepton asymmetries generated in the early Universe. Cosmology can also be used to constrain the existence
of non-standard neutrino interactions, possibly related to the mechanism of mass generation. Even though they 
are not the focus of this review, we will briefly touch  some of this aspects in the final sections of the review.
 
Cosmological data have reached a very good level of maturity over the last decades. Measurements of the CMB anisotropies from the Planck satellite have put the tightest constraints ever on cosmological parameters from a single experiment~\cite{PlanckXIII}, dramatically improving the constraints from the predecessor satellite WMAP~\cite{WMAP}. From the ground, the Atacama Cosmology Telescope (ACT) polarization-sensitive receiver and the South Pole Telescope (SPT) have been measuring with incredible accuracy CMB anisotropies at the smallest scales in temperature and polarization~\cite{Louis:2016ahn,Henning:2017nuy}. At degree and sub-degree scales, the BICEP/Keck collaboration~\cite{Array:2016afx,Array:2015xqh} and the POLARBEAR telescope~\cite{Ade:2017uvt} are looking at the faint CMB ``B-mode'' signal, containing information about both the early stages of the Universe (\textit{primordial} B-modes) and the late time evolution (\textit{lensing} B-modes). The Cosmology Large Angular Scale Surveyor~\cite{Essinger-Hileman:2014pja} is working at mapping the CMB polarization field over 70\% of the sky. The SPIDER balloon~\cite{spider} successfully completed its first flight and is in preparation for the second launch likely at the end of 2018. The CMB data are often complemented with information from the large-scale structure of the universe. The SDSS III-BOSS galaxy survey have recently released its last season of data~\cite{BOSSDR12}. Extended catalogues of galaxy clusters have been completed from several surveys (see e.g.~\cite{PlanckXXIV} and references therein). In addition, weak lensing surveys (Canada-France-Hawaii Telescope Lensing Survey~\cite{CFHTLens}, Kilo-Degree Survey~\cite{Kohlinger:2017sxk}, Dark Energy Survey~\cite{DES}) are mature enough to provide constraints on cosmological parameters that are competitive with those from other observables. They also allow to test the validity of the standard cosmological paradigm by comparing results obtained from high redshift observables to those coming from measurements of the low redshift universe. 

The current scenario is just a taste of the constraining power of cosmological observables that will be available with the next generation of experiments,
that will be taking measurements in the next decade. Future CMB missions -- including Advanced ACTPol~\cite{Henderson:2015nzj}, SPT-3G~\cite{Benson:2014qhw}, CMB Stage-IV~\cite{S4}, Simons Observatory~\cite{SO}, Simons Array~\cite{SA}, CORE~\cite{Delabrouille:2017rct}, LiteBIRD~\cite{Matsumura:2013aja}, PIXIE~\cite{pixie}-- will test the universe over a wide range of scales with unprecedented accuracy. The same accuracy will enable the reconstruction of the weak lensing signal from the CMB maps down to the smallest scales and with high sensitivity, providing an additional probe of the distribution and evolution of structures in the universe. On the other hand, the new generation of large scale structure surveys -- including the Dark Energy Spectroscopic Instrument~\cite{DESI}, the Large Synoptic Survey Telescope~\cite{LSST}, Euclid~\cite{EuclidStudyRep} and the Wide Frequency InfraRed Spectroscopic Telescope~\cite{WFIRST} -- will also probe the late-time universe with the ultimate goal of shedding light on the biggest unknown of our times, namely the nature of dark energy and dark matter. 

The aim of this review is to provide the state of the art of the current knowledge of neutrino masses from cosmological probes and give an overview of future prospects. 
The review is organized as follows: in Sec. \ref{sec:prelim}, we outline the framework of this review, introducing some useful notation and briefly reviewing the basics of neutrino cosmology. Sec.~\ref{sec:eff} is devoted to discussing, from a broad perspective, cosmological effects induced by massive neutrinos. In Sec.~\ref{sec:obs}, we will describe in detail how the effects introduced in Sec.~\ref{sec:eff} affect cosmological observables, such as the CMB anisotropies, large scale structures and cosmological distances. Sec.~\ref{sec:current} and Sec.~\ref{sec:future} present a detailed collection of the current and future limits on $\sumnu$ from the measurements of the cosmological observables discussed in Sec.~\ref{sec:obs}, mostly derived in the context of the $\lcdm$ cosmological model. Constraints derived in more extended scenarios are summarized in Sec.~\ref{sec:extended}. Sec.~\ref{sec:hierarchy} briefly deals with the  issue of whether cosmological probes are able to provide information not only on $\sumnu$, but also on its distribution among the mass eigenstates, i.e. about the neutrino hierarchy. In Sec.~\ref{sec:lab}, we will briefly go through the complementarity between cosmology and laboratory searches in the quest for constraining neutrino properties. Finally, Sec.~\ref{sec:beyond} offers a summary of the additional information about neutrino properties beyond their mass scale that we can extract from cosmological observables. We derive our conclusions in Sec.~\ref{sec:concl}. The impatient reader can access the summary of current and future limits from Tables~\ref{tab:current}, \ref{tab:extended}, \ref{tab:future} and~\ref{tab:neff}.
 
%%%%%%%

\section{Notation and preliminaries \label{sec:prelim}}

\subsection{Basic equations \label{sec:cosmobas}}

Inferences from cosmological observations are made under the assumption that the Universe is homogeneous 
and isotropic, and as such it is well described, in the context of general relativity, by a Friedmann-Lemaitre-Robertson-Walker (FLRW)
metric. Small deviations from homogeneity and isotropy are modelled as perturbations over of a FLRW background.

In a FLRW Universe, expansion is described by the Friedmann equation\footnote{All throughout this review, we take $c=\hbar=k_B=1$.} for the Hubble parameter $H$:
\begin{equation}
H(a)^2 = \frac{8\pi G}{3}\rho(a) - \frac{K}{a^2} \, ,
\end{equation}
where $G$ is the gravitational constant, $K$ parameterizes the spatial curvature\footnote{We choose not to rescale $K$ to make it equal to $\pm 1$ for an open or closed Universe,
so that we are left with the freedom to rescale the scale factor today $a_0$ to unity.}, $a$ is the cosmic scale factor and the $\rho$ is the 
total energy density. This is given by the sum of the energy densities of the various components of the cosmic fluid.

Considering cold dark matter ($c$), baryons ($b$), photons ($\gamma$),
dark energy ($DE$) and massive neutrinos ($\nu$), and introducing the redshift $1+z = a^{-1}$, the Friedmann equation can be recast as:
\begin{align}
H(z)^2 = H_0^2 \Bigg[ & \left( \Omega_c + \Omega_b \right) (1+z)^3 + \Omega_\gamma (1+z)^4  + \nonumber \\
& +\Omega_{DE} (1+z)^{3(1+w)} + \Omega_k (1+z)^2+ \frac{\rho_{\nu,\mathrm{tot}}(z)}{\rho_{\mathrm{crit},0}}\Bigg ] \, ,
\label{eq:hubble}
\end{align}
where we have introduced the present value of the critical density required for flat spatial geometry $\rho_{\mathrm{crit},0} \equiv 3H_0^2/8\pi G$
(in general, we use a subscript $0$ to denote quantities evaluated today), and the present-day density parameters 
$\Omega_i = \rho_{i,0}/\rho_{\mathrm{crit},0}$ (since we will be always referring to the density parameters today, we omit the subscript $0$ in this case). The scalings with $(1+z)$ come from the fact that the energy densities of nonrelativistic matter and radiation
scale with $a^{-3}$ and $a^{-4}$, respectively. For DE, in writing Eq. (\ref{eq:hubble}) we have left open the possibility for an arbitrary
(albeit constant) equation-of-state parameter $w$. In the case of neutrinos, since the parameter of their equation of state is not constant, 
we could not write a simple scaling with redshift, although this is possible in limiting regimes (see Sec. \ref{sec:cosmonu}). We use $\rho_{\nu,\mathrm{tot}}$
to denote the total neutrino density, i.e., summed over all mass eigenstates. Finally, we have defined a ``curvature density parameter'' $\Omega_k = - K/H_0^2$.
From Eq. (\ref{eq:hubble}) evaluated at $z=0$ it is clear that the density parameters, including curvature, satisfy the constrain $\sum_i \Omega_i = 1$. 

Let us also introduce some extra notation and jargon that will be useful in the following. We will use $\Omega_m$ to refer to the 
total density of nonrelativistic matter today. Thus, this in general includes dark matter, baryons and those neutrinos species that are
heavy enough to be nonrelativistic today. 
In such a way we have that $\Omega_m + \Omega_{DE} = 1$ in a flat Universe (or $\Omega_m + \Omega_{DE} = 1 - \Omega_k$ in general),
since the present density of photons and other relativistic species is negligible.
Since many times we will have to consider the density of matter that is nonrelativistic at all the redshifts that are probed by cosmological observables,
i.e. dark matter and baryons but not neutrinos, we also introduce $\Omega_{c+b}$, with obvious meaning. When we consider
dark energy in the form of a cosmological constant ($w=-1$) we use $\Omega_\Lambda$ in place of $\Omega_{DE}$ to make this fact clear.
Finally, we also use the physical density parameters $\omega_i \equiv \Omega_i h^2$, with $h$ being the present value of the Hubble parameter in units
of 100 km s$^{-1}$ Mpc$^{-1}$.

As we shall discuss in more detail in the following, cosmological observables often carry the imprint of particular length scales, related to
specific physical effects. For this reason we recall some definitions that will be useful in the following. 
The \emph{causal horizon} $r_h$ at time $t$ is defined as the distance traveled by a photon from the Big Bang ($t=0$) until time $t$. This is given by:
\begin{equation}
r_h(t) = \int_0^t \frac{dt'}{a(t')} = \int_{z(t)}^\infty \frac{dz'}{H(z')} \, .
\end{equation}
Note that this is actually the \emph{comoving} causal horizon; in the following, unless otherwise noted, we will always use comoving distances.
We also note that the comoving horizon is equal to the conformal time $\eta(t)$ (defined through $dt = a d\eta$ and $\eta(t=0) = 0$).
In a Friedmann Universe (i.e., one composed only by matter and radiation), the physical causal horizon is proportional,
by a factor of order unity, to the Hubble length $d_H(t) \equiv H(t)^{-1}$. For this reason, we shall sometimes indulge in the
habit of calling the latter the Hubble horizon, even though this is, technically, a misnomer. 

A related quantity is the sound horizon $r_s(t)$, i.e., the distance traveled in a certain time by an acoustic wave in the baryon-photon plasma. The expression
for $r_s$ is very similar to the one for the causal horizon, just with the speed of light (equal to 1 in our units) replaced by the speed of sound  $c_s$
in the plasma:
\begin{equation}
r_s(t) = \int_0^t \frac{c_s(t')}{a(t')} dt' = \int_{z(t)}^\infty \frac{c_s(z')}{H(z')} dz'\, .
\label{eq:rs}
\end{equation}
The speed of sound is given by $c_s = 1/\sqrt{3(1+R)}$, with $R=(p_b+\rho_b)/(p_\gamma+\rho_\gamma)$ being the baryon-to-photon momentum density ratio. When the baryon density is negligible relative to the photons, $c_s\simeq 1/\sqrt{3}$ and $r_s \simeq r_h /\sqrt{3}=\eta/\sqrt{3}$.

Imprints on the cosmological observables of several physical processes usually depend on the value of those scales at some particular time.
For example, the spacing of acoustic peaks in the CMB spectrum is reminescent of the sound horizon at the time of hydrogen recombination;
the suppression of small-scale matter fluctuations due to neutrino free-streaming is set by the causal horizon at the time neutrinos become nonrelativistic; and so on. Moreover, since today we see those scales through their projection on the sky, what we observe is actually a combination of the scale itself
and the distance to the object that we are observing. We find then useful also to recall some notions related to cosmological distances. The comoving distance $\chi$ between us and an object at redshift $z$ is
\begin{equation}
\chi(z) = \int_0^z \frac{dz'}{H(z')} \, ,
\end{equation}
and this is also equal to $\eta_0 - \eta(z)$. The comoving angular diameter distance $d_A(z)$ is given by 
\begin{equation}
d_A(z) = \frac{\sin\left(\sqrt{K}\chi\right)}{\sqrt{K}} \, ,
\label{eq:rAcurv}
\end{equation}
so that 
\begin{equation}
d_A(z) = \chi(z) = \int_0^z \frac{dz'}{H(z')} \qquad \mathrm{for }\; \Omega_k = 0 \, .
\end{equation}
The angular size $\theta$ of an object is related to its comoving linear size $\lambda$ through $\theta = \lambda/d_A(z)$. 
This justifies the definition of an object of known linear size as a \emph{standard ruler} for cosmology. In fact, knowing $\lambda$, 
we can use a measure of $\theta$ to get $d_A$ and make inferences on the cosmological parameters that determine its value through 
the integral in Eq. (\ref{eq:rAcurv}).

Another measure of distance is given by the luminosity distance $d_L(z)$, that relates the observed flux $F$ to the intrinsic luminosity $L$ of an object at redshift $z$:
\begin{equation}
d_L(z) \equiv \sqrt{\frac{L}{4\pi F}} = (1+z) d_A(z) \, .
\end{equation}
Similarly to what happened for the angular diameter distance, this allows to use \emph{standard candles} - objects of known intrinsic luminosity - as a 
mean to infer the values of cosmological parameters, after their flux has been measured.

\subsection{Neutrino mass parameters}

According to the standard theory of neutrino oscillations, the observed neutrino flavours $\nu_\alpha$ ($\alpha = e,\,\mu,\,\tau$)
are quantum superpositions of three mass eigenstates $\nu_i$ ($i=1,\,2,\,3$):
\begin{equation}
\left|\nu_\alpha\right\rangle = \sum_i U_{\alpha i}^* \left| \nu_i \right\rangle , \,
\end{equation}
where $U$ is the Pontecorvo-Maki-Nakagawa-Sasaka (PMNS) mixing matrix. The PMNS matrix
is parameterized by three mixing angles $\theta_{12}, \,\theta_{13}, \, \theta_{23}$, and three CP-violating phases: one Dirac, $\delta$, and two Majorana phases, $\alpha_{21}$ and $\alpha_{31}$. The Majorana phases 
are non-zero only if neutrinos are Majorana particles. They do not affect oscillation phenomena,
but enter lepton number-violating processes like $0\nu2\beta$ decay. The actual form of the PMNS matrix is:
\begin{equation}
U=
\begin{bmatrix}
   c_{12}c_{13}  & s_{12} c_{13}  & s_{13}e^{-i\delta} \\
   -s_{12}c_{23}-c_{12}s_{23}s_{13}e^{i\delta}  & c_{12}c_{23}-s_{12}s_{23}s_{13}e^{i\delta}  &s_{23}c_{13}\\
   s_{12}s_{23}-c_{12}c_{23}s_{13}e^{i\delta}  & -c_{12}s_{23}-s_{12}c_{23}s_{13}e^{i\delta}  &c_{23}c_{13}
\end{bmatrix}
\times\mathrm{diag}\left(1,\,e^{i\alpha_{21}/2},\, e^{i\alpha_{31}/2}\right) \, ,
\end{equation}
where $c_{ij}\equiv \cos\theta_{ij}$ and $s_{ij}\equiv \sin\theta_{ij}$.

In addition to the element of the mixing matrix, the other parameters of the neutrino sectors are the mass eigenvalues $m_i$ ($i=1,\,2,\,3$).
Oscillation experiments have measured with unprecedented accuracy the three mixing angles and
the two mass squared differences relevant for the solar and atmospheric transitions, namely
the solar splitting $\Delta m^2_\mathrm{sol} = \Delta m_{21}^2 \equiv m^2_2 - m^2_1 \simeq 7.6\times 10^{-5}\,\eV^2$,
and the atmospheric splitting $\Delta m^2_\mathrm{atm}= |\Delta m_{31}^2| \equiv |m^2_3 - m^2_1|\simeq 2.5\times 10^{-3}\,\eV^2$ (see~\cite{deSalas:2017kay,Capozzi:2017ipn,Esteban:2016qun} for a global fit of the neutrino mixing parameters and mass splittings).
We know, because of matter effects in the Sun, that, of the two eigenstates involved, the one with the smaller mass has the largest electron fraction.
By convention, we identify this with eigenstate ``1'', so that the solar splitting is positive. On the other hand, we do not know
the sign of the atmospheric mass splitting, so this leaves open two possibilities: the normal hierarchy (NH), where $\Delta m^2_{31} > 0$  
and $m_1 < m_2 < m_3$, or the inverted hierarchy, where $\Delta m^2_{31} < 0$ and $m_3 < m_1 < m_2$.

Oscillation experiments are unfortunately insensitive to the absolute scale of neutrino masses. In this review, we will mainly 
focus on cosmological observations as a probe of the absolute neutrino mass scale. To a very good approximation, cosmological
observables are mainly sensitive to the sum of neutrino masses $\sumnu$, defined simply as
\begin{equation}
\sumnu\equiv \sum_i m_i \, .
\label{eq:sumnu}
\end{equation}

Absolute neutrino masses can also be probed by laboratory experiments. These will be reviewed in more detail in Sec.~\ref{sec:lab},
where their complementarity with cosmology will be also discussed. For the moment, we just recall the definition of the
mass parameters probed by laboratory experiments. The effective (electron) neutrino mass $m_\beta$
\begin{equation}
m_\beta= \left( \sum_i \left|{U_{ei}}\right|^2 m_i^2 \right)^{1/2}\, ,
\label{eq:mb}
\end{equation}
can be constrained by kinematic measurements like those exploiting 
the $\beta$ decay of nuclei. The effective Majorana mass of the electron neutrino $\mbb$:
\begin{equation}
\mbb = \left|\sum_i U_{ei}^2 m_i \right|\, ,
\label{eq:mbb}
\end{equation}
can instead be probed by searching for $0\nu2\beta$ decay.

%As explained in more detail in Sec.~\ref{sec:lab}, 
%there are essentially three ways to probe the absolute neutrino masses. One is to measure the endpoint of the spectrum of electrons produced in the $\beta$-decay of nuclei. This is sensitive to the effective (electron) neutrino mass 
%\begin{equation}
%m_\beta= \left( \sum_i \left|{U_{ei}}\right|^2 m_i^2 \right)^{1/2}\, .
%\label{eq:mb}
%\end{equation}
%Current constraints from the $\beta$ decay of tritium are $m_\beta<2.05\,\eV$~\cite{Aseev:2011dq} and $m_\beta<2.3\,\eV$~\cite{Kraus:2004zw} at 95\% CL from the Troitsk and 
%Mainz experiments, respectively. Although they are not as sensitive as other probes, kinematic measurements like those based
%on $\beta$ decay are very robust as they only rely on energy conservation, and as such are model-independent.
%
%Another technique is to look for $0\nu2\beta$ decay, that allows to measure the effective Majorana mass of the electron neutrino $\mbb$:
%\begin{equation}
%\mbb = \left|\sum_i U_{ei}^2 m_i \right|\, .
%\label{eq:mbb}
%\end{equation}
%
%Finally, cosmological observations, the focus of this review, are mainly sensitive to the sum of neutrino masses $\sumnu$:

\subsection{The standard cosmological model}

Our best description of the Universe is currently provided by the spatially flat $\Lambda$CDM model with adiabatic, nearly scale-invariant
initial conditions for scalar perturbations. With the exception of some mild (at the $\sim 2 \sigma$ level) discrepancies that will be discussed in the
part devoted to observational limits, all the available data can be fit in this model, that in its simplest (``base'') version is described by just six parameters. 
In the base $\Lambda$CDM model, the Universe is spatially flat ($\Omega_k=0$), 
and the matter and radiation content is provided by cold dark matter, baryons, photons and neutrinos, while dark energy is in the form of a cosmological constant ($w=-1$). The energy density of photons is fixed by measurements of the CMB temperature, while neutrinos are assumed to be very light, usually fixing the sum of the masses to $\sumnu = 0.06\,\eV$, the minimum value allowed by oscillation experiments. In this 
way, the energy density of neutrinos is also fixed at all stages of the cosmological evolution (see Sec. \ref{sec:cosmonu}). From Eq. (\ref{eq:hubble}), and taking into account the flatness constraint, it is clear then that the background evolution in such a model is described by three parameters, for example\footnote{In the analysis of CMB data, the angle subtended by the sound horizon at recombination is normally used in place 
of $h$, as it is measured directly by CMB observations, see Sec \ref{sec:obscmb}} $h$, $\omega_c$ and $\omega_b$, with $\Omega_\Lambda$
given by $1-\Omega_m$. The initial scalar fluctuations are adiabatic and have a power-law, nearly scale invariant, spectrum, that is thus parameterized
by two parameters, an amplitude $A_s$ and a logarithimc slope $n_s-1$ (with $n_s = 1$ thus corresponding to scale invariance). Finally,
the optical depth to reionization $\tau$ parameterises the ionization history of the Universe. 

This simple, yet very successful, model can be extended in several ways. The extension that we will be most interested in, given the topic of this review,
is a one-parameter extension in which the sum of neutrino masses is considered as a free parameter. We call this seven-parameter model $\Lambda$CDM+$\sumnu$. This is also in some sense the best-motivated extension of $\Lambda$CDM, as we actually know from oscillation experiments that neutrinos have a mass, and from $\beta$-decay experiments that this can be as large as 2~eV. In addition to this minimal extension, we will also discuss how relaxing some of the assumptions of the $\Lambda$CDM model affects estimates of the neutrino mass. 
Among the possibilities that we will consider, there are those of varying the curvature ($\Omega_k$), the equation-of-state parameter of dark energy ($w$)
or the density of radiation in the early Universe ($\Neff$, defined in sec. \ref{sec:cosmonu}).

There are many relevant extensions to the $\Lambda$CDM model that however we will not consider here (or just mention briefly). The most important
one concerns the possibility of non-vanishing tensor perturbations, i.e. primordial gravitational waves, that, if detected,
would provide a smoking gun for inflation. This scenario is parameterized through an additional parameter, the tensor-to-scalar ratio $r$.
In the following, we will always assume $r=0$. In any case, this assumption will not affect the estimates reported here, as the effect of 
finite neutrino mass and of tensor modes on the cosmological observables are quite distinct. Similarly, we will not consider the possibility of non-adiabatic initial perturbations, nor of more
complicated initial spectra for the scalar perturbations, including those with a possible running of the scalar spectral index.

\subsection{Short thermal history}

Given that cosmological observables carry the imprint of different epochs in the history of the Universe, we find it useful to shortly recall
some relevant events taking place during the expansion, and their relation to the cosmological parameters. For our purposes, it is enough to start just when the temperature of the Universe 
was $T\sim 1\,\MeV$, i.e. around the time of Big Bang Nucleosynthesis (BBN) and neutrino decoupling. At these early times ($z\sim 10^{10}$), since matter and radiation densities scale as $(1+z)^3$ and $(1+z)^4$, respectively, the Universe is radiation-dominated.

\begin{itemize}
\item At $T\sim 1\,\MeV$ ($z\sim 10^{10}$), the active neutrinos decouple from the rest of the cosmological plasma. Before this time,
neutrinos were kept in equilibrium by weak interactions with electrons and positrons, that were in turn coupled electromagnetically to the photon bath. After this time, their mean free path becomes much larger the the Hubble length,
so they essentially move along geodesics, i.e., they free-stream. Shortly after neutrino decouple, electrons and positrons in the Universe annihilate, heating the photon-electron-baryon plasma, and, to a much lesser extent, the neutrino themselves (in the Sec. \ref{sec:cosmonu} we shall discuss in more detail the neutrino thermal history). After this time, the Universe can essentially be thought as composed of photons, electrons, protons and neutrons (either free, or, after BBN, bound together into the light nuclei), neutrinos, dark matter and dark energy.
\item Soon after, at $T\sim 0.1\,\MeV$, primordial nucleosynthesis starts, and nuclear reactions bind nucleons into light nuclei.
After this time, nearly all of the baryons in the Universe are in the form of $^1\mathrm{H}$ and $^4\mathrm{He}$ nuclei, with small traces of $^2\mathrm{H}$ and $^7\mathrm{Li}$. The yields of light
elements strongly depend on the density of baryons, on the density and energy spectrum of electron neutrinos and antineutrinos (as those set the equilibrium of the nuclear reactions
through which the nuclei are built) and on the total radiation density (as this sets the expansion rate at the time of nucleosynthesis).
\item As said above, at early times (high $z$) the Universe is radiation-dominated, given that the ratio of radiation to matter scales like $(1+z)$.
However, the radiation density decreases faster than that of matter, and, at some redshift $\zeq$, the matter and radiation contents of the Universe will be equal: $\rho_m(\zeq) = \rho_r(\zeq)$. This is called the epoch of matter-radiation equality, that marks the beginning of the matter-dominated era in the history of the Universe. From the scaling of the two densities, it is easy to see that $1+\zeq = \Omega_m/(\Omega_\gamma+\Omega_\nu)$ in a Universe with massless neutrinos (so that their density always scale as $(1+z)^4$; see Sec. \ref{sec:cosmonu} for further discussion on this point) . Given the current estimates of cosmological parameters, $\zeq \simeq 3400$~\cite{PlanckXIII}.
\item At $T\simeq 0.3\,\eV$, electrons and nuclei combine to form neutral hydrogen and helium, that are transparent to radiation. This recombination epoch thus roughly corresponds to the time of decoupling of radiation from matter. This is the time at which the cosmic microwave background (CMB) radiation is emitted. After decoupling, the CMB photons free-stream until the present time (with some caveats, see below). Most of the features that we observe in the CMB anisotropy pattern are created at this time. Given the current estimates of cosmological parameters, $\zrec \simeq 1090$~\cite{PlanckXIII}.
Note that in fact the temperature at recombination is basically fixed by thermodynamics, so once the present CMB temperature is determined through observations,
$\zrec = T(z=\zrec)/T(z=0)$  depends very weakly on the other cosmological parameters. 
\item Even if photons decoupled from matter shortly after recombination, the large photon-to-baryon ratio keeps baryons coupled to the photon bath for some time after that. The drag epoch $\zdrag$ is the time at which baryons stop feeling the photon drag.
A good fit to numerical results in a CDM cosmology is given by~\cite{Eisenstein:1997ik}
\begin{eqnarray}
\zdrag&=&1291\frac{(\omega_c+\omega_b)^{0.251}}{1+0.659(\omega_c+\omega_b)^{0.828}}[1+b_1(\omega_c+\omega_b)^{b_2}],\nonumber\\
b_1&=&0.313(\omega_c+\omega_b)^{-0.419}[1+0.607(\omega_c+\omega_b)^{0.674}],\nonumber\\
b_2&=&0.238(\omega_c+\omega_b)^{0.223}
\end{eqnarray}
Given the current estimates of cosmological parameters, $\zdrag \simeq 1060 $~\cite{PlanckXIII}.
\item For a long time after recombination, the Universe stays transparent to radiation. These are the so-called ``dark ages''. However, in the late history of the Universe, the neutral hydrogen gets ionized again due to UV emission of the first stars, that puts an end to the dark ages. This is called the reionization epoch. After reionization, the CMB photons are scattered again by the free electrons. Given the current estimates of cosmological parameters, $\zre \simeq 8$~\cite{PIPXLVI}.
\item At some point during the recent history of the Universe, that we denote with $z_\Lambda$, the energy content of the Universe starts to be dominated by the dark energy component. The end of matter domination, and the beginning of this DE domination is set by $\rho_{DE}(\zde) = \rho_m(\zde)$. 
For a cosmological constant ($w=-1$), $1+\zde = (\Omega_\Lambda/\Omega_m)^{1/3}$. Around this time, the cosmological expansion becomes accelerated.  

\end{itemize}

\subsection{Evolution of cosmic neutrinos \label{sec:cosmonu}}

In this section, we discuss the thermal history of cosmic neutrinos. 

As anticipated above, in the early Universe neutrinos are kept in equilibrium with the cosmological plasma by weak interactions.
The two competing factors that determine if equilibrium holds are the expansion rate, given by the Hubble parameter $H(z)$, and the
interaction rate $\Gamma(z) = n\langle\sigma v\rangle$, where $n$ is the number density of particles, $\sigma$ is the
interaction cross section, and $v$ is the velocity of particles (brackets indicate a thermal average). In fact, neutrino interactions become too weak to keep them in equilibrium once $\Gamma < H$.
The left-hand side of this inequality is set by the standard model of particle physics, as the interaction rate at a given temperature only depends on the cross-section for weak interactions, and thus, ultimately, on the value of the Fermi constant ($\sigma_w \sim G_F^2 T^2$). The right-hand side is instead set, through Eq. (\ref{eq:hubble}) by the total radiation density (the only relevant component at such early times): $H^2=(8\pi G/3) (\rho_\gamma+\rho_\nu)$. This is also fixed at any given temperature, in the framework of the minimal $\lcdm$ model, so that the temperature of neutrino decoupling, defined through $\Gamma(T_{\nu,\mathrm{dec}}) = H(T_{\nu,\mathrm{dec}})$ does not depend on 
any free parameter in the theory. A quite straightforward calculation shows that $T_{\nu,\mathrm{dec}} \simeq 1\,\MeV$~\cite{Kolb:1990vq}.

While they are in equilibrium, the phase-space distribution $f(p)$ of neutrinos is a Fermi-Dirac distribution\footnote{We are assuming a vanishing chemical potential for neutrinos and antineutrinos, i.e., a vanishing lepton asymmetry.}:
\begin{equation}
f(p,\, t) = \frac{1}{e^{p/T_\nu(t)}+1} \, , 
\label{eq:FD}
\end{equation}
where it has been taken into account that at $T\gtrsim 1\,\MeV$, the active neutrinos are certainly ultrarelativistic (i.e., $T_\nu \gg m_\nu$) and
thus $E(p) \simeq p$. The distribution does not depend on the spatial coordinate $\vec x$, nor on the direction of momentum $\hat p$, due to
the homogeneity and isotropy of the Universe. Before decoupling, the neutrino temperature $T_\nu$ is the common temperature of all the species in the cosmological plasma, that we denote generically with $T$, so that $T_\nu =T$. We recall that the temperature of the plasma evolves
according to $g_{*s}^{1/3} a T = \mathrm{const.}$, where $g_{*s}$ counts the effective number of relativistic degrees of freedom that are relevant 
for entropy \cite{Kolb:1990vq}.

Since decoupling happens while neutrinos are ultrarelativistic, it can be shown that, as a consequence of the Liouville theorem, 
the shape of the distribution function is preserved by the expansion. In other words, the distribution function still has the form Eq. (\ref{eq:FD}),
with an effective temperature $T_\nu(z)$ (that for the sake of simplicity we will continue to refer to as the neutrino temperature) 
that scales like $a^{-1}$ (i.e., $aT=\mathrm{const}$). We stress that this means that, when computing integrals over the distribution function,
one still neglects the mass term in the exponential of the Fermi-Dirac function, even at times when neutrinos are actually nonrelativistic.

Shortly after neutrino decouple, electrons and positrons annihilate and transfer their entropy to the rest of the plasma, but not to neutrinos. In other words, while the neutrino temperature scales like $a^{-1}$, the photon temperature scales like $a^{-1} g_{*s}^{-1/3}$, and thus decreases slightly more slowly during $e^+e^-$ annihilation, when $g_{*s}$ is decreasing. In fact, applying entropy conservation one finds that the ratio between the neutrino and photon temperatures after electron-positron annihilation is $T_\nu/T = (4/11)^{1/3}$. The photon temperature has been precisely determined by measuring 
the frequency spectrum of the CMB radiation: $T_0=(2.725\pm0.002)$~K~\cite{mather,fixen}, so that the present temperature of relic neutrinos should be $T_{\nu,0} \simeq 1.95$~K$\simeq 1.68\times10^{-4}\,\eV$.

The number density $n_\nu$ of a single neutrino species (including both neutrinos and their antiparticles) is thus given by:
\begin{equation}
n_\nu(T_\nu) = \frac{g}{(2\pi)^3}\int  \frac{d^3p}{e^{p/T_\nu}+1} = \frac{3\zeta(3)}{4\pi^2} T_\nu^3 \, ,
\label{eq:nnu}
\end{equation}
where $\zeta(3)$ is the Riemann zeta function of 3, and in the last equality we have taken into account that $g=2$ for neutrinos. This corresponds
to a present-day density of roughly $113$ particles$/\mathrm{cm}{^3}$.

The energy density of a single neutrino species is instead 
\begin{equation}
\rho_\nu(T_\nu) = \frac{g}{(2\pi)^3}\int  \frac{\sqrt{p^2+m^2}}{e^{p/T_\nu}+1} d^3p \, .
\end{equation}
This is the quantity that appears, among other things, in the right-hand side of the Friedmann equation (summed over all mass eigenstates). In the ultrarelativistic ($T_\nu\gg m$) and
nonrelativistic ($T_\nu\ll m$) limits,  the energy density takes simple analytic forms:
\begin{equation}
\rho_\nu(T_\nu) = \left\{ 
\begin{array}{ll}
\displaystyle{\frac{7\pi^2}{120} T_\nu^4} & \mathrm{(UR)}\\[0.2cm]
m_\nu n_\nu &\mathrm{(NR)}
\label{eq:rhonu}
\end{array}
\right.
\end{equation}
These scalings are consistent with the fact that one expects neutrinos to behave as pressureless matter, $\rho_\nu \propto (1+z)^3$, in the nonrelativistic regime, and as radiation, $\rho_\nu \propto (1+z)^4$, in the ultrarelativistic regime.

Given that the present-day neutrino temperature is fixed by measurements of the CMB temperature and by considerations of entropy conservation, it is clear from the above formulas how the present energy density of neutrinos depends only on one free parameter, namely the sum of neutrino masses $\sumnu$ defined in Eq. (\ref{eq:sumnu}).
%\begin{equation}
%\sumnu \equiv \sum_i m_i \, .
%\end{equation}
Introducing the total density parameter of massive neutrinos $\Omega_\nu \equiv \sum_{i} \rho_{\nu_i,0}/\rho_{\mathrm{crit},0}$, one easily finds from Eq. (\ref{eq:nnu}):
\begin{equation}
\Omega_\nu h^2 = \frac{\sumnu}{93.14 \, \eV} \, .
\label{eq:omeganu}
\end{equation}
where we have already included the effects of non-instantaneous neutrino decoupling, see below. In the instantaneous decoupling approximation, the quantity at denominator would be $94.2\,\eV$. 

On the other hand, the neutrino energy density in the early Universe only depends on the temperature, and thus it is completely fixed in the framework of 
$\lcdm$ model. Using the fact that for photons $\rho_\gamma = (\pi^2/15) T^4$, together with the relationship between the photon and neutrino temperatures, one can write for the total density in relativistic species in the early Universe, after $e^+e^-$ annihilation:
\begin{equation}
\rho_{\gamma + \nu} = \rho_\gamma\left[1 + \frac{7}{8}\left(\frac{4}{11}\right)^{4/3}N_\nu\right] \, ,
\end{equation}
where $N_\nu$ is the number of neutrino families. In the framework of the standard model of particle physics, considering the active neutrinos, one has $N_\nu=3$. However, the above formula slightly underestimates the total density at early times; the main reason is that neutrino are still weakly coupled to the plasma when $e^+e^-$ annihilation occurs, so that they share a small part of the entropy transfer. Moreover, finite temperature QED radiative corrections and flavor oscillations also play a role. This introduces nonthermal distortions at the subpercent level in the neutrino energy spectrum; the integrated effect is that at early times the combined energy densities of the three neutrino species are not exactly equal to $3\rho_\nu$, with $\rho_\nu$ given by the upper row of Eq. \ref{eq:rhonu}, but instead are given by $(3.046 \rho_\nu)$~\cite{Dolgov:2002wy,Mangano:2005cc}.
A recent improved calculation, including the full collision integrals for both the diagonal and off-diagonal elements of the neutrino density
matrix, has refined this value to $(3.045 \rho_\nu)$~\cite{deSalas:2016ztq}. It is then customary to introduce an effective number of neutrino families $\Neff$ and rewrite the energy density at early times as: 
\begin{equation}
\rho_{\gamma + \nu} = \rho_\gamma\left[1 + \frac{7}{8}\left(\frac{4}{11}\right)^{4/3}\Neff\right] \, .
\label{eq:Neff}
\end{equation}
In this review, we will consider $\Neff = 3.046$ as the ``standard'' value of this parameter in the $\Lambda$CDM model, and not
the more precise value found in Ref.~\cite{deSalas:2016ztq}, since most of the literature still makes use of the former value.
This does not make any difference, however, from the practical point of view, given the sensitivity of present and next-generation instruments. 

It is also customary to consider extensions of the minimal $\lcdm$ model in which one allows for the presence of additional light species in the early Universe (``dark radiation''). In this kind of extension, the total radiation density of the Universe is still given by the right-hand side
of Eq. (\ref{eq:Neff}), where now however $\Neff$ has become a free parameter. In other words, Eq. (\ref{eq:Neff}) becomes a \emph{definition} for $\Neff$,
that is, just a way to express the total energy density in radiation. The effect on the expansion history of this additional radiation component 
can be taken into account by the substitution
\begin{equation}
\Omega_\gamma \to \Omega_\gamma \left[1 + \frac{7}{8}\left(\frac{4}{11}\right)^{4/3}\Delta\Neff\right] 
\end{equation}
in the rhs of the Hubble equation (\ref{eq:hubble}), with $\Delta\Neff \equiv \Neff - 3.046$. Note that this substitution
fully captures the effect of the additional species only if this is exactly massless, and not just very light (as in the case of a light massive sterile neutrino, for example - see Sec. \ref{sec:beyond}).

It is often useful, to understand some of the effects that we will discuss in the following, to have a feeling for the time at which neutrinos of a given mass become nonrelativistic, or, thinking the other way around, for the mass of a neutrino that becomes nonrelativistic at a given redshift. The average momentum of neutrinos at a temperature $T_\nu$ is $\langle p \rangle = 3.15 T_\nu$. We take as the moment of transition from the relativistic to the nonrelativistic regime the time when $\langle p \rangle = m_\nu$. Then, using the fact that $T_\nu(z) = (4/11)^{1/3} T_0 (1+z) = 1.68\times 10^{-4} (1+z)\,\eV$, one has
\begin{equation}
1+z_\mathrm{nr} \simeq 1900 \left( \frac{m_\nu}{\eV} \right) \, .
\end{equation}
This relation can be used to show e.g. that neutrinos with mass $m \simeq 0.6\,\eV$ turn nonrelativistic at recombination. 
In the following, when discussing the effect of neutrino masses on the CMB anisotropies, we will assume that this is the case. Note however
that the actual statistical analyses from which bounds on neutrino masses are derived do not make such an assumption.
We also note that, given the current measurements of the neutrino mass differences, only the lightest mass eigenstate can still be relativistic today. Thus at least two out of the three active neutrinos become nonrelativistic at some time between recombination and the present.

We conclude this section with a clarification on the role of neutrinos in determining the redshift of matter-radiation equality. Given the present bounds on neutrino masses, we know that equality likely takes place when neutrino are relativistic. In fact, observations of the CMB anisotropies 
constrain $\zeq\simeq3400$, so that neutrinos with mass $m \simeq 1.8\,\eV$, just below the current bound from tritium beta-decay, turn nonrelativistic at equality. Thus, for masses sufficiently below the tritium bound, the total density of matter at those times is proportional to 
$\Omega_{c+b}$. The radiation density is instead provided by photons and by the relativistic neutrinos (and as such does not depend on the neutrino mass), plus any other light species present in the early Universe. So the redshift of equivalence is given by
\begin{equation}
1+\zeq = \frac{\Omega_c+\Omega_b}{\Omega_\gamma \left[1 + \frac{7}{8}\left(\frac{4}{11}\right)^{4/3}\Neff\right]} = \frac{\omega_c+\omega_b}{\omega_\gamma \left[1 + \frac{7}{8}\left(\frac{4}{11}\right)^{4/3}\Neff\right]} \, ,
\end{equation}
where the last equality makes it clear that, in the framework of the minimal $\lcdm$ model, the redshift of equivalence only depends
on the quantity $\omega_c+\omega_b$, since $\Neff$ is fixed and $\omega_\gamma$ is determined through observations (it is basically the CMB energy density).

%%%%%%%%

\section{Cosmological effects of neutrino masses \label{sec:eff}}

The impact of neutrino masses - and in general of neutrino properties - on the cosmological evolution
can be divided in two broad categories: \emph{background} effects, and \emph{perturbation} effects. 
The former class refers to modifications in the expansion history, i.e. in changes to the evolution
of the FLRW background. The latter class refers instead to modifications
in the evolution of perturbations in the gravitational potentials and in the different components of the cosmological fluid.
We shall now briefly review both classes; we refer the reader who is interested in a more detailed analysis to the excellent review
by Lesgourgues \& Pastor~\cite{Lesgourgues:2006nd}.

To start, we shall consider a spatially flat Universe, i.e. $\Omega_k=0$, in which dark energy is in the form of a cosmological constant
($w=-1$) and there are no extra radiation components ($\Neff=3.046$). Let us also consider a particular realization of this scenario, that we refer to
as our reference model, in which the sum of neutrino masses is very small; for definiteness, we can think that $\sumnu$ is equal to the
minimum value allowed by oscillations, $\sumnu=0.06\,\eV$. When needed, we will take the other parameters as fixed to their $\Lambda$CDM best-fit values from
Planck 2015 \cite{PlanckXIII} . Our aim is to understand what happens when we change the value of $\sumnu$.  Increasing the sum of neutrino masses $\sumnu$
will increase $\omega_\nu = \Omega_\nu h^2$ according to Eq. (\ref{eq:omeganu}). Remember that the sum of the density parameters $\sum_i \Omega_i = 1$; this constraint can be recast in the form:
\begin{equation}
\omega_c + \omega_b + \omega_\Lambda + \omega_\gamma + \omega_\nu+\omega_k = h^2 \, .
\end{equation}
Since $\omega_\gamma$ is constrained by observations, and $\omega_k$ is zero by assumption, we have four degrees of freedom
that we can use to compensate for the change in $\omega_\nu$, namely: increase $h$, or decrease any of $\omega_c$, $\omega_b$ or 
$\omega_\Lambda$. For the moment, for simplicity, we will not distinguish between baryons and cold dark matter, pretending that as nonrelativistic components they have the same effect on cosmological observables. This is of course not the case, but we will come back to this later.
Then we are left with three independent degrees of freedom that we can use to compensate for the change in $\omega_\nu$: $h$, $\omega_{b+c}$, and $\omega_\Lambda$. We prefer to use $\Omega_\Lambda$ in place of $\omega_\Lambda$, so that in the end our parameter basis for this discussion will be $\left\{h,\,\omega_{c+b},\,\Omega_\Lambda\right\}$.

The first option, increasing the present value of the Hubble constant while keeping $\Omega_\Lambda$ and $\omega_{b+c}$ has the effect of making 
the Hubble parameter at any given redshift after neutrinos become nonrelativistic larger with respect to the reference model. This can be understood 
by looking at Eq. (\ref{eq:hubble}), that we rewrite here in this particular case
\begin{align}
H(z)^2 = H_0^2 \Bigg[ & \left( \Omega_c + \Omega_b \right) (1+z)^3 + \Omega_\gamma (1+z)^4   +\Omega_{\Lambda}  + \frac{\rho_{\nu,\mathrm{tot}}(z)}{\rho_{\mathrm{crit},0}}\Bigg ] \, .
\label{eq:hubble2}
\end{align}
With respect to the reference model, the first three terms in the RHS are unchanged, while the fourth increases because $\Omega_\Lambda$ is fixed but $h$ is larger. The last term does not depend on $h$ (because the factor $H_0^2$ in front
of the square brackets cancel the one in the critical density) but yet increases because $\rho_\nu = \sumnu n_\nu$ is larger as long as neutrinos are in the nonrelativistic regime. On the other hand, before neutrino become nonrelativistic, $\rho_\nu$ is the same in the two models, and the change in the $\Omega_\Lambda h^2 $ term is irrilevant, because the DE density is only important at very low redshift. So we can conclude that at $z \gg z_\mathrm{nr}$, the two models share the same expansion history, while for $ z \lesssim z_\mathrm{nr}$ the model with ``large'' neutrino mass is always expanding faster (larger $H$), or equivalently, is always younger, at those redshifts. In terms of the length scales and of the distance measures introduced in Sec. \ref{sec:cosmobas}, 
it is easily seen that the causal and sound horizons at both equality and recombination (as well as at the drag epoch) are unchanged, because
the expansion history between $z=\infty$ and $z\simeq z_\mathrm{nr}$ is unchanged. On the other hand,
distances between us and objects at any redshift - for example, the angular diameter distance to recombination - are always smaller than in the reference model, because $H$ is always larger between $z\simeq z_\mathrm{nr}$ and $z=0$. $H$ increases with the extra neutrino density, so this effect increases with larger neutrino masses (and moreover, $z_\mathrm{nr}$ also gets larger for larger masses). Given this, we expect for example the angle subtended by the sound horizon at recombination, $\theta_s = r_s(\zrec)/d_A(\zrec)$ to become smaller when we increase $\sumnu$. We conclude this part of the discussion that in this case the redshift of equality $\zeq$ does not change, since $\omega_{b+c}$ is being kept constant, and neutrinos contribute to the radiation density at early times (see discussion at the end of the previous section).

If we instead choose to pursue the second option, i.e., we keep $h$ and $\Omega_\Lambda$ constant while lowering $\omega_{c+b}$,
we are again changing the expansion history, but this time on a different range of redshifts. In fact, when neutrinos are nonrelativistic,
the RHS of Eq.~(\ref{eq:hubble2}) is unchanged, because the changes in the present-day densities of neutrinos and nonrelativistic matter perfectly compensate; this continues to hold as long as both densities scale as $(1+z)^3$, i.e., roughly for $z<z_\mathrm{nr}$. On the other hand, at $z>z_\mathrm{nr}$ the
neutrino density is the same as in the reference model, while the matter density is smaller, so $H(z)$ is smaller as well. Finally, when the Universe is radiation dominated, the two models share again the same expansion history. Then in this scenario we change the expansion history, decreasing $H$, for $z_\mathrm{nr} \lesssim z \lesssim \zeq$. The sound horizon at recombination increases, and so does the angular diameter distance, so one cannot 
immediately guess how their ratio varies. However, a direct numerical calculation shows that, starting from the Planck best-fit model, the net effect is to increase $\theta_s$, meaning
that the sound horizon will subtend a larger angular scale on the sky when $\sumnu$ increases. For what concerns instead the redshift of matter-radiation equality, it is immediate to see that it decreases proportionally to $\omega_{c+b}$, i.e., equality happens later in the model with larger $\sumnu$.

Finally, when $\Omega_\Lambda$ is decreased, the main effect is to delay the onset of acceleration and make the matter-dominated era last longer. This has  some effect on the evolution of perturbations, as we shall see in the following. For what concerns the expansion history, since the model under consideration and the reference model only differ in the neutrino mass and in the DE density, they are identical when neutrinos are relativistic and DE is negligible, i.e., 
at $z > z_\mathrm{nr}$. For $z < z_\mathrm{nr}$, instead, starting as usual from Eq. (\ref{eq:hubble2}) one finds, with some little algebra, that $H(z)$ is always
larger in the model with smaller $\Omega_\Lambda$ and larger $\sumnu$. As in the previous case, both $r_s$ and $r_A$ at recombination vary in the same direction (decreasing in this case); the net effect is again that $\theta_s$ becomes larger with $\sumnu$. Also, since the matter density at early times is not changing in this case, the redshift of equivalence is the same in the two models.

We know comment briefly about $\omega_b$. One could choose to modify $\omega_b$ in place of $\omega_c$ in order to compensate for the
change in $\omega_\nu$. From the point of view of the background expansion, both choices are equivalent, since the baryon and cold dark matter density
only enter through their sum $\omega_{b+c}$ in the RHS of Eq. (\ref{eq:hubble2}). However, changing the baryon density also produces some peculiar effects, mainly related to the fact that i) it determines the BBN yields, and ii) it affects the evolution of photon perturbations prior to recombination. Thus the density of baryons is quite well constrained by the observed abundances of light elements and by the relative ratio between the heights of odd and even peaks in the CMB, (see Sec. \ref{sec:obscmb}) and there is little room for changing it without spoiling the agreement with observations.

Let us now turn to discuss the effects on the evolution of perturbations. Given that we have observational access to the fluctuations in the radiation and matter fields, it is useful to discuss separately these two components. The photon perturbations are sensitive to time variations in the gravitational potentials along the line of sight from us up to the last-scattering surface; this is the so-called \emph{integrated Sachs-Wolfe} (ISW) effect. The gravitational potentials are constant in a purely matter-dominated Universe, so that the observed ISW gets an early contribution right after recombination, when the radiation component is not yet negligible, and a late contribution, when the dark energy density begins to be important. Coming back to our previous discussion, it is clear to see how delaying the time of equality will increase the amount of early ISW, while anticipating dark energy domination will increase the late ISW, and viceversa. For what concerns matter inhomogeneities, a first effect is again related to the time of matter-radiation equality. Changing $\zeq$ affects the growth of perturbations, since most of the growth happens during the matter dominated era. Apart from that, a very peculiar effect is related to the clustering properties of neutrinos. In fact, while neutrinos are relativistic, they tend to \emph{free stream} out of overdense regions, damping out all perturbations below the horizon scale. The net effect is that neutrino clustering is exponentially suppressed below a certain critical scale, the free-streaming scale, that corresponds to the size of the horizon at the time of the transition from the ultrarelativistic to the nonrelativstic regime. If the transition happens during matter domination, this is given by:
\begin{equation}
k_\mathrm{fs} \simeq 0.018\, \Omega_m^{1/2} \left(\frac{m}{1\eV}\right)^{1/2}\, h \mathrm{Mpc}^{-1} \, .
\label{eq:kfs}
\end{equation}
On the contrary, above the free-streaming scale neutrinos cluster as dark matter and
baryons do. Thus, increasing the neutrino mass and consequently the neutrino energy density will suppress small-scale matter fluctuations relative to the large scales. It will also make small-scale perturbations in the other components grow slower, since neutrino do not source the gravitational potentials at those scales. It should also be noted that the free-streaming scale depends itself on the neutrino mass - specifically, heavier neutrinos will become nonrelativistic earlier and the free-streaming scale will be correspondingly smaller. Moreover, there is actually a free-streaming scale for each neutrino species, each depending on the individual neutrino mass. In principle one could think to go beyond observing just the small-scale suppression and try to access instead the scales around the nonrelativistic transition(s),
in order to get more leverage on the mass and perhaps also on the mass splitting. We shall see however in the following that this is not the case.

The suppression of matter fluctuations due to neutrino free-streaming also affects the path of photons coming from distant sources, since
those photons will be deflected by the gravitational potentials along the line of sight, resulting in a gravitational lensing effect. This is relevant for the CMB, as it modifies the anisotropy pattern by mixing photons that come from different directions. Another application of this effect, of particular importance for estimates of neutrino masses, is to use the distortions of the shape of distant galaxies due to lensing, to reconstruct the intervening matter distribution.

%%%%%%%%

\section{Cosmological observables \label{sec:obs}}

In this section we review the various cosmological observables, and explain how the effects described in the previous section propagate
to the observables.

\subsection{CMB anisotropies \label{sec:obscmb}}

The CMB consists of polarized photons that, for the most part, have been free-streaming from the time of recombination
to the present. The pattern of anisotropies in both temperature (i.e., intensity) and polarization thus encodes a wealth of information about the early Universe, down to $z=\zrec \simeq 1100$. Moreover, given that the propagation of photons from decoupling to the present is also affected by the cosmic environment, 
the CMB also has some sensitivity to physics at $z < \zrec$. Two relevant examples for the topic under consideration are the CMB sensitivity to the
redshift of reionization (because the CMB photons are re-scattered by the new population of free electrons) and to the integrated matter distribution along the line of sight (because clustering at low redshifts modifies the geodesics with respect to an unperturbed FLRW Universe, resulting in a gravitational lensing of the CMB, see next section). However, the CMB sensitivity to these processes is limited due to the fact these are integrated effects.

The information in the CMB anisotropies is encoded in the power spectrum coefficients $C^{TT}_{\ell}$, i.e., the coefficients of the expansion in Legendre polynomials of the two-point correlation function. In the case of the temperature angular fluctuations $\Delta T(\hat n)/\overline T$:
\begin{equation}
\left\langle\frac{\Delta T(\hat n)}{\overline T}\frac{\Delta T(\hat n')}{\overline T}\right\rangle = \sum_{\ell = 0}^{\infty}\frac{2\ell+1}{4\pi} C^{TT}_\ell P_\ell(\hat n \cdot\hat n') \, .
\end{equation}
For Gaussian fluctuations, all the information contained in the anisotropies can be compressed without loss in the two-point function,
or equivalently in its harmonic counterpart, the power spectrum. A similar expression holds for the polarization field and for its cross-correlation with temperature. In detail, the polarization field can be decomposed into two independent components, known as $E-$  (parity-even and curl-free) and $B-$ (parity-odd and divergence-free) modes. Given that, it is clear that we can build a total of six spectra 
$C_\ell^{XY}$ with $X,\,Y = T,\,E,\,B$; however, if parity is not violated in the early Universe, the $TB$ and $EB$ correlations are bound to vanish.
Let us also recall that, in linear perturbation theory, $B$ modes are not sourced by scalar fluctuations. Thus, in the framework of the standard inflationary paradigm, primordial $B$ modes can only be sourced in the presence of tensor modes, i.e., gravitational waves.

The shape of the observed power spectra is the result of the processes taking place in the primordial plasma around the time of recombination. In brief, 
in the early Universe, standing, temporally coherent acoustic waves set in the coupled baryon-photon fluid, as
a result of the opposite action of gravity and radiation pressure ~\cite{Hu:2001bc}. Once the photons decouple after hydrogen recombination, the waves are ``frozen'' and thus we observe a series of peaks and throughs in the temperature power spectrum, corresponding to oscillation modes that were caught at an extreme of compression or rarefaction (the peaks), or exactly in phase with the background (the throughs). The typical scale of the oscillations is set by the sound horizon at recombination $r_{s}(\zrec)$, i.e. the distance travelled by an acoustic wave from some very early time until recombination, see Eq. \ref{eq:rs}. The position of the first peak in the CMB spectrum is set by the value of this quantity and corresponds to a perturbation wavenumber that had exactly the time to fully compress once. The second peak corresponds to the mode with
half the wavelength, that had exactly the time to go through one full cycle of compression and rarefaction, and so on. Thus, smaller scales (larger multipoles) than the first peak correspond to scales that could go beyond one full compression, while larger scales (smaller multipoles) did not have the time to do so. In fact,
scales much above the sound horizon are effectively frozen to their initial conditions, provided by inflation. This picture is complicated a little bit by the presence of baryons, that shift the zero of the oscillations, introducing an asymmetry between even and odd peaks. Finally, the peak structure is further modulated by an exponential suppression, due to the Silk damping of photon perturbations (further related to the fact that the tight coupling approximation breaks down at very small scales). This description also holds for polarization pertubations, with some differences, like the fact that the polarization perturbations have opposite phase with respect to temperature perturbations. 

As noted above, the large-scale temperature fluctuations, that have entered the horizon very late and did not have time to evolve, trace the power spectrum of primordial fluctuations, supposedly generated during inflation. On the contrary, since there are no primordial polarization fluctuations, but those are instead generated at the time of recombination and then again at the time of reionization, the polarization spectra at large scales are expected to vanish, with the exception of the so-called reionization peak. 

We can now understand how the CMB power spectra are shaped by the cosmological parameters, in a minimal model with fixed neutrino mass. The overall amplitude and slope of the spectra
are determined by $A_s$ and $n_s$, since these set the initial conditions for the evolution of perturbations. The height of the first peak strongly depends on the redshift of equivalence $\zeq$ (that sets the enhancement in power due to the early ISW), while its position is determined by the angle $\theta_s$ subtended by the sound horizon at recombination. As we have discussed before, $\zeq$ and $\theta_s$ are in turn set by the values of the background densities and of the Hubble constant. 
The baryon density further affects the relative heights of odd and even peaks, and also the amount of damping at small scales, through its effect on the Silk scale.
The ratio of the densities of matter and dark energy fixes the redshift of dark energy domination and the amount of enhancement of large-scale power due to the
late ISW.
Finally, the optical depth at reionization $\tau$ induces an overall power suppression, proportional to $e^{-2\tau}$, in all spectra, at all but the largest scales.
This can be easily understood as the effect of the new scatterings effectively destroying the information about the fluctuation pattern at recombination, at the scales that are inside the horizon at reionization. Reionization also generates the large-scale peak in the polarization spectra, described above. Measuring the power spectra gives a precise determination of all these parameters:
simplifying a little bit, the overall amplitude and slope give $A_s e^{-2\tau}$ and $n_s$ (the latter especially if we can measure a large range of scales),
the ratio of the peak heights and the amount of small-scale damping fix $\omega_b$, while the position and height of the first peak
fix $\theta_s$ and $\zeq$, and thus $h$ and $\omega_{b+c}$. The polarization spectra further help in that they are sensitive to $\tau$ directly, allowing
to break the $A_s - \tau$ degeneracy, and that the peaks in polarization are sharper and thus allow, in principle, for a better determination of their position
\cite{Galli:2014kla}.
It is clear that adding one more degree of freedom to this picture, for example considering curvature, the equation of state parameter of dark energy, or
the neutrino mass as a free parameter, will introduce parameter degeneracies and degrade the constraints.

\begin{figure*}
\begin{center}
\includegraphics[width=0.8\textwidth]{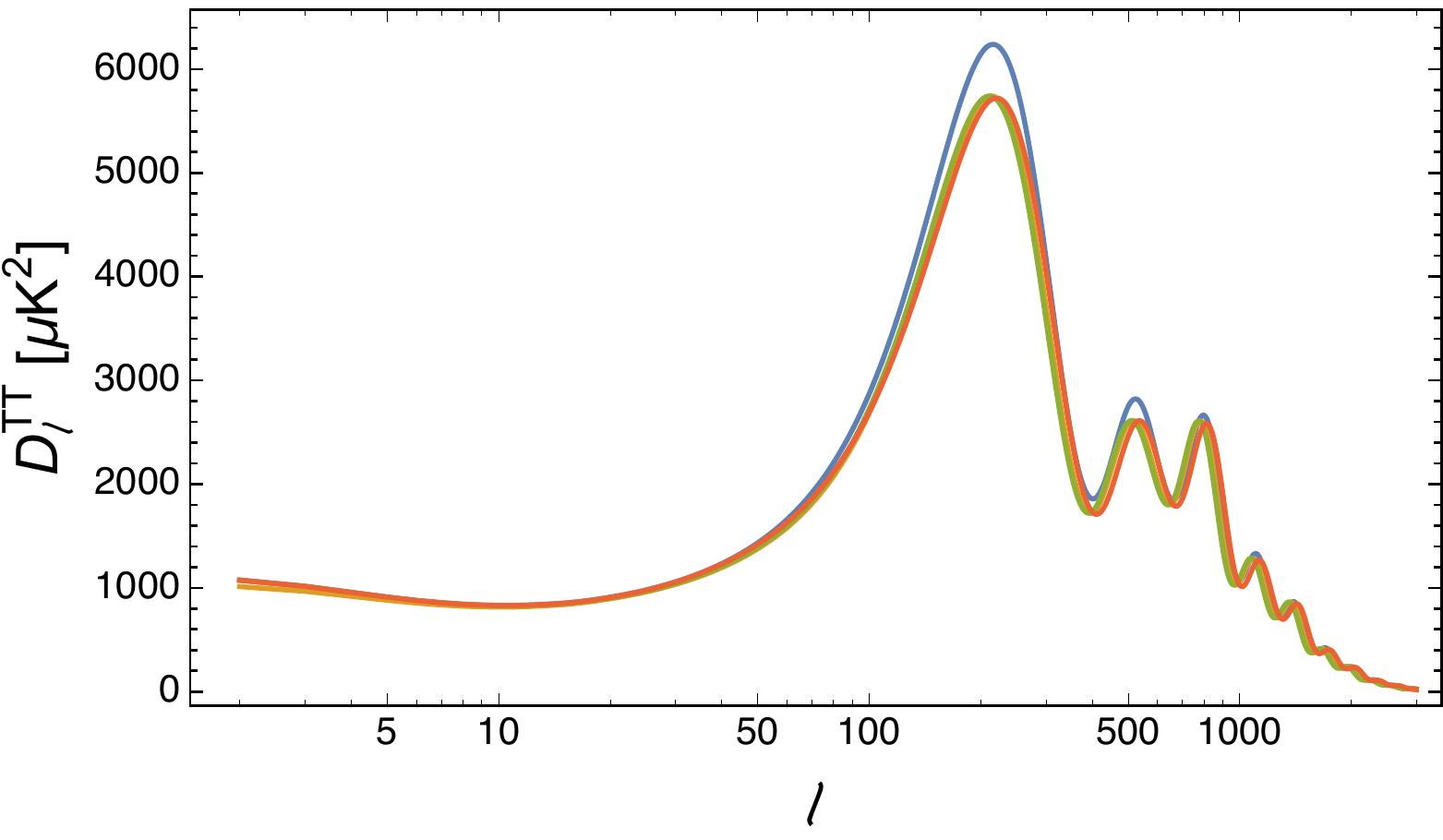}
\includegraphics[width=0.8\textwidth]{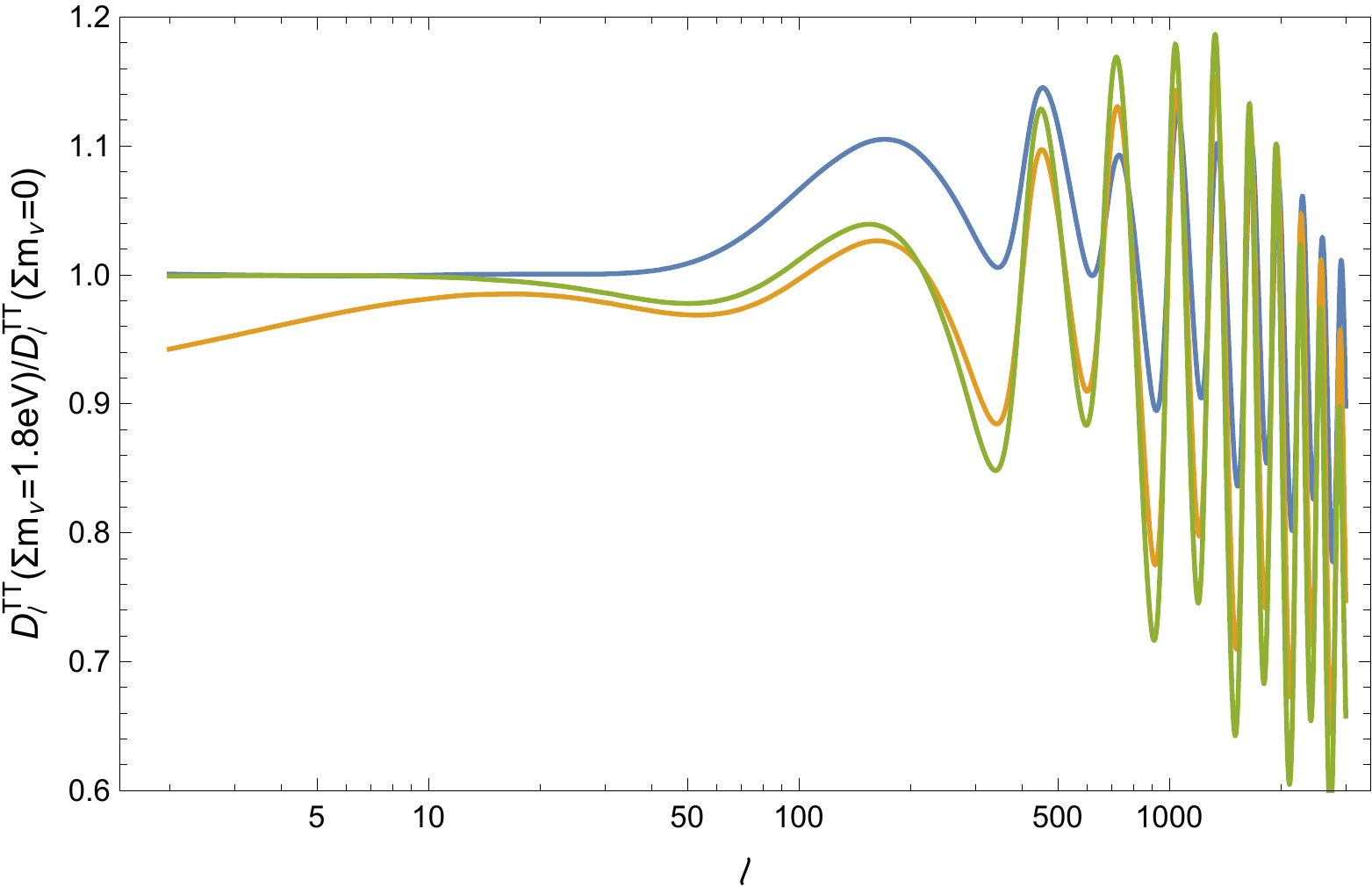}
\end{center}
\caption{\textit{Left:} CMB TT power spectra for different values of $\sumnu$. The quantity on the vertical axis is $D^{TT}_\ell \equiv \ell(\ell+1) C^{TT}_\ell/2\pi$. The red curve is a cosmological model with $\sumnu = 0.06\,\eV$ and all other parameters fixed to the Planck best-fit. The other curves
are for models with $\sumnu = 1.8\, \eV$, in which the curvature is kept vanishing by changing $h$ (green), $\Omega_\Lambda$ (yellow, always below the green apart from the lowest $\ell$'s) or $\omega_c$ (blue). The model in blue has a smaller $\zeq$ with respect to the reference; the models in yellow and green have a larger $\theta_s$; in addition, the yellow model also has a smaller $z_\mathrm{\Lambda}$. \textit{Right:} Ratio between the models with $\sumnu = 1.8\, \eV$ and the reference model\label{fig:cmb}.}
\end{figure*}

Coming to massive neutrinos, as we have discussed in Sec \ref{sec:eff}, there is a combination of the following effects when $\sumnu$, and consequently $\omega_\nu$, is increased, depending
on how we are changing the other parameters to keep $\Omega=1$: i) an increase in $\theta_s$; ii) a smaller $\zeq$  and thus a longer radiation-dominated era; iii) a delay of the time of dark energy domination. These changes will in turn result in: i) a shift towards the left of the  position of the peaks; ii) an increased height of the first peak, that is set by the amount of early ISW; iii) less power at the largest scales, due to the smaller amount of late ISW. 
A more quantitative assessment of this effects can be obtained using a Boltzmann code, like CAMB~\cite{CAMB} or CLASS~\cite{CLASS}, to get a theoretical prediction for the CMB power spectra in presence of massive neutrinos. These are shown in Fig. \ref{fig:cmb}. In the left panel we plot the unlensed CMB temperature power spectra for a reference model with $\sumnu = 0.06\,\eV$
($\omega_\nu\simeq 6.4 \times 10^{-4}$) (the other parameters are fixed to their best-fit values from Planck 2015) and for three models with $\sumnu = 1.8\,\eV$ ($\omega_\nu\simeq 1.9 \times 10^{-2}$), where either $h$, $\omega_c$ or $\Omega_\Lambda$ are changed to keep $\Omega = 1$. We consider three degenerate neutrinos with $m=0.6\,\eV$ each, so that they become nonrelativistic around recombination. We also show the ratio between these spectra and the reference spectrum in the right panel of the same figure.

 These imprints are in principle detectable in the CMB, especially the first two, since the position and height of the first peak are very well measured; much less so the redshift of DE domination, due to the large cosmic variance at small $\ell$'s. However, following the above discussion, it is quite easy to convince oneself that these effects can be pretty much canceled due to parameter degeneracies. In fact, simplifying again a little bit, in standard $\Lambda$CDM we use the very precise determinations of the height and position of the
first peak to determine $\theta_s$ and $\zeq$, and from them $\omega_{c+b}$ and $h$. In an extension with massive neutrinos, we still have the same determination of 
$\theta_s$ and $\zeq$, but we have to use them to fix three parameters, namely $\omega_{c+b}$, $h$ and $\omega_\nu$, so that the system is underdetermined.
One could argue that the amount of late ISW, as measured by the large-scale power, could be used to break this degeneracy, as it would provide a further
constraint on the matter density (given that the DE density is fixed by the flatness condition). Unfortunately, measurements of the large-scale CMB power are plagued
by large uncertainties, due to cosmic variance, so they are of little help in solving this degeneracy. Given the experimental uncertainties, then, it is clear that, when trying to fit a theory to the data, there will be a strong degeneracy direction corresponding to models having the same $\theta_s$ and $\zeq$, and thus with identical predictions for the first peak, and slightly different values of $z_\Lambda$, with very low statistical weight due to the large uncertainties in the corresponding region of the spectrum. In other words, the effects of neutrino masses will be effectively ``buried' in the small-$\ell$ plateau, where experimental uncertainties are large. The situation is even worse in extended models, for example if we allow the spatial curvature or the equation of state of dark energy to vary~\cite{Lesgourgues:2006nd}. In any case, the degeneracy between $h$ and $\omega_{c+b}$ is not completely exact, so that
the unlensed CMB still has some degree of sensitivity to neutrinos that were relativistic at recombination. For example, the Planck 2013 temperature data, 
in combination with high-resolution observations from ACT and SPT, were able to constrain $\sumnu < 1.1\,\eV$ after marginalizing over the effects
of lensing.

\subsubsection{Secondary anisotropies and the CMB Lensing \label{sec:cmblens}}
As observed above, in addition to the features that are generated at recombination, the so-called \emph{primary anisotropies}, the CMB spectra also carry the imprint of effects that are generated along the line of sight. We have already given an example of one of these \emph{secondary anisotropies} when we have mentioned the re-scattering of photons over free electrons at low redshift, that creates the distinctive ``reionization bump'' in the low-$\ell$ region of the polarization spectra.  Another important secondary anisotropy is the gravitational lensing of the CMB (see~\cite{LewisChallinor, Smith:2006nk}): photon paths are distorted by the presence of matter inhomogeneities along the line of sight. In the context of General Relativity, the deflection angle $\alpha$ for a CMB photon is 
\begin{equation}
\alpha=-2\int_0^{\chi_{*}} \der \chi \frac{f_K(\chi_{*}-\chi)}{f_K(\chi_{*}) f_K(\chi)} \nabla \Psi(\chi \textbf{n},\eta_0-\chi)
\end{equation}\label{eq:alphalens}
where $\chi_*$ is the comoving distance to the last scattering surface, $f_K(\chi)$ is the angular-diameter distance (Eq. \ref{eq:rAcurv}) thought 
as a function of the comoving distance, $\Psi$ is the gravitational potential, $\eta_0-\chi$ is the conformal time at which the photon was along the direction \textbf{n}. If we then define the lensing potential as
\begin{equation}
\phi(\hat{\textbf{n}})\equiv-2\int_0^{\chi_{*}} \der \chi \frac{f_K(\chi_{*}-\chi)}{f_K(\chi_{*}) f_K(\chi)}\Psi(\chi \textbf{n},\eta_0-\chi) \, ,
\label{eq:lenspotential}
\end{equation}
it is straightforward to see that the deflection angle is the gradient of the lensing potential, $\alpha=\nabla\phi$. From the harmonic expansion of the lensing potential, we can build an angular power spectrum\footnote{We are assuming that the lensing field is isotropic.} as $<\phi_{\ell m}\phi^{*}_{\ell'm'}>\equiv\delta_{\ell\ell'}\delta_{mm'}C^{\phi\phi}_\ell$. The lensing power spectrum $C_\ell^{\phi\phi}$ is therefore proportional to the integral along the line of sight of the power spectrum of the gravitational potential $P_\Psi$, which in turn can be expressed in terms of the power spectrum of matter fluctuations $P_m$ (see the next section for its definition).

The net effect of lensing on the CMB is that photons coming from different directions are mixed, somehow ``blurring'' the anisotropy pattern. This effect is mainly sourced by inhomogeneities at $z<5$ and has a typical angular scale of $2.5'$. In the power spectra, this translates in a several percent level smoothing of the primary peak structure ($\ell\gtrsim1000$), while the lensing effect becomes dominant at $\ell \gtrsim 3000$. We stress that lensing only alters the spatial distribution of CMB fluctuations, while leaving the total variance unchanged. Lensing, being a non-linear effect, creates some amount of nongaussianity in the anisotropy pattern. Thus, other than through its indirect effect on the temperature and polarization power spectra (i.e. on the two-point correlation functions), lensing can be detected and measured by looking at higher order correlations, in particular at the four-point correlation function. In fact, in such a way it has been possible to directly measure the power spectrum $C_\ell^{\phi\phi}$ of the lensing potential $\phi$. Another consequence of the nonlinear nature of lensing is that it is able to source ``spurious'' $B$ modes by converting some of the power in $E$ polarization, thus effectively creating $B$ polarization also in the absence of a primordial component of this kind. The latter effect represents an additional tool to enable the reconstruction of the lensing potential, especially for future CMB surveys. An alternative reconstruction technique is based on the possibility to cross-correlate the CMB signal with tracers of large-scale structures, such as Cosmic Infrared Background (CIB) maps, therefore leading to an ``external'' reconstruction~\cite{Hanson:2013hsb} (opposite to the ``internal'' reconstruction performed with the use of CMB-based only estimators~\cite{HuOkamoto,OkamotoHu}). 

The lensing power spectrum basically carries information about the integrated distribution of matter along the line of sight. Given the peculiar effect of neutrino free-streaming on the evolution of matter fluctuations, CMB lensing offers an important handle for estimates of neutrino masses. 
Since a larger neutrino mass implies a larger neutrino density and less clustering on small scales, because of neutrino free-streaming, 
the overall effect of larger neutrino masses is to decrease lensing. In the temperature and polarization power spectra, the result is that
the peaks and throughs at high-$\ell$'s are sharper.  Concerning the shape of the lensing power spectrum, for light massive neutrinos the net effect is a rescaling of power at intermediate and small scales (see e.g.~\cite{Kaplinghat}). Thus the lensing power spectrum is a powerful tool for constraining $\sumnu$ and will probably drive even better constraints on $\sumnu$ in future. In fact, it is almost free from systematics coming from poorly understood astrophysical effects, it directly probes the (integral over the line of sight of the) distribution of the total matter fluctuations (as opposed to what galaxy surveys do, as we will see in the next section) at scales that are still in the linear regime.

Given a cosmological model, it is quite straightforward, using again CAMB or CLASS, to get a theoretical prediction for the lensing power spectrum, as well as for the lensing BB power spectrum. Note that non-linear corrections (see next section for further details) to the lensing potential are important in this case to get accurate large-scale BB spectrum coefficients~\cite{LewisChallinor}. Additional corrections that take into account modifications to the CMB photon emission angle due to lensing can further modify the large-scale lensing BB spectrum~\cite{Lewis:2017ans}.

\subsection{Large scale structures}

\subsubsection{Clustering}

The clustering of matter at large scales is another powerful probe of cosmology. The clustering
can be described in terms of the two-point correlation function, or, equivalently, of the power spectrum of matter density fluctuations:
\begin{equation}
\left\langle \delta_m (\vec k,\, z) \delta_m(\vec k',\,z) \right \rangle = P_m(k,z) \delta^{(3)} \left(\vec k - \vec k'\right) \, ,
\end{equation}
where $\delta_m(\vec k,\,z)$ is the Fourier transform of the matter density perturbation at redshift $z$. Note that,
contrarily to the CMB, that we are bound to observe at a single redshift (that of recombination), the matter power spectrum can,
in principle, be measured at different times in the cosmic history, thus allowing for a tomographic analysis.

As for the CMB, the large-scale (small $k$'s) part of the power spectrum traces the primordial fluctuations generated during inflation, while
smaller scales reflects the processing taking place after a given perturbation wavenumber enters the horizon. A relevant distinction in this regard is whether a given mode enters the horizon before or after matter-radiation equality. Since subhorizon perturbations grow faster during matter domination, the matter power spectrum shows a turning point at a characteristic scale, corresponding to the horizon at $\zeq$. Given that perturbations grow less efficiently also during DE domination, increasing $z_\Lambda$ produces a suppression in the power spectrum. Also, increasing $h$ will make the horizon at a given redshift smaller; so the mode $k$ that is entering the horizon at that redshift will be larger.

\begin{figure*}
\begin{center}
\includegraphics[width=0.9\textwidth]{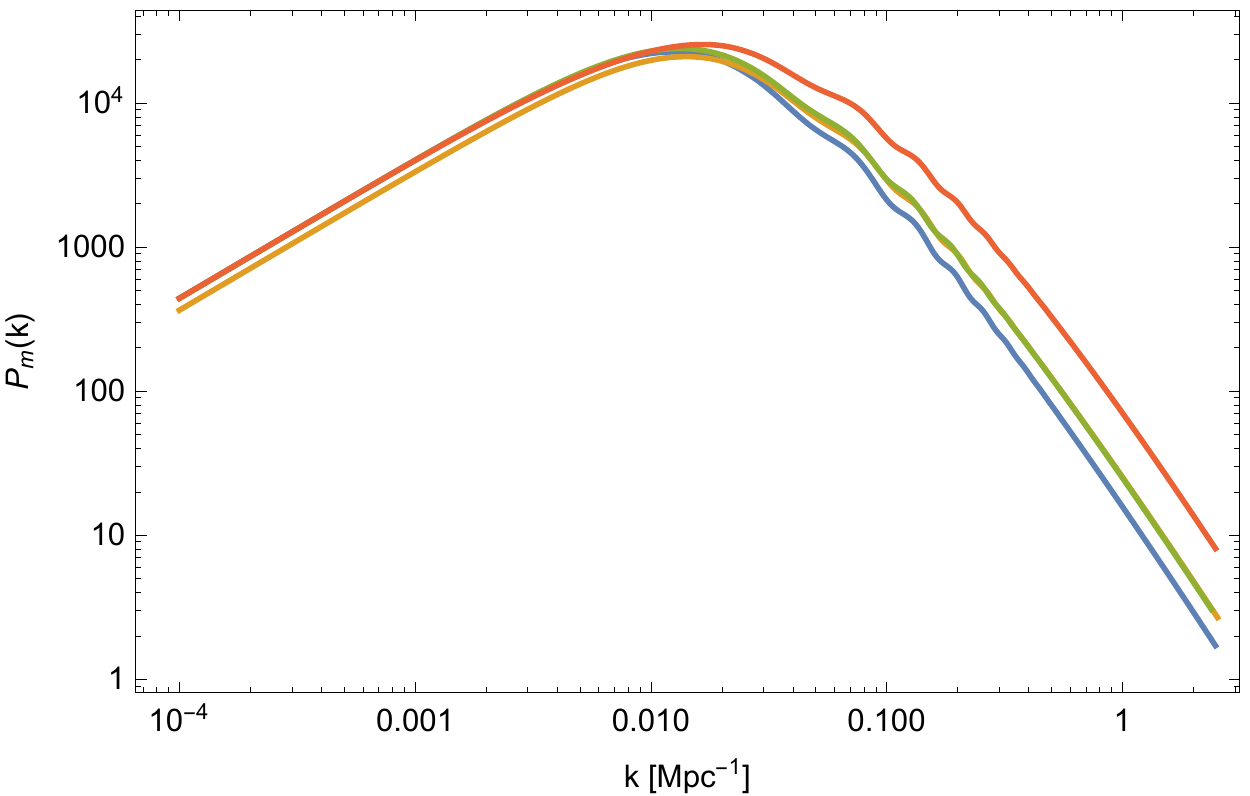}
\end{center}
\caption{Total matter power spectrum $P_m$ for the same models shown in Fig. \ref{fig:cmb}}\label{fig:pk}
\end{figure*}

Varying the sum of neutrino masses has some indirect effects on the shape of matter power spectrum, related to induced changes in background quantities, similarly to what happens for the CMB.
As explained in Sec. \ref{sec:eff}, increasing $\sumnu$ while keeping the Universe flat has to be compensated by changing (a combination of) $\omega_m$, $\Omega_\Lambda$ or $h$. 
This will in turn result in a shift of the turning point and/or in a change in the global normalization of the spectrum. This can be seen in Fig. \ref{fig:pk}, 
where we show the matter power spectra for the same models considered when discussing the background effects of neutrino masses on the CMB.

As it is for the CMB, these effects can be partly canceled due to parameter degeneracies. 
Neutrinos, however, have also a peculiar effect on the evolution of matter perturbations. This is due to the fact that neutrinos possess large thermal velocities for a considerable part of the cosmic history, so they can free-stream out of overdense regions, effectively canceling perturbations on small scales. In particular, one can define the free-streaming length at time $t$ as the distance that neutrinos can travel from decoupling until $t$. The comoving free-streaming length reaches a maximum at the time of the nonrelativistic transition. This corresponds to a critical wavenumber $k_\mathrm{fs}$, given in equation (\ref{eq:kfs}) for transitions happening during matter-domination, above which perturbations in the neutrino component are erased.

A first consequence of neutrino free-streaming is that, below the free-streaming scale, there is a smaller amount of matter that can cluster. This results
in an overall suppression of the power spectrum at small scales, with respect to the neutrinoless case. Secondly, subhorizon perturbations in the non-relativistic (i.e., cold dark matter and baryons) components 
grow more slowly. In fact, while in a perfectly matter-dominated Universe, the gravitational potential is constant and the matter perturbation grows linearly with the scale factor, $\delta_m \propto a$, in a mixed matter-radiation Universe the gravitational potential decays slowly inside the horizon. Below the free-streaming scale, neutrinos effectively behave as radiation; then in the limit in which the neutrino fraction $f_\nu = \Omega_\nu/\Omega_m$  is small, one has for $k\gg k_\mathrm{fs}$
\begin{equation}
\delta_m(k\gg k_\mathrm{fs}) \propto a^{1-(3/5)f_\nu} \, ,
\end{equation}
while $\delta_m \propto a$ for $k \ll k_\mathrm{fs}$.
These two effects can be qualitatively understood as follows: if one considers a volume with linear size well below the free-streaming scale, this region will resemble 
a Universe with a smaller $\Omega_m$ and a larger radiation-to-matter fraction than the ``actual'' (i.e., averaged over a very large volume) values. This yields
a smaller overall normalization of the spectrum, as well as a larger radiation damping; the two effects combine to damp the matter perturbations inside the region. So,
looking again at the full power spectrum, the net effect is that, in the presence of free-streaming neutrinos, power at small-scales is suppressed with respect to the case of no neutrinos.
A useful approximation is $P_m(k\gg k_\mathrm{fs},\,f_\nu)/P_m(k\gg k_\mathrm{fs},\, f_\nu =0) \simeq 1 -8 f_\nu$ at $z=0$ \cite{Hu:1997mj}.

It is useful to stress that since $f_\nu$ is linear in  $\sumnu$, we have the somehow counterintuitive result that the effects of free-streaming 
are more evident for \emph{heavier}, and thus colder, neutrinos. The reason is simply that the asymptotic suppression of the spectrum depends 
only on the total energy density of neutrinos, as this determines the different amount of non-relativistic matter between small and large scales. 

Until now, we have somehow ignored the role of baryons in shaping the matter power spectrum. In fact, on scales that enter the horizon after $\zdrag$,
the baryons are effectively collisionless and behave exactly like cold dark matter. On the other hand, baryon perturbations at smaller scales, entering the horizon before
$\zdrag$ exhibit acoustic oscillations due to the coupling with photons. This causes the appearance of an oscillatory structure in the matter power spectrum.
These wiggles in $P_m(k)$, that go under the name of \emph{baryon acoustic oscillations} (BAO), have a characteristic frequency, related to the value of the sound horizon at $\zdrag$.
Thus they can serve as a standard ruler and can be used very effectively in order to constrain the expansion history.

In more detail, the acoustic oscillations that set up in the primordial Universe produce a sharp feature in the two-point correlation function of luminous matter at the scale of the
sound horizon evaluated at the drag epoch, $r_s(z_d) \equiv r_d$; this sharp feature translates in (damped) oscillations in the Fourier transform of the two-point correlation function, i.e., the power spectrum.
Measuring the BAO feature at redshift $z$ allows in principle to separately constrain the combination $d_A(z)/r_d$, for measurements in the transverse direction with respect to the line of sight, or $r_d H(z)$ for measurements along the line of sight. An isotropic analysis instead measures, approximately, the ratio between the combination 
\begin{equation}
d_V(z) = \left[ \frac{z d_A^2(z)}{H(z)}\right]^{1/3} \, ,
\end{equation}
called the volume-averaged distance, and the sound horizon $r_d$. Given that the value of the sound horizon is well constrained by CMB observations, measuring the BAO features, possibly 
at different redshifts, allows to directly constrain the expansion history, as probed by the evolution of the angular diameter distance $d_A(z)$ and of the Hubble function $H(z)$, 
or of their average $d_V(z)$. In particular, it is straightforward to see that BAO measurements put tight constraints on the $\Omega_m-H_0 r_d$ plane, along a different degeneracy direction that it is instead probed by CMB~\cite{AddisonBAO,Aubourg}. Therefore, when estimating neutrino masses, the addition of BAO constraints to CMB data helps breaking the parameter degeneracies discussed in the previous section, yielding in general tighter constraints on this quantity.

The linear matter power spectrum for a given cosmological model can be computed using a Boltzmann solver. However, comparison with observations is complicated by the nonlinear evolution of cosmic structures. Note that both CAMB and CLASS are able to handle non-linearities in the evolution of cosmological perturbations with the inclusion of non-linear corrections from the Halofit model~\cite{Halofit} calibrated over numerical simulations. In particular, for cosmological models with massive neutrinos, the preferred prescription is detailed in~\cite{BirdHalofit}.

From the observational point of view, $P_m(k, z)$ can be probed in different ways. In galaxy surveys, the 3-D spatial distribution of galaxies is measured, allowing to measure the two-point correlation function and to obtain an estimate of the power spectrum of galaxies $P_g(k,z)$. Since in this case one is measuring the distribution of luminous matter only, and not of all matter (including dark matter), this does not necessarily coincide with the quantity for which we have a theoretical prediction, i.e., $P_m$; in other words, galaxies are a biased tracer of the matter distribution. To take this into account, one relates the two quantities through a bias $b(k,\,z)$:
\begin{equation}
P_g(k,\,z) = b^2(k,\,z) P_m(k,\,z) \,.
\end{equation}
The bias is in general both a function of redshift and scale. If it is approximated as a scale-independent factor, then the presence of the bias only
amounts to an overall rescaling of the matter power spectrum (at a given redshift). In this case, one marginalizes over the amplitude of the
matter spectrum, effectively only using the information contained in its shape. A scale-independent bias is considered to be a safe approximation for the larger scales: as an example, for Luminous Red Galaxies sampled at an efficient redshift of $0.5$ (roughly corresponding to the CMASS sample of the SDSS III-BOSS survey), a scale-independent bias is a good approximation up to $k\lesssim 0.2\, h$~Mpc$^{-1}$~\cite{Okumura}. On the other hand, scale-dependent features are expected to appear on smaller scales.
In this case, the bias can still be described using a few ``nuisance'' parameters, that are then marginalized over. In any case the exact functional form of the bias function, the range of scales considered, as well as prior assumptions on the bias parameters, are delicate issues that should be treated carefully.
An additional complication arises from the fact that massive neutrinos themselves induce a scale-dependence feature in the bias parameter, due to the scale-dependent growth of structures in cosmologies with massive neutrinos~\cite{Loverde,Castorina}. 

It has to be mentioned that, at any given redshift, there exist a certain scale $k_\mathrm{NL}$ below which the density contrast approaches the limit $\delta~\sim1$. In this regime, the evolution of cosmic structures cannot be completely captured by a linear theory of perturbations. The modelling of structures in the non-linear regime relies on numerical N-body simulations that must take into account the astrophysical and hydrodynamical processes at play at those scales. The level of complexity of N-body simulations has been increasing over the years,
%\footnote{See e.g. the EAGLE project: \url{http://icc.dur.ac.uk/Eagle/index.php}, the APOSTLE collaboration \url{http://blogs.helsinki.fi/sawala/the-apostle-collaboration/}, the Illustris project \url{http://www.illustris-project.org}, the Millennium Run \url{http://wwwmpa.mpa-garching.mpg.de/millennium/}.}, 
so that the physical processes included in the simulations and the final results are much closer to the observations than they used to be at the beginning. A recent example is given by the MassiveNuS simulations~\cite{Liu:2017now}, based on the Gadget-2 code~\cite{Springel:2005mi} modified to include the effects of massive neutrinos, and the nuCONCEPT simulations~\cite{Dakin:2017idt}.\footnote{Prescriptions for the matter power spectrum in the non-linear regime are also provided by the Halofit model~\cite{BirdHalofit}, the Coyote Universe emulator~\cite{Heitmann:2013bra}, the semi-analytical approach of PINOCCHIO~\cite{Rizzo:2016mdr} and additional methods referenced in~\cite{Rizzo:2016mdr}}. Nevertheless, the uncertainties related to the non-linear evolutions of cosmological structures are still higher than those affecting the linear theory, therefore reducing the constraining power coming from the inclusion of those scales in cosmological analysis. In fact, the conservative choice of not including measurements at $k<k_\mathrm{NL}$ is usually made when performing cosmological analysis. It is easy to understand that the scale entering the non-linear regime is smaller for higher redshifts. 

Additional probes of $P_m$ are measurements of Lyman-$\alpha$ ($\Lya$) forests  and 21-cm fluctuations (see e.g.~\cite{Weinberg:2003eg} and \cite{Furlanetto} for reviews). Although they are promising avenues since they can probe the matter distribution at higher redshifts and smaller scales than those usually accessible with typical galaxy samples, they still have to reach the level of maturity required to take full advantage of their constraining power. 
The observation of high-redshift ($z\sim2$) quasars and in particular the measurement of their flux provides a powerful tool for cosmological studies. Indeed, the absorption of the $\Lya$ emission from quasars by the intervening intergalactic medium -- an observational feature known as ``$\Lya$ forest'' -- constitutes a tracer of the total matter density field at higher redshifts and smaller scales than those usually probed by galaxy surveys. Similarly to what is done for galaxy samples, one can compute a correlation function of the measured flux variation, or equivalently its power spectrum $P_{\Lya}$. The latter is again proportional to the total $P_m$ via a bias parameter $b_{\Lya}$. The $\Lya$ bias factor is in general different from the galaxy bias, as each tracer of the underlying total matter distribution exhibits its own characteristics. The $\Lya$ forest is ideally a powerful cosmological tool, being able to access high redshift. Therefore, at fixed scale $k$, the physics governing the $\Lya$ spectrum is much closer to the linear regime than that related to the galaxy power spectrum. Furthermore, the redshift window probed by $\Lya$ is complementary to that probed by traditional galaxy surveys, in a sense that at higher redshift the relative impact of dark energy on the cosmic inventory is much less. However, a reliable description of the astrophysics at play in the intergalactic medium is essential for deriving the theoretical model for the $\Lya$ absorption features along the line of sight. This description heavily depends on hydrodynamical simulations that reproduce the behaviour of baryonic gas and on poorly known details of the reionization history. In addition, uncertainties in the theory of non-linear physics of the intergalactic medium at small scales can play a non-negligible role.

Finally, another tracer of the total matter fluctuations is represented by fluctuations in the 21-cm signal. The 21-cm line is due to the forbidden transition of neutral hydrogen (HI) between the two hyperfine levels of the ground state (spin flip) of the hydrogen atom. The observational technique resides in the possibility to measure the brightness temperature relative to the CMB temperature. Fluctuations in the 21-cm brightness are related to fluctuations in HI (or equivalently to the fraction of free electrons $x_e$), which in turn trace the matter fluctuations. Therefore, one can infer $P_m$ observationally by measuring the power spectrum of 21-cm fluctuations $P_\mathrm{21-cm}$. Apart from the technological challenges associated with the detection of the 21-cm signal, the main source of systematics come from the difficulties to separate the faint 21-cm signal from the much brighter foreground contamination, mostly due to synchrotron emission from our own galaxy.

\subsubsection{Cluster abundances}
The variation of the number of galaxy clusters of a certain mass $M$ with redshift $\der N(z,M)/\der z$ is also a valid source of information about the evolution of the late time Universe (see e.g.~\cite{Carlstrom:2002na} for a review). The expected number of clusters to be observed in a given redshift window is an integral over the redshift bin of the quantity
\begin{equation}
\frac{\der N}{\der z}=\int \der \Omega \int \der M \hat{\chi} \frac{\der N}{\der M \der z \der \Omega} 
\label{eq:clusterN}
\end{equation}
where $\Omega$ is the solid angle, $\hat{\chi}$ is the so-called completeness of the survey (a measure of the probability that the survey will detect a cluster of a given mass $M$ at a given redshift $z$) and $\frac{\der N}{\der M} (z,M)$ is the mass function giving the number of clusters per unit volume. The latter can be predicted once a cosmological model has been specified. The quantity in Eq.(\ref{eq:clusterN}) is  thus directly sensitive to the matter density $\Omega_m$ and to the current amplitude of matter overdensities, usually parametrized in terms of $\sigma_8$, the variance of matter fluctuations within a sphere of $8\,h^{-1}\mathrm{Mpc}$. As a result, this probe can be highly beneficial for putting bounds on $\sumnu$.

Extended catalogues of galaxy clusters have been published in the last decade by the Atacama Cosmology Telescope (ACT)~\cite{Hilton:2017gal,Hasselfield:2013wf}, the South Pole Telescope (SPT)~\cite{deHaan:2016qvy} and the Planck~\cite{PlanckXXIV} collaborations. CMB experiments are in fact able to perform searches for galaxy clusters by looking for the thermal Sunyaev-Zeldovich (SZ) effect, the characteristic upward shift in frequency of the CMB signal induced by the inverse-Compton scattering of CMB photons off the hot gas in clusters. The redshift of cluster candidates is identified with follow-up observations, whereas their mass is usually inferred with X-ray observations or, more recently, calibrated through weak lensing. Regardless of how it is calibrated, the determination of the cluster mass is the largest source of uncertainty for the cluster count analysis, due to possibly imprecise assumptions about the dynamical state of the cluster and/or survey systematics. A common way to factorise the uncertainties related to the mass calibration is to introduce a mass bias parameter that relates the true cluster mass to the mass inferred with observations.

\subsubsection{Weak lensing}
The weak gravitational lensing effect is the deflection of the light emitted by a source galaxy caused by the foreground large-scale mass distribution (lens).  The shape of the source galaxy therefore appears as distorted, i.e. it acquires an apparent ellipticity. The cosmic shear is the weak lensing effect of all the galaxies along the line of sight (see e.g.~\cite{Kilbinger} for a review). Weak lensing surveys offer the possibility to directly test the distribution of intervening matter at low redshifts, thus providing a powerful tool to investigate the late-time evolution of the Universe. By correlating the apparent shapes of source galaxies at different redshifts, one can compute the shear field $\gamma(\hat{n},z)$ as a function of the angular position $\hat{n}$ and redshift $z$. The shear field is usually decomposed in two components: the curl-free $E$-modes and the divergence-free $B$-modes. It can be shown that, in absence of systematics, the $B$-modes are expected to vanish, whereas the power spectrum of the $E$-modes is equivalent to the lensing power spectrum $C^{\phi\phi}(\ell)$. The integrated lensing potential has been defined in Eq. (\ref{eq:lenspotential}) for a source located at recombination. The corresponding expression for a source at a generic redshift $z$ can
be obtained simply by substituting $\chi_*$ with the comoving distance of the source.

Thus, the power spectrum of the lensing potential -- which is due to intervening matter along the line of sight -- is recovered from the measurements of the lensing-induced ellipticity of background galaxies; in a similar way, the lensing power spectrum is recovered from the redistribution of CMB photons due to the forming structures along the line of sight. As we have seen in Sec. \ref{sec:cmblens}, the spectrum of the lensing potential is a function of the matter power spectrum integrated along the line of sight. Therefore, it carries information about the distribution and growth of structures, representing a powerful tool for constraining $\sumnu$. It should be mentioned that the observed shear signal $\gamma_\mathrm{obs}$ is a biased tracer of the true shear $\gamma_\mathrm{true}$. This effect, mostly due to noise in the pixels when galaxy ellipticity is measured, is usually taken into account by introducing a multiplicative bias $m$ that relates $\gamma_\mathrm{true}$ and $\gamma_\mathrm{obs}$: $\gamma_\mathrm{obs}=(1+m)\gamma_\mathrm{true}+c$, where c is the additional noise bias~\cite{Heymans}. 

In addition, the shear signal can be cross-correlated with the angular distribution of foreground (lens) galaxies (the so-called \textit{galaxy-shear} or \textit{galaxy-galaxy lensing} cross-correlation). This cross-correlation is a powerful way to overcome the limitations induced in the galaxy-galaxy auto-correlation by the unknown galaxy bias. Indeed, the galaxy-galaxy lensing is basically a cross-correlation between the galaxy field and the total matter fluctuation field. Measurements of the galaxy-galaxy lensing cross spectrum can therefore help determine the form of the bias.

Cosmological constraints from weak lensing survey are often summarized in terms of bounds on $\Omega_m$ and $\sigma_8$. As an additional probe of the large scale structure in the Universe, weak lensing can be profitably used to constrain $\sumnu$.

\subsection{Supernovae Ia and direct measurements of the Hubble constant }
Measurements of the distance-redshift relation of Supernovae Ia (SNIa) have provided the compelling evidence of the accelerated Universe~\cite{Riess:1998cb,Perlmutter:1998np}. SNIa are produced in binary stellar systems in which one of the stars is a white dwarf. Accreting matter from its companion, the white dwarf explodes once it reaches the Chandrasekhar mass limit. Therefore, SNIa are standard candles, because their absolute magnitude can be theoretically inferred from models of stellar evolution. A comparison between the absolute magnitude and the apparent luminosity yields an estimate of their luminosity distance $d_L(z)$. The expected value of $d_L$ in turn depends on the underlying cosmological model. The constraints coming from SNIa in the $\Omega_m-\Omega_\Lambda$ plane are orthogonal to those obtained from CMB. As a result, the combination of the two probes is extremely efficient in breaking the degeneracy between the two parameters. For this reason, SNIa are very useful for constraining models of dark energy and/or arbitrary curvature. Nonetheless, constraints on $\sumnu$ can benefit from the use of SNIa data, thanks to the improved bounds on $\Omega_m$. 

As already discussed, the effect of light massive neutrinos on the background evolution of the universe can be also compensated by a change in the value of the Hubble constant $H_0$. Therefore, it is clear that any direct measurements of $H_0$ can be highly beneficial for putting bounds on $\sumnu$. Direct measurements only rely on local distance indicators (i.e. redshift $z\ll$1), therefore they are little or not-at-all sensitive to changes in the underlying cosmological model. In contrast, indirect estimates from high-redshift probes, such as primary CMB, can suffer from model dependency.

Direct measurements of $H_0$ are based on the geometric distance calibration of nearby Cepheids luminosity-period relation and the subsequent calibration of SNIa over Cepheids observed in the same SNIa galaxy hosts (see e.g.~\cite{Riess} and references therein). The goal is to connect the precise geometric distances measured in the nearby universe (usually referred to as ``anchors'') with the distant SNIa magnitude-redshift relation in order to extract the estimate of $H_0$. The main systematics are of course related to the calibration procedure. Further improvements on the precision of direct measurements of $H_0$ are expected to come once the precise parallaxes measurements from the Gaia satellite will be available.

Local measurements of $H_0$ are not directly sensitive to $\sumnu$. Besides, their results, in combination with cosmological probes, can break the degeneracy between cosmological parameters and improve constraints on $\sumnu$. The main example is in fact the possibility to break the strong (inverse) degeneracy between $H_0$ and $\sumnu$ that affects CMB constraints.

Indirect estimates of $H_0$ can be obtained from CMB and BAO measurements. We have already seen in Sec.~\ref{sec:obs} that the position and amplitude of the first acoustic peak in the CMB spectrum depends on $H_0$ in combination with other parameters. In addition, we shall mention that, once the BAO are calibrated with the precise determination of $r_d$ from CMB, measurements of $d_A/r_d$ and $H r_d$ (or $d_V/r_d$) yields bounds on $H_0$ that are competitive with CMB estimates and direct measurements. 

We finally mention an additional independent measurement of $H_0$. The detection of gravitational wave (GW) signals emitted by merging compact objects (standard sirens) in combination with the observation of an electromagnetic counterpart has been proposed as a standard siren~\cite{Holz:2005df,Chen:2017rfc}. The GW waveform reconstruction allows for a determination of the luminosity distance to the source. Precise determinations of the source localization can lead to percent accuracy in the luminosity distance estimation. The observation of the electromagnetic counterpart of the GW event is then essential to determine the redshift to the source. The full combination of distance-redshift pair can finally be employed to constrain $H_0$. In the absence of the detection of an electromagnetic counterpart, methods to infer the redshift from the GW event have been proposed, see e.g.~\cite{DelPozzo:2011yh}.

\subsection{Summary of the effects of neutrino masses \label{sec:effsum}}

Before moving to report the current observational constraints, we find it useful to summarize the constraining power of different cosmological observables 
with respect to the neutrino mass. The discussion is somehow qualitative, also given the high-level complexity of the cosmological models. The purpose is also to underline the importance of combining different cosmological probes.

We start from the CMB. For the present discussion, it is useful to consider separately the information coming from the unlensed CMB (i.e. the primary CMB plus all the secondary effect with the exclusion of lensing) and that coming from the weak lensing of CMB photons. For what concerns the former, the sensitivity of the unlensed CMB to neutrino masses is somehow limited. This is mainly due to a geometrical degeneracy between $h$ and $\omega_\nu$ thanks to which one can simultaneously change the two parameters (decreasing $h$ and increasing $\omega_\nu$) to keep $\theta_s$ constant, thus preserving the position of the first peak, with only limited changes to other part of the spectrum (especially changes in the low-$\ell$ region , where the sensitivity is limited by cosmic variance, induced by variations in $\Omega_\Lambda$). The height of the first peak is preserved by keeping $\omega_c$ fixed. Having access to the information contained in the CMB lensing, either 
through its effect on the temperature and polarization power spectra, or through a direct estimation of the lensing power spectrum, helps because $\sumnu$ also affects the matter distribution and then the amplitude of the lensing potential at small scales. This helps breaking the degeneracy described above.

\begin{figure*}
\begin{center}
\includegraphics[width=0.8\textwidth]{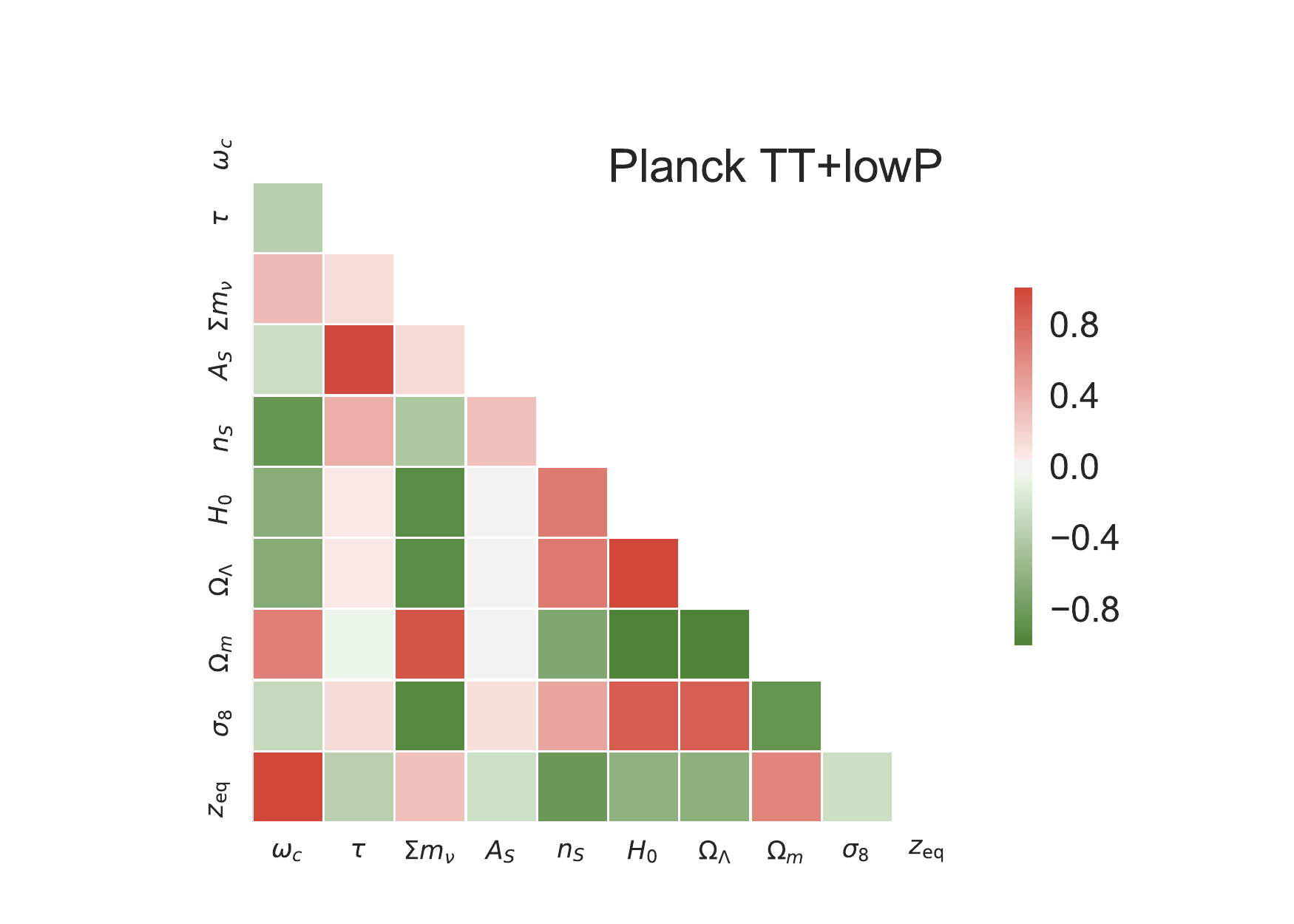}
\includegraphics[width=0.8\textwidth]{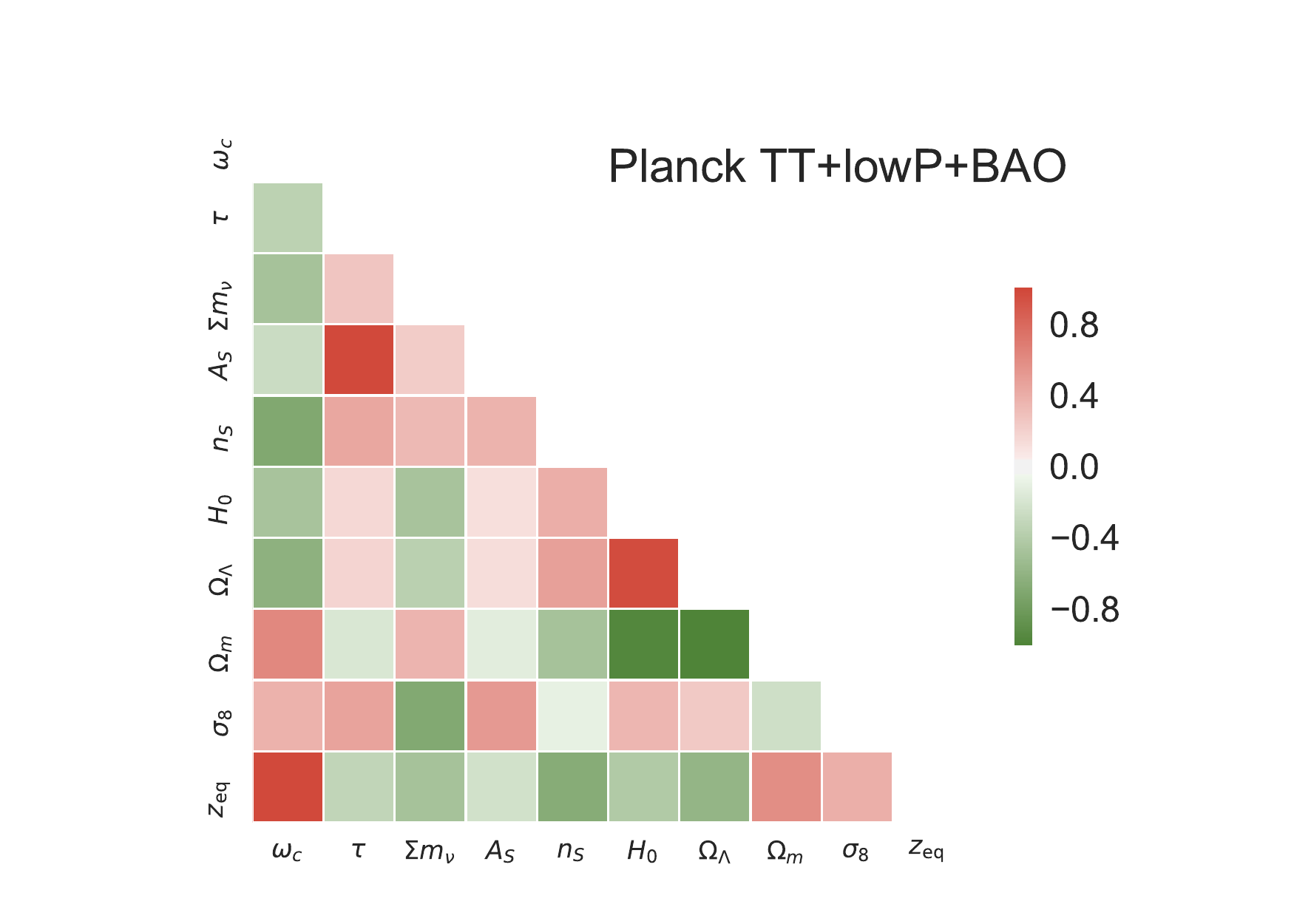}
\end{center}
\caption{Correlation matrices of a selection of cosmological parameters for the combinations of Planck TT+lowP (upper panel) and Planck TT+lowP+BAO (lower panel). See Sec.~\ref{sec:currcmb} for the description of these datasets. The darker the color shade, the stronger the degeneracy between the corresponding parameter pair. In both panels, the third row and the third column correspond to the correlation coefficients between $\sumnu$ and the remaining cosmological parameters. From the comparison between the two panels, it is clear that the inclusion of BAO data helps reduce the degeneracy between parameters (see e.g. the correlation between $\sumnu$ and $H_0$, $\Omega_\Lambda$); in a few cases, in fact, the inclusion of BAO reverts the degeneracy (see e.g. the correlation between $\sumnu$ and $n_s$).}\label{fig:corr}
\end{figure*}

To illustrate this point, in the upper panel of Fig. \ref{fig:corr} we show the parameter correlations derived by an analysis of the Planck observations of the temperature, over a wide range of scale, and large-scale polarization anisotropies. We remember that this dataset contains some information about lensing through the high-$\ell$ part of the temperature power spectrum. The negative degeneracy between $\sumnu$ and
$H_0$ is particularly evident. Given that $\omega_c$ and $\omega_b$ are both measured quite well from the CMB, this also translates into a strong degeneracy with $\Omega_m = (\omega_c +\omega_b)/h^2$ and $\Omega_\Lambda = 1- \Omega_m$. Among the other parameters, one
can notice mild correlations with $A_s$ and $\tau$. These are related to the small-scale effects related to the increased lensing in models
with larger $\sumnu$. 
The overall amplitude of the spectrum $A_s e^{-2\tau}$ is very
precisely determined by CMB observations. On the other hand, the lensing amplitude depends on $A_s$ but not on $\tau$. So, the 
lensing amplitude can be kept constant by increasing both $A_s$ and $\omega_\nu$. At this point $\tau$ has to be increased as well to preserve the
scalar amplitude $A_s e^{-2\tau}$.

Geometric measurements, like those coming from BAO, SNIa or direct measurements of $H_0$, greatly help solving the geometrical degeneracy 
between $H_0$ and $\sumnu$. This is evident by comparing the $(H_0, \, \sumnu)$  square in the lower panel of Fig. \ref{fig:corr}, where we
show parameter correlations from an analysis of the same dataset as above, with the addition of BAO data, with the corresponding square in the upper panel. Measurements
of large scale structures, and especially those that are directly sensitive to the total matter distribution at small scales, are very helpful, in that 
on the one hand they allow to further constrain $\Omega_m$, $A_s$ and $n_s$ and thus reduce degeneracies with these parameters;
on the other hand, they allow to probe the regime in which neutrino free-streaming is important. Finally, it is also clear that a precise measurement of $\tau$ from a CMB experiment that is sensitive to the large-scale 
polarization (meaning that it can access a large fraction of the sky) will be highly beneficial.

We have focused our attention to the $\Lambda$CDM+$\sumnu$ model. In extended dark energy models (as well as modified gravity models), for example for arbitrary equations of state of the dark energy fluid, the degeneracy between $\sumnu$ and $\Omega_\Lambda$ is amplified. Both massive neutrinos and dark energy-modified gravity affect the late time evolution of the universe, so that the individual effects on cosmological observables (mostly structures) can be reciprocally cancelled.

%%%%%%%%

\section{Current observational constraints on $\sumnu$}\label{sec:current}

In this section we report current constraints on $\sumnu$ from cosmological and astrophysical observations. These constraints are also
summarized in Tab.~\ref{tab:current} for the reader's convenience. Unless otherwise stated,
the results are obtained in the framework of a minimal one-parameter extension of the $\Lambda$CDM model with varying neutrino mass, dubbed $\Lambda$CDM$+\sumnu$, in which the three mass eigenstates are degenerate ($m_i=\sumnu/3$). Given the sensitivity of current experiments, 
the degenerate approximation is appropriate. See Sec.~\ref{sec:hierarchy} for a more detailed discussion on this point.

\subsection{CMB \label{sec:currcmb}}
CMB observations are probably the most mature cosmological measurements. The frequency spectrum is known with great accuracy~\cite{fixen}. Measurements of the power spectrum of CMB anisotropies in temperature are cosmic-variance limited down to very small scales ($\ell\sim1500$) and the quality of current CMB data in polarization is already good enough to tighten constraints on cosmological parameters ~\cite{PlanckXIII,Louis:2016ahn,Henning:2017nuy,Ade:2017uvt,Array:2015xqh}. The next generation of CMB experiments will further improve our knowledge of CMB polarization anisotropies~\cite{Essinger-Hileman:2014pja,S4,SO,SA,Delabrouille:2017rct}. The main systematics involved in CMB measurements are due to foreground contamination (atmospheric, galactic, extragalactic), calibration uncertainties and spurious effects induced by an imprecise knowledge of the instrument (see e.g.~\cite{shimon,polocalc,Rosset:2004jj,hwp,dust} for a sample list of references). 

The tightest constraints on $\sumnu$ from a single experiment come from the measurements of the Planck satellite~\cite{PlanckXIII}. In the context of a one-parameter extension of the $\lcdm$ cosmological background, the state of the art after the 2015 data release was as follows. The combination of the measurements of the CMB temperature anisotropies up to the multipole $\ell\simeq2500$ (hereafter, ``Planck TT'') and the large scale ($\ell<30$) polarization anisotropies (hereafter ``lowP'') leads to an upper bound of $\sumnu<0.72\,\eV$ at 95\% CL. The inclusion of the small scale ($\ell\ge30$) polarization measurements (which we globally label as ``Planck TE,EE'') provides a tighter upper bound of $\sumnu<0.49\,\eV$ at 95\% CL. This latter bound should be regarded as less conservative, as a small level of residual systematics could still affect the small scale polarization data. 

The Planck collaboration also provides the most significant measurements of the CMB lensing potential power spectrum for the multipole range $40<L<400$ (labeled as ``lensing'')~\cite{PlanckXV}. When this dataset is included in the analysis, the constraints on $\sumnu$ become: $\sumnu<0.68\,\eV$ for Planck TT+lowP+lensing and $\sumnu<0.59\,\eV$ for Planck TT,TE,EE+lowP+lensing~\cite{PlanckXIII}. When combining the lensing reconstruction data from Planck with the measurements of the CMB power spectra, it should be kept in mind that CMB power spectra as measured by Planck prefer a slightly higher lensing amplitude than that estimated with the lensing reconstruction. As a result, the bounds on $\sumnu$ obtained by their combination have less weight for smaller values of $\sumnu$ than the corresponding bounds obtained from CMB power spectra only. Nevertheless, higher values of $\sumnu$ are still disfavoured. 

In 2016, new estimates of the reionization optical depth $\tau$ have been published by the Planck collaboration~\cite{PIPXLVI}, obtained from the analysis of the high frequency CMB maps, in 2015 still affected by unexplained systematics effects at large scales. The estimated 68\% credible interval for $\tau$ coming from the $EE-$only low-$\ell$ data is $\tau=0.055\pm0.009$. This estimate is lower than the corresponding interval obtained in 2015 from the analysis of the low frequency maps ($\tau=0.067\pm0.023$), though the two estimates are well in agreement with each other. The lower value of $\tau$ has an impact on the constraints on $\sumnu$, due to the degeneracy between the optical depth and the amplitude of primordial perturbation $A_S$, as they together fix the normalization amplitude $A_S\,e^{-2\tau}$. A lower $\tau$ implies a lower $A_S$ and thus a lower lensing amplitude, leaving less 
room for large values of $\sumnu$ (that would further reduce lensing). If the ``lowP'' dataset is replaced by the new estimate of $\tau$ (labeled as ``SimLow''), the bounds improve as follows: $\sumnu<0.59\,\eV$ for Planck TT+SimLow and $\sumnu<0.34\,\eV$ for Planck TT,TE,EE+SimLow~\cite{PIPXLVI}.

\subsection{Large-Scale Structure Data}
Although the CMB is an extremely powerful dataset, multiple degeneracies between cosmological parameters limit the constraining power on $\sumnu$ from CMB only, as seen in Sec~\ref{sec:effsum}.
Measurements of the large scale structures (LSS) can help solving these degeneracies. LSS surveys map the distribution and clustering properties of matter at later times (or equivalently at lower redshift) than those accessible with CMB data and are directly sensitive to cosmological parameters that CMB data can only constrain indirectly, such as the total matter abundance at late times (see e.g.~\cite{Weinberg} for a review). In this section, we gather constraints on $\sumnu$ from different LSS probes alone and in combination with CMB data.

\subsubsection{Baryon acoustic oscillations and the full shape of the matter power spectrum from the clustering of galaxies}
BAO measurements, obtained by mapping the distribution of matter at relatively low redshifts ($z<3$) if compared to the redshifts relevant for CMB, constrain the geometry of the expanding universe, providing estimates of the comoving angular diameter distance $d_A(z)$ and the Hubble parameter $H(z)$ at different redshifts (or an angle-averaged combination of the two parameters, $d_V(z)=[zd_A^2(z)/H(z)]^{1/3}$~). Therefore, BAO constrain cosmological parameters which are relevant for the late-time history of the Universe, helping break the degeneracy between those parameters and $\sumnu$.

BAO extraction techniques rely on the ability to localise the peak of the two-point correlation function of some tracer of the baryon density, or equivalently the locations of the acoustic peaks in the matter power spectrum, thus neglecting the information coming from the broad band shape of the matter power spectrum itself. In principle, the full shape (FS) of the matter power spectrum is a valuable source of information about clustering properties of the different constituents of the universe and their reciprocal interactions. In particular, full shape measurements of the power spectrum also provide estimates of the growth of structures at low redshifts through the anisotropies induced by the redshift-space distortions (RSD), usually encoded in the parameter $f(z)\sigma_8(z)$, where $f(z)$ is the logarithmic growth rate and $\sigma_8(z)$ is the normalization amplitude of fluctuations at a given redshift in terms of rms fluctuations in a $8 h^{-1}$~Mpc sphere.

In 2016, the final galaxy clustering data from the Baryon Oscillation Spectroscopic Survey (BOSS) were released, as part of the Sloan Digital Sky Survey (SDSS) III~\footnote{Recently, the DES collaboration has reported a 4\% measurement of the angular diameter distance from the distribution of galaxies to redshift z=1~\cite{Abbott:2017wcz}. Cosmological constraints are derived in the LCDM framework, with $\sumnu$ fixed to the minimal value of $0.06$~eV. Therefore, no bounds on $\sumnu$ have been extracted from the BAO measurements from DES yet.}. Joint consensus constraints on $d_A(z)$, $H(z)$ and $f(z)\sigma_8(z)$ from BAO and FS measurements at three different effective redshifts ($z_\mathrm{eff}=0.38,0.51,0.61$) are employed to derive constraints on $\sumnu$\footnote{Note that the authors follow the assumption that all the mass is carried by only one of three neutrino species, i.e. $m_1=\sumnu,\, m_{2,3}=0\,\eV$, instead of the more widely used fully-degenerate approximation of $m_i=\sumnu/3,\,i=1,2,3$ for each of the three neutrino species.} in combination with Planck TT,TE,EE+lowP~\cite{BOSSDR12}. The 95\% upper bound is $\sumnu<0.16\,\eV$. When relaxing the constraining power coming from CMB weak lensing (through the rescaling of the lensing potential with the lensing amplitude $A_L$) and the RSD (through the rescaling of the $f\sigma_8$ parameter with the amplitude $A_{f\sigma_8}$), the bound degrades up to $\sumnu<0.25\,\eV$. 

When using the FS measurements, it has to be noted that the constraining power of this dataset is highly reduced if one considers that 1) the majority of the pieces of information encoded in the FS usually comes from the small-scale region of the power spectrum, where the still imprecisely known non-linearities play a non-negligible role; 2) the exact shape and scale-dependence of the bias $b$ between the observed galaxy clustering and the underlying total matter distribution is still debated. Therefore, it is useful to disentangle BAO and FS measurements, to gauge the relative importance of the two measurements in constraining $\sumnu$. For a thorough comparison between the constraining power of the two datasets, we refer the reader to \cite{Vagnozzi} (see also~\cite{Cuesta:2015iho,Hamann} for analyses using older data), where the authors focus on recent BAO measurements and FS measurements. Here, we summarise the conclusion of the paper: \textit{``The analysis method commonly adopted [for FS measurements] results in their constraining power still being less powerful than that of the extracted BAO signal''}. 

\subsubsection{Weak lensing}

The most recent weak lensing datasets have been released by the Kilo-Degree Survey (KiDS~\cite{Kohlinger:2017sxk,Hildebrandt:2016iqg}) and the Dark Energy Survey (DES~\cite{DES,Krause}). It is interesting to note that all of the aforementioned datasets provide results in terms of cosmological parameters which are slightly in tension with the corresponding estimates coming from CMB data (which we remind is a high-redshift probe). In particular, the values of $\Omega_m$ and $S_8=\sigma_8(\Omega_m/0.3)^{0.5}$ inferred from weak lensing data are lower than the best fit obtained with CMB data. The significance of this tension is at $\sim2\sigma$ level for KiDS and more than $1\sigma$ level for the 1-D marginalized constraints on $\Omega_m$ and $S_8$ for DES (even though a more careful measure of the consistency between the two datasets in the full parameter space provides ``substantial'' evidence for consistency, see~\cite{DES} for details). 

Weak lensing data tend to favour higher values of $\sumnu$ than those constrained by CMB power spectrum data. In fact, lower values of $\Omega_m$ and $S_8$ imply a reduced clustering amplitude, an effect that can be obtained by increasing the sum of neutrino masses. In~\cite{DES}, the combination of DES shear, galaxy and galaxy-shear spectra with Planck TT+lowP and other cosmological datasets in agreement with CMB results (i.e. BAO from 6dFGS~\cite{6dfgs}, SDSS DR7 MGS~\cite{mgs} and BOSS DR12~\cite{BOSSDR12}, and luminosity distances from the Joint Lightcurve Analysis (JLA) of distant SNIa~\cite{Betoule:2012an,Betoule:2014frx}) yields an upper bound at 95\% CL on the sum of the neutrino masses of $\sumnu<0.29\,\eV$, almost 20\% higher than the corresponding bound obtained dropping DES data ($\sumnu<0.245\,\eV$). Interestingly enough, the DES collaboration shows that a marginal improvement in the agreement between DES and Planck data is obtained when the sum of the neutrino masses is fixed to the minimal mass allowed by oscillation experiments $\sumnu=0.06\,\eV$.

To conclude this section, we also report the upper bound on $\sumnu$ obtained by weak lensing only data from the tomographic weak lensing power spectrum as measured by the KiDS collaboration~\cite{Kohlinger:2017sxk}. They found $\sumnu<3.3\,\eV$ and $\sumnu<4.5\,\eV$ at 95\% CL depending on the number of redshift bins retained in the analysis. These bounds are significantly broader than the constraints coming from CMB only data. Nevertheless, they come from independent cosmological measurements and still tighter than the constraints coming from kinematic measurements of $\beta$ decay.

\subsubsection{Cluster counts}
An additional low-redshift observable is represented by measurements of the number of galaxy clusters as a function of their mass at different redshifts. Cluster number counts provide a tool to infer the present value of the matter density $\Omega_m$ and the clustering amplitude $\sigma_8$, to be compared with the equivalent quantities probed at higher redshift by the primary CMB anisotropies.

Depending on the prior imposed on the mass bias, cluster counts tend to prefer lower values of $\Omega_m$ and $\sigma_8$ than the corresponding values obtained with primary CMB. The tension between the two datasets can be as high as $3.7\sigma$ for the lowest value of the mass bias as quantified by the Planck collaboration in 2015~\cite{PlanckXXIV}. Again, this preference for less power in the matter distribution favours higher values of the sum of the neutrino masses. Indeed, the Planck collaboration reports~\cite{PlanckXXIV} a upper bound of $\sumnu<0.20\,\eV$ at 95\% CL when Planck TT,TE,EE+lowP+BAO is combined with the SZ cluster count dataset (with a prior on the mass bias $(1-b)=0.780\pm0.092$ from the gravitational shear measurements of the Canadian Cluster Comparison Project, CCCP~\cite{Hoekstra:2015gda}), to be compared with the corresponding 95\% upper bound $\sumnu<0.17\,\eV$ without the SZ cluster count dataset~\cite{PlanckXIII}. 

Recently, \cite{Salvati:2017rsn} updated constraints on cosmological parameters, including $\sumnu$, from the SZ clusters in the Planck SZ catalogue, considering cluster count alone and in combination with the angular power spectrum of SZ sources. A comparison with bounds coming from primary CMB anisotropies is also performed. The combination of the two SZ probes (complemented with BAO measurements from Anderson et al 2014 to fix the underlying cosmology) confirms the discrepancy in $\Omega_m$ and $\sigma_8$ at the level of $2.1\sigma$ and provides an independent upper limit on the sum of the neutrino masses of $\sumnu<1.47\,\eV$ at 95\% CL. When combined with primary CMB, the bound reduces to $\sumnu<0.18\,\eV$. This bound is slightly higher than $\sumnu<0.12\,\eV$ found by~\cite{Vagnozzi} in absence of SZ data, as we should expect due to the aforementioned tension between SZ and primary CMB estimates of matter density and power.

\subsubsection{Lyman-$\alpha$ forests}
As all the datasets that probe the clustering of matter over cosmological distances, the $\Lya$ power spectrum is sensitive to $\sumnu$ primarly through the power suppression induced by massive neutrinos at small scales. The $\Lya$ spectrum alone can  constraint $\sumnu$ at a level of 1~eV (see e.g.~\cite{PalanqueDelabrouille}). The constraining power of the $\Lya$ spectrum is evident when it is combined with CMB data. In this case, the $\Lya$ data are used for setting the overall normalization of the spectrum through their sensitivity to $\Omega_m$ and $\sigma_8$, whereas the CMB fixes the underlying cosmological parameters and helps break degeneracies between $\Omega_m$, $\sigma_8$ and $\sumnu$. Recently, Y\`eche et al~\cite{Yeche} reported constraints on $\sumnu$ from the combination of the one-dimensional (i.e. angle-averaged) $\Lya$ power spectra from the SDSS III-BOSS collaboration and from the VLT/XSHOOTER legacy survey (XQ- 100). When the power spectra are used alone (complemented with a gaussian prior on $H_0=(67.3\pm1.0)\,\kmsmpc$), the authors obtain $\sumnu<0.8\,\eV$ at 95\% CL. The bounds dramatically improves to $\sumnu<0.14\,\eV$ when CMB power spectrum data from Planck TT+lowP are added to the analysis. The tightest bound on $\sumnu$ from $\Lya$ power spectrum comes from~\cite{PalanqueDelabrouille}, with $\sumnu<0.12\,\eV$ from Planck TT+lowP in combination with the $\Lya$ flux power spectrum from BOSS-DR12. Interestingly enough, in both analyses, the limit set by $\Lya$+Planck TT+lowP does not further improve when the $\Lya$ spectra are combined instead with the full set of CMB data from Planck, including small scale CMB polarization (Planck TT,TE,EE+lowP), and with BAO data from 6dFGS, SDSS MGS, BOSS-DR11.

The BAO signal can be also extracted from the $\Lya$ spectrum (see~\cite{McDonald:2006qs} for a pivotal study), providing estimates of the comoving angular diameter distance $d_A(z)$ and of the Hubble parameter $H(z)$ at redshift $z\simeq2$. Recently, the SDSS III-BOSS DR12 collaboration reported measurements of the BAO signal at $z=2.33$ from $\Lya$ forest \cite{Lya}. The estimated values of $d_A$ and $H$ are in agreement with a $\lcdm$ model (even though a slight tension with Planck primary CMB is present), although their precision is smaller than the precision obtained with galaxy-derived BAO measurements. Therefore, at present, the impact of $\Lya$-BAO data on simple extensions of the $\lcdm$ model is minimal.

We conclude that it is a conservative choice to take the constraints coming from $\Lya$ with some caution (a similar comment applies to constraints coming from aggressive analysis of the broadband shape of the matter power spectrum from galaxy surveys), until this probe will reach the level of maturity comparable with other traditional cosmological probes. 

\subsection{Local measurements of the Hubble constant and Supernovae Ia}
The most recent estimate of the Hubble constant has been reported in~\cite{Riess}. The authors improved over their previous measurement of $H_0$ from 3.3\% to 2.4\% thanks to an increased sample of reliable SNIa in nearby galaxies calibrated over Cepheids. Their final estimate, based on the combination of three different anchors, is $H_0=(73.24\pm1.74)\,\kmsmpc$, $3.2\sigma$ higher than the indirect estimate of $H_0$ from Planck TT+SimLow ($3.4\sigma$ higher than Planck TT,TE,EE+SimLow) in the context of a $\lcdm$ cosmology with $\sumnu=0.06\,\eV$. Previous analyses from the same authors also pointed to a $\sim2\sigma$ tension between direct measurements of $H_0$ and indirect estimate from primary CMB anisotropies from Planck (although see~\cite{Efstathiou:2013via} for a re-analysis of the same dataset which slightly reduces the discrepancy to within $1\sigma$ agreement). A discussion about the possible reasons behind this discrepancy and ways to alleviate it invoking non-standard cosmological scenarios are beyond the scope of this work. We refer the reader to the dedicated works~\cite{AddisonBAO,Aylor:2017haa,Aghanim:2016sns} for further reading. 

Since the Hubble constant and the sum of neutrino masses are anti-correlated, given the tension between the two probes it is clear that the combination of direct measurements of $H_0$ with CMB data leads to a preference for smaller values of $\sumnu$ with respect to CMB-only constraints. Indeed, several authors have pointed out the tight constraints on $\sumnu$ for such a combination. As an example,~\cite{Vagnozzi} showed that constraints on $\sumnu$ can be as tight as $\sumnu<0.148\,\eV$ at 95\% CL when Planck TT+lowP+BAO are complemented with a gaussian prior on $H_0$ equal to the estimate of the Hubble constant in~\cite{Riess}, to be compared with $\sumnu<0.186\,\eV$ from Planck TT+lowP+BAO only.
When lowP is replaced by a gaussian prior on $\tau$ compatible with the new estimates from SimLow, these numbers change to $\sumnu<0.115\,\eV$ ($\sumnu<0.151\,\eV$) with (without) the $H_0$ prior. 

For the sake of completeness, we shall also mention that independent estimates of $H_0$ from BAO measurements conducted by the SDSS III-BOSS DR12 collaboration~\cite{BOSSDR12} are in agreement with CMB estimates (see also~\cite{AddisonBAO} for a recent discussion). See also Ref.~\cite{Abbott:2017smn} for an additional independent estimate of $H_0$ with a combination of clustering and weak lensing measurements from DES-Y1 with BAO and BBN data. A discussion about the combination of five independent measurements of $H_0$ from cosmological probes and local measurements is also reported in~\cite{Abbott:2017smn,Vega-Ferrero:2017yqr}.

Finally, we report that a standard siren measurement of $H_0$ has been performed after the detection of the neutron star-neutron star merger GW170817~\cite{Abbott:2017xzu,Guidorzi:2017ogy,DiValentino:2017clw}. The Hubble constant has been constrained as $H_0=70.0^{+12.0}_{-8.0}\,\kmsmpc$ at 68\% CL. The accuracy of this determination is not comparable with the precise estimates of direct measurements and other cosmological constraints. However, the standard siren approach represents an additional independent estimate of $H_0$ and appears as a promising avenue as more GW events with electromagnetic counterparts are detected.

Concerning the inclusion of SNIa, the bounds from Planck TT+lowP improve from $\sumnu<0.72\,\eV$ to $\sumnu<0.33\,\eV$ at 95\% CL when data from the Joint Lightcurve Analysis~\cite{Betoule:2012an,Betoule:2014frx} are included\footnote{Bounds from the Planck Legacy Archive: \url{https://wiki.cosmos.esa.int/planckpla2015/index.php/Cosmological_Parameters}}. The most relevant systematics that affect SNIa measurements are related to the way in which SNIa light curves are standardized, with issues mostly arising from photometric calibrations and lightcurve fitting procedures. 

\begin{table*}
\begin{center}
\begin{tabular}{l|c|c}
Dataset	&$\sumnu [\eV]$	&Reference\\
\hline
\hline
Planck TT+lowP	&$<0.72$	&\cite{PlanckXIII}\\
Planck TT+lowP+lensing	&$<0.59$	&\cite{PlanckXIII}\\
Planck TT,TE,EE+lowP	&$<0.49$	&\cite{PlanckXIII}\\
Planck TT+SimLow	&$<0.59$	&\cite{PIPXLVI}\\
Planck TT,TE,EE+lowP+BAO+FS	&$<0.25$	&\cite{BOSSDR12}\\
Planck TT+lowP+BAO	&$<0.19$	&\cite{Vagnozzi}\\
Planck TT,TE,EE+lowP+BAO	&$<0.15$	&\cite{Vagnozzi}\\
Planck TT+lowP+FS	&$<0.30$	&\cite{Vagnozzi}\\
Planck TT+lowP+BAO+JLA	&$<0.25$	&\cite{DES}\\
Planck TT+lowP+BAO+JLA+WL	&$<0.29$	&\cite{DES}\\
Planck TT,TE,EE+BAO+SZ	&$<0.20$	&\cite{PlanckXXIV}\\
Planck TT+lowP+$\Lya$-FS	&$<0.14$	&\cite{Lya}\\
\hline
\end{tabular}
\caption{Constraints on $\sumnu$ from different combination of current cosmological data. Bounds given in this table are 95\% CL. BAO data for rows no. 1-4 and no. 11 are from 6dFGS~\cite{6dfgs}, SDSS MGS~\cite{mgs}, BOSS LOWZ DR11 and BOSS CMASS DR11~\cite{dr11} (see~\cite{PlanckXIII} for details). BAO+FS for row 5 are from SDSS BOSS DR12~\cite{BOSSDR12}. BAO data for rows no. 6-7 are from 6dFGS~\cite{6dfgs}, WiggleZ~\cite{wigglez}, SDSS BOSS DR11 LOWZ and SDSS BOSS DR11 CMASS~\cite{dr11} (see~\cite{Vagnozzi} for details). FS for row no. 8 is from SDSS BOSS DR12 CMASS~\cite{Gil-Marin:2015sqa} (see~\cite{Vagnozzi} for details). BAO for row no. 9-10 are from 6dFGS~\cite{6dfgs}, SDSS MGS~\cite{mgs}, BOSS DR12~\cite{BOSSDR12} (see~\cite{DES} for details). JLA for row no. 9-10 is the catalogue of luminosity distance measurements from the Joint Lightcurve Analysis~\cite{Betoule:2012an,Betoule:2014frx}. WL for row no. 10 is the combination of galaxy, shear and galaxy-galaxy lensing spectra from DES Year1~\cite{DES}. SZ in row no. 11 is the SZ cluster count dataset from~\cite{PlanckXXIV}. $\Lya$-FS in the last row is the $\Lya$ power spectrum measurement from BOSS~\cite{PalanqueDelabrouille}.}\label{tab:current}
\end{center}
\end{table*}

\section{Constraints on $\sumnu$ from future surveys}\label{sec:future}
In this section, we will discuss the expected improvements in the constraints on $\sumnu$ from the upcoming generation of CMB and LSS surveys. 
These constraints are also summarized in Tab.~\ref{tab:future} for the reader's convenience
\subsection{CMB surveys: CORE and CMB Stage-IV} 
The tightest bounds on $\sumnu$ from a single CMB experiment are those from the Planck satellite, reported in Sec. \ref{sec:currcmb}. As already explained, this sensitivity mostly come from the ability 1) to detect, at the level of CMB power spectrum, the smoothing effect of gravitational lensing of CMB photons, and,  2) to directly reconstruct the lensing power spectrum itself. These effects arise at small angular scales (higher multipoles $\ell$), therefore it is crucial to observe this region of the power spectrum with high accuracy in order to improve the sensitivity on $\sumnu$. Improved measurements of the polarization power spectra at all scales are also important to break degeneracies between cosmological parameters. The main example is the effect that a better estimate of the reionization optical depth $\tau$ from the large scale polarization spectrum has on $\sumnu$. Concerning, the lensing power spectrum, this is internally reconstructed by the Planck collaboration with high statistical significance up to intermediate scales $L$. However, the full power of this probe will be definitively unveiled when better measurements of polarization maps are available, enabling reconstruction from E-B estimators with lower variance and up to smaller scales \cite{HuOkamoto}.

A detailed summary of the expected sensitivity to cosmological parameters, including $\sumnu$, of all pre-2020 and post-2020 CMB missions can be found in~\cite{Errard:2015cxa}. As relevant examples, in this section we focus on two classes of future (post 2020) CMB experiments: a space mission and a ground based telescope. 

Recently, a proposal for a future CMB space mission has been submitted to the European Space Agency (ESA) in response to a call for medium-size mission proposals (M5). The mission, named Cosmic ORigin Explorer (CORE), is designed to have 19 frequency channels in the range $60-600\,\GHz$ for simultaneously solving for CMB and foreground signals, angular resolution in the range $2'-18'$ depending on the frequency channel and aggregate sensitivity of $2\,\uka$~\cite{Delabrouille:2017rct} (for comparison, the Planck satellite has 9 frequency channels in the range $30-900\,\GHz$, angular resolution in the range $5'-33'$ and the most sensitive channel shows a temperature noise of $0.55\,\mu\mathrm{K\cdot deg}$ at $143\,\GHz$~\cite{PlanckI}). This experimental setup would enable to constrain $\sumnu=(0.072^{+0.037}_{-0.051})\,\eV$ at 68\% CL assuming a $\lcdm$ model with a fiducial value of the sum of the neutrino masses $\sumnu=0.06\,\eV$, for the combination of CORE TT,TE,EE,PP (temperature and E-polarization auto and cross spectra and lensing power spectrum PP)~\cite{COREPar}. This roughly corresponds to a sensitivity of $\sigma(\sumnu)\sim0.044\,\eV$ (note that the target threshold for a $3\sigma$ detection in the minimal mass scenario is $\sigma(\sumnu)=0.020\,\eV$; for comparison, a simulated Planck-like experiment could only put an upper limit of $\sumnu<0.315\,\eV$ at 68\% CL for the same model). Other than to the capability of measuring
with high precision the small scale polarization (also in order to reconstruct the lensing potential),  part of this high sensitivity also comes from the improved limits that a science mission like CORE can put on $\tau$: compared to Planck, CORE would achieve an almost cosmic-variance-limited (CVL) detection of the reionization optical depth ($\sigma_\mathrm{CVL}(\tau)\simeq0.002$). 

A roadmap towards a Stage-IV (S4) generation of CMB ground-based experiments\footnote{\url{https://cmb-s4.org}} has been also developing~\cite{S4}. The goal is to set a definitive CMB experiment with $\sim250000$ detectors surveying half of the sky, with angular resolution of $1'-2'$ and a sensitivity of $1\,\uka$ at $150\,\GHz$. The greatest contaminant for a ground-based experiment is the atmospheric noise, which highly reduces the accessible frequencies for CMB observations to a total of four windows, roughly 35, 90, 150, and 250 GHz. The main advantages with respect to a space-borne mission are a larger collecting area with an incredibly higher number of detectors (for a comparison, the CORE proposal accounts for a total of 2100 detectors\cite{Delabrouille:2017rct}, the Planck satellite has 74 detectors~\cite{PlanckI}) and subsequent suppression of experimental noise. 
At large scales, the Stage-IV target is the recombination bump at $\ell>20$. The reduced sky fraction accessible from ground, foreground contaminations and atmospheric noise are the main issues that limit the possibility to target also the range $\ell<20$. Therefore, it is likely that S4 would be complemented by balloon-based and satellite based measurements at the largest scales.
%Compared to a space mission, a ground-based telescope has access to a smaller fraction of the sky, of order $f_\mathrm{sky}\sim50\%$ or less depending on the latitude of the observational site. Therefore, it is unlikely that a single ground-based mission could observe a multipole range down to $\ell=2$, as opposed to a space mission, and simultaneously constrain $\tau$ and the other cosmological parameters. 
As a result, forecasts for S4 relies on external measurements of $\tau$. The sensitivity $\sigma({\sumnu})$ of S4 TT,TE,EE,PP complemented with a gaussian prior on the optical depth of $\tau=0.060\pm0.01$ (roughly corresponding to the latest estimate from Planck-HFI~\cite{PIPXLVI}) is in the range $[0.073-0.110]\,\eV$, depending on the angular resolution and noise level, for $f_\mathrm{sky}=40\%$~\cite{S4}.

Neither of the two classes of future CMB mission proposals can achieve alone the necessary sensitivity to claim a detection of $\sumnu=0.06\,\eV$ at the $3$-$\sigma$ level. Nevertheless, we will see in the next section that the combination of future CMB missions with future galaxy surveys could possibly lead to the first detection of neutrino masses from cosmology.

\subsection{Future LSS surveys: DESI, Euclid, LSST, WFIRST}
Improved performances from future galaxy surveys with respect to the current status can be achieved by mapping a larger volume of the sky, therefore increasing the number of samples observed and going deeper in redshift. In this section, we will briefly review the expected performances of the main Stage-IV LSS surveys.

The successor to SDSS III-BOSS survey will be the ground-based Dark Energy Spectroscopic Instrument\footnote{\url{http://desi.lbl.gov}} (DESI). It is designed to operate for 5 years and cover roughly a $14000\,\mathrm{deg}^2$ survey area. The extension in redshift is expected to be up to $z=1$ for Luminous Red Galaxies (LRG), $z=1.7$ for Emission Line Galaxies (ELG) and $z=3.5$ for $\Lya$ forests, for a total of over 20 million galaxy and quasar redshifts. With these numbers, DESI will improve over the BOSS survey by an order of magnitude in both volume covered and number of objects observed. It can achieve a $3.49\%$ and $4.78\%$ determination of the BAO signal across ($d_A/r_d$) and along ($H r_d$) the line-of-sight, respectively, at $z=1.85$, and $16\%$ and $9\%$ determination of the same quantities at the highest redshift achievable with $\Lya$ forest $z=3.55$~\cite{DESI}.  
Even in the most conservative scenario when DESI BAO only (i.e. without including information from the broadband shape of the matter power spectrum and $\Lya$ forests) are combined with future CMB experiments, the sensitivity on $\sumnu$ greatly improves. It goes down to $\sigma(\sumnu)=0.021\,\eV$ for CORE TT,TE,EE,PP+DESI BAO, forecasting a $\sim 3\sigma$ detection of $\sumnu$ in the minimal mass scenario~\cite{COREPar}. In the case of S4+DESI BAO~\cite{S4}, $\sigma(\sumnu)$ is in the range $[0.023-0.036]\,\eV$ ( or $[0.020-0.032]\,\eV$) with a prior of $\tau=0.06\pm0.01$ (or $\tau=0.060\pm0.006$, the expected sensitivity from Planck-HFI~\cite{PlanckBlue}) and $f_\mathrm{sky}=0.40$, depending on the S4 angular resolution and noise level. For a $1'$ resolution and a noise level lower than $2.5\,\uka$, $\sigma(\sumnu)$ could be further improved with a better measurement of $\tau$ down to the level of $\sigma(\sumnu)<0.015\,\eV$, that would guarantee a $>4\sigma$ detection of $\sumnu$ in the minimal mass scenario. 

The DESI mission will be complementary to the science goals of the Large Synoptic Survey Telescope\footnote{\url{https://www.lsst.org}} (LSST), a Stage-IV ground-based optical telescope. The main science fields in which LSST will mostly operate are~\cite{LSST}: \textit{``Inventory of the Solar System, Mapping the Milky Way, Exploring the Transient Optical Sky, and Probing Dark Energy and Dark Matter''}. These goals will be achieved by surveying a $\sim30000\,\mathrm{deg}^2$ area (2/3 of which in a ``deep-wide-fast'' survey mode) over 10 years, in six bands (\textit{ugrizy}), with incredible angular resolution ($\sim0.7''$), producing measurements of roughly 10 billion stars and galaxies. Thanks to its peculiar observational strategy, LSST will provide multiple probes of the late-time evolution of the universe with a single experiment, namely, weak lensing cosmic shear, BAO in the galaxy power spectrum, evolution of the mass function of galaxy clusters, and a compilation of SNIa redshift-distances. The expected sensitivity on $\sumnu$~\cite{LSST} is in the range $\sigma(\sumnu)=[0.030-0.070]\,\eV$, depending on the fiducial value of $\sumnu$ assumed when performing forecasts ($\sumnu^\mathrm{fid}=[0-0.66]\,\eV$). Larger fiducial values for the mass yield better sensitivity. These numbers include a marginalization over the uncertainties coming from an extended cosmological scenario, where a number of relativistic species different than $3.046$, a non-zero curvature and a dynamical dark energy $w_0-w_a$ component are allowed. They also take into account the combination of the three-dimensional cosmic shear field as measured by a LSST-like survey with Planck-like CMB data and can be improved by a factor of $2$ if either BAO or SNIa measurements are also considered, whereas a factor of $\sqrt{2}$ degradation could come from systematic effects. Interestingly enough, the observational strategy of LSST (large and deep survey) could provide the necessary sensitivity to explore the faint effects that the distinct neutrino mass eigenstates have on cosmological probes. This is a highly debated topic and we refer the reader to Sec.~\ref{sec:hierarchy} for related discussion.

Synergy between these large ground-based observatories and future space missions is expected. We consider here the ESA Euclid satellite\footnote{\url{https://www.euclid-ec.org}} and the NASA Wide Field Infrared Survey Telescope\footnote{\url{https://wfirst.gsfc.nasa.gov}} (WFIRST) as representative space-borne missions.
Euclid will be a wide-field satellite that operates with imaging and spectroscopic instruments for 6 years and covers roughly $15000\,\mathrm{deg}^2$ in the optical and near-infrared bands, observing a billion galaxies and measuring $\sim100$ million galaxy redshifts~\cite{EuclidStudyRep}. The redshift depth will be up to $z\sim2$ for galaxy clustering and up to $z\sim3$ for cosmic shear. The combination of the galaxy power spectrum measured with Euclid and primary CMB from Planck is expected to give $\sigma(\sumnu)=0.04\,\eV$; if instead the weak lensing dataset produced by Euclid is considered in combination with primary CMB, we expect $\sigma(\sumnu)=0.05\,\eV$~\cite{Euclid}. Both combinations provide a $\sim1\sigma$ evidence in the minimal mass scenario. Some authors have also pointed out that weak lensing data as measured by Euclid could discriminate between the two neutrino hierarchies if the true value of $\sumnu$ is small enough (i.e. far enough from the degenerate region of the neutrino mass spectrum), see~\cite{Euclid} and references therein.

WFIRST is an infrared telescope with a primary mirror as wide as the Hubble Space Telescope's primary ($2.4\,\mathrm{m}$) and will operate for 6 years~\cite{WFIRST}. The primary instrument on board, the Wide Field Instrument, will be able to operate both in imaging and spectroscopic mode, observing a billion galaxies. The instrumental characteristics of WFIRST will more than double the surface galaxy density measured by Euclid. With this setup, WFIRST will test the late expansion of the universe with great accuracy employing supernovae, weak lensing, BAO, redshift space distortions (RSD), and clusters as probes. From the BAO and broadband measurements of the matter power spectrum, WFIRST in combination with a Stage-III CMB experiment could provide $\sigma(\sumnu)<0.03\,\eV$~\cite{WFIRST}. 

We want to conclude this section by pointing out that the aforementioned missions will be extremely powerful if combined together. Indeed, they are quite complementary~\cite{Boyle:2017lzt}. A significant example concerning the improvement of constraints on massive neutrinos is the combination of all the previously discussed surveys with the lensing reconstruction from CMB. The cross correlation of weak lensing (optical), CMB lensing power spectrum and galaxy clustering (spectroscopic) can highly reduce the systematics affecting each single probe, in particular the multiplicative bias in cosmic shear~\cite{Das:2013aia}. For example, a combination of WFIRST, Euclid, LSST and CMB Stage-III can achieve $\sigma(\sumnu)<0.01\,\eV$~\cite{WFIRST}.
Another example is the calibration of the cluster mass for SZ cluster count analyses. This calibration can be performed through optical surveys such as LSST or through CMB lensing calibration, with comparable results. In Ref.~\cite{Madhavacheril:2017onh}, the authors show that lensing-calibrated SZ cluster counts can provide a detection of the minimal neutrino mass $\sumnu$ at $>3\sigma$ level, also in extended cosmological scenarios. 

\subsection{21-cm surveys}
In this section, we will briefly comment about the possibility to use 21-cm survey data to constrain $\sumnu$. We refer the reader to the relevant papers for further readings. 
Measurements of the 21-cm signal such as those expected from the Square Kilometer Array\footnote{\url{http://skatelescope.org}} (SKA) and the Canadian Hydrogen Intensity Mapping Experiment\footnote{\url{https://chime-experiment.ca}} (CHIME) can shed light on the Epoch of Reionization, including a better determination of the reionization optical depth $\tau$. In addition, they map the distribution of neutral hydrogen in the universe, a tracer of the underlying matter distribution. Therefore, constraints on $\sumnu$ can benefit from 21-cm measurements in two ways: by breaking the degeneracy between $\sumnu$ and $\tau$ (see e.g.~\cite{Archidiacono:2016lnv}, where the authors report $\sigma(\sumnu)=0.012\,\eV$ for a combination of CORE+Euclid lensing and FS+ a prior on $\tau$ compatible with expectations from future 21-cm surveys); by detecting the effect of $\sumnu$ on the evolution of matter perturbations (see e.g.~\cite{Pritchard:2008wy,Oyama:2015gma,Villaescusa}).

\begin{table*}
\begin{center}
\begin{threeparttable}
\begin{tabular}{l|c|c}
Dataset	&$\sigma(\sumnu) [\mathrm{meV}]$	&Reference\\
\hline
\hline
CORE TT,TE,EE,PP	&$44$	&\cite{COREPar}\\
S4 TT,TE,EE,PP	&$73$	&\cite{S4}\\
CORE TT,TE,EE,PP+DESI	&$21$	&\cite{COREPar}\\
S4 TT,TE,EE,PP\,\tnotex{tnote:tau1} +DESI	&$23$	&\cite{S4}\\
S4 TT,TE,EE,PP\,\tnotex{tnote:tau2} +DESI 	&$15$	&\cite{S4}\\
Planck CMB+LSST-shear	&$30\tnotex{tnote:lsst}$	&\cite{LSST}\\
Planck+Euclid-FS	&$40$	&\cite{Euclid}\\
Stage-III CMB (ACTPol)+WFIRST BAO+FS	&$30$	&\cite{WFIRST}\\
Stage-III CMB+WFIRST+Euclid+LSST	&$8$	&\cite{WFIRST}\\
\hline
\end{tabular}
\begin{tablenotes}
\item\label{tnote:tau1}The combination assumes a gaussian prior on $\tau=0.06\pm0.01$ roughly corresponding to the new estimate from~\cite{PIPXLVI}.
\item\label{tnote:tau2}The combination assumes $\sigma(\tau)=0.002$ and noise level of $2.5\uka$.
\item\label{tnote:lsst}For a fiducial value $\sumnu=0\,\eV$ and marginalising over $w_0-w_a$ dark energy, arbitrary curvature and $N_\mathrm{eff}$.
\end{tablenotes}
\end{threeparttable}
\caption{Expected sensitivity on $\sumnu$ from different combination of future cosmological data. Unless otherwise stated, the sensitivity $\sigma(\sumnu)$ is forecasted assuming a standard cosmological model with $\sumnu=0.06\,\eV$. DESI refers to the simulated DESI-BAO dataset based on expected experimental performances~\cite{DESI} (see~\cite{COREPar,S4} for details). FS refers to the use of the (simulated) measurements of the full shape of the matter power spectrum. The last line implies the use of CMB lensing, Euclid and WFIRST to calibrate the multiplicative bias in the shear measurements from LSST~\cite{WFIRST}.}\label{tab:future}
\end{center}
\end{table*}

\section{Constraints on $\sumnu$ in extended cosmological scenarios}\label{sec:extended}
The constraints derived so far apply to the simple one-parameter extension of the standard cosmological model, $\lcdm+\sumnu$. When derived in the context of more complicated scenarios, such as models that allow arbitrary curvature and/or non-standard dark energy models and/or modified gravity scenarios etc., constraints on $\sumnu$ are expected in general to degrade (although tighter constraints on $\sumnu$ can be also possible in particular extended scenarios) with respect to those obtained in a $\lcdm+\sumnu$ cosmology. This effect is due to the multiple degeneracies arising between cosmological parameters that describe the cosmological model under scrutiny. In other words, when more degrees of freedom are available -- in terms of cosmological parameters that are not fixed by the model --, more variables can be tuned in order to adapt the theoretical model to the data. For example, CMB data measure with incredible accuracy the location (expressed by the angular size of the horizon at recombination $\theta$) and amplitude (basically driven by the exact value of $z_\mathrm{eq}$) of the first acoustic peak. Therefore, we want to preserve this feature in any cosmological model.
As explained before, $h$, $\Omega_m$ and $\sumnu$ can be varied together in order to do this. Adding other degrees of freedom, like
curvature or evolving dark energy, allows for even more freedom, thus making the degeneracy worse. Of course, the addition of different cosmological data, which are usually sensitive to different combinations of the aforementioned parameters, is extremely helpful in tightening the constraints on $\sumnu$ (and, in general, on any other cosmological parameter) in complex scenarios.

In more detail, constraints on the sum of neutrino masses are particularly sensitive to the so-called ``geometric degeneracy''. This term refers to the possibility of
adjusting the parameters in order to keep constant the angle subtended by the sound horizon at last scattering, that controls the position of the first peak of the CMB anisotropy spectrum. 
The degeneracy is worsened in models with a varying curvature density $\Omega_k$ or parameter of the equation of state of dark energy $w$. 
Constraints on the expansion history, like those provided by BAO or by direct measurements of the Hubble constant, are particularly helpful in breaking the geometric degeneracy.
In principle, one could also expect a degeneracy between the effective number of degrees of freedom $\Neff$ and $\sumnu$, but for a different reason: both parameters can be varied in order to keep constant the redshift of matter-radiation equality. However, this can be done only at the expense of changing the CMB  damping scale (see Sec.~\ref{sec:beyond} for further details). High-resolution measurements of the CMB anisotropies are therefore a key to partially break the degeneracy. Finally, a nonstandard relation between the matter density distribution and the lensing potential can be modelled by introducing a phenomenological parameter $A_L$, which modulates the amplitude of the lensing signal \cite{Calabrese:2008rt}. Most of the current constraining power of CMB experiments on $\sumnu$ comes from CMB lensing. Therefore, it is clear that in models with varying $A_L$ the limits on neutrino masses are strongly degraded. However, it should also be noted that $A_L$ is usually introduced as a proxy for instrumental systematics; if considered as an actual physical parameter, its value is fixed by general relativity to be $A_L=1$.

\begin{figure*}
\begin{center}
\includegraphics[width=0.9\textwidth]{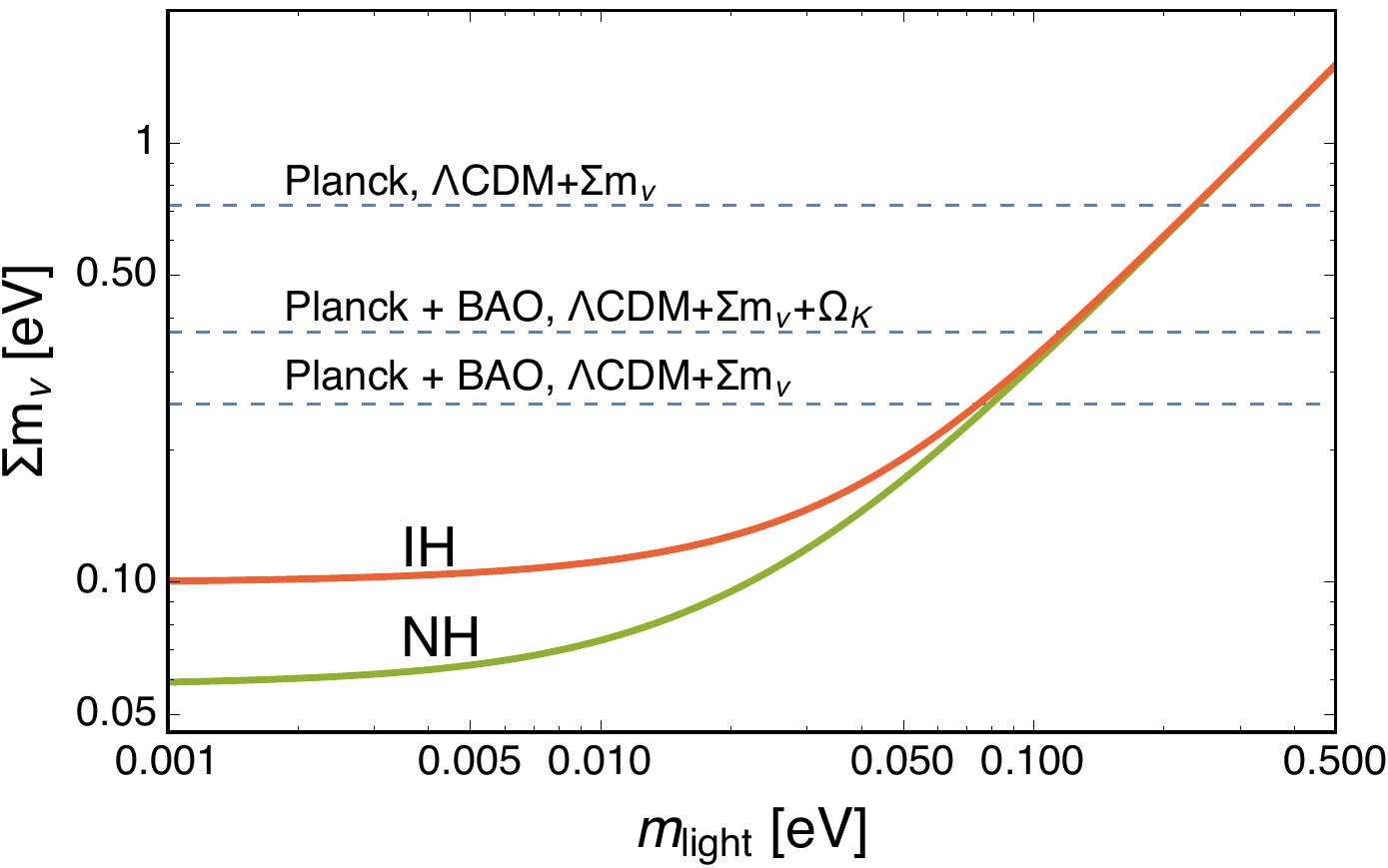}
\end{center}
\caption{Sum of neutrino masses $\sumnu$ as a function of the mass $m_\mathrm{light}$ of the lightest neutrino eigenstate, for normal (green) or inverted (red) hierarchy. The horizontal dashed lines show 95\% CL upper limits for different dataset combinations, from top to bottom: PlanckTT+lowP in the $\lcdm+\sumnu$ model, PlanckTT+lowP+BAO in the $\lcdm+\sumnu+\Omega_K$ model, PlanckTT+lowP+BAO in the $\lcdm+\sumnu$ model.  \label{fig:ext}}
\end{figure*}

To make the discussion more quantitative, we see how this applies to the constraints obtained with present data and future data. In Tab.~\ref{tab:extended}, we report a comparison of the constraints on $\sumnu$ for some extensions of the $\lcdm$ model. In the upper part of the table, we report constraints obtained from the PlanckTT+lowP+lensing+BAO dataset combination, described in Sec.~\ref{sec:currcmb}. These are taken from the  full grid of results made available by the Planck collaboration\footnote{The full grid can be downloaded from the \href{https://wiki.cosmos.esa.int/planckpla2015/index.php/Cosmological_Parameters}{Planck Legacy Archive}
.} and have obtained with the same statistical techniques used for the $\lcdm$ model. We see that the constraints are degraded by $30\%$ in models with varying $\Neff$, by $50\%$ in models with varying $\Omega_K$ or $w$, and by $65\%$ in models with varying $A_L$. 
This information is also conveyed, for an easier visual comparison, in Fig.~\ref{fig:ext},
 where we show the mass the sum of neutrino masses as a function of the mass $m_\mathrm{light}$ of the lightest eigenstate. The green and red curves are for normal and inverted hierarchy, respectively. We show 95\% constraints on $\sumnu$ for different models and dataset combinations as horizontal lines.
 In the lower section of Tab.~\ref{tab:extended} we instead report a similar comparison, based on the expected sensitivities of future CMB and LSS probes~\cite{COREPar}. The pattern is very similar to that observed for present data, although it should be noted that the increased precision of future experiments will allow to further reduce the degeneracies. In particular, it is found that the constraints on $\sumnu$ are degraded by $\sim30\%$ in models with varying $\Omega_K$ or $w$, and not degraded at all in models with varying $\Neff$ (models with varying $A_L$ have not been considered in Ref.~\cite{COREPar}).

The cases reported in Tab.~\ref{tab:extended} hardly exhaust all the possible, well-motivated extensions to the $\lcdm+\sumnu$ model. To make a few examples of more complicated extension, without the aim of being complete, the interplay between inflationary parameters and the neutrino sector has been investigated in Refs.~\cite{Gerbino:2016sgw,DiValentino:2016ikp}. In Refs.~\cite{DiValentino:2015ola,DiValentino:2016hlg,DiValentino:2017zyq} ``extended parameter spaces'' are considered, in which $12$ parameters, including $\sumnu$, are varied simultaneously. Neutrino-dark matter interaction are discussed in Ref.~\cite{DiValentino:2017oaw}, while low-reheating scenarios are studied in Ref.~\cite{deSalas:2015glj}. Finally, constraints on $\sumnu$ in the context of cosmological models with time-varying dark energy are derived in Ref.~\cite{Lorenz:2017fgo}.

\begin{table*}
\begin{center}
\begin{threeparttable}
\begin{tabular}{l|c|c}
Extension to $\lcdm$		&$\sumnu [\mathrm{meV}]$	&Dataset\\
\hline
\hline
$\lcdm+\sumnu$	&$<254\tnotex{tnote:95}$	&Planck TT+lowP+lensing+BAO\tnotex{tnote:pla}\\
$\lcdm+\sumnu+\Omega_K$	&$<368\tnotex{tnote:95}$	&Planck TT+lowP+lensing+BAO\tnotex{tnote:pla}\\
$\lcdm+\sumnu+w$	&$<372\tnotex{tnote:95}$	&Planck TT+lowP+lensing+BAO\tnotex{tnote:pla}\\
$\lcdm+\sumnu+N_\mathrm{eff}$	&$<323\tnotex{tnote:95}$	&Planck TT+lowP+lensing+BAO\tnotex{tnote:pla}\\
$\lcdm+\sumnu+A_\mathrm{lens}$	&$<413\tnotex{tnote:95}$	&Planck TT+lowP+lensing+BAO\tnotex{tnote:pla}\\
\hline
$\lcdm+\sumnu$	&$62\pm16\,\tnotex{tnote:68}$	&CORE TT,TE,EE,PP+BAO~\cite{COREPar}\\
$\lcdm+\sumnu+\Omega_K$	&$63\pm21\,\tnotex{tnote:68}$	&CORE TT,TE,EE,PP+BAO~\cite{COREPar}\\
$\lcdm+\sumnu+w$	&$48^{+22}_{-17}\,\tnotex{tnote:68}$	&CORE TT,TE,EE,PP+BAO~\cite{COREPar}\\
$\lcdm+\sumnu+N_\mathrm{eff}$	&$68^{+15}_{-17}\,\tnotex{tnote:68}$	&CORE TT,TE,EE,PP+BAO~\cite{COREPar}\\
$\lcdm+\sumnu+Y_\mathrm{He}$	&$62\pm16\,\tnotex{tnote:68}$	&CORE TT,TE,EE,PP+BAO~\cite{COREPar}\\
$\lcdm+\sumnu+r$	&$60^{+15}_{-17}\,\tnotex{tnote:68}$	&CORE TT,TE,EE,PP+BAO~\cite{COREPar}\\
\hline
\end{tabular}
\begin{tablenotes}
\item\label{tnote:95} 95\% CL
\item\label{tnote:pla}From the \href{https://wiki.cosmos.esa.int/planckpla2015/index.php/Cosmological_Parameters}{Planck 2015 Explanatory Supplement Wiki}
\item\label{tnote:68} 68\% CL
\end{tablenotes}
\end{threeparttable}
\caption{Constraints on $\sumnu$ from different extensions to the $\lcdm$ model for the indicated datasets. $\Omega_K$ is the curvature density parameter, $w$ is the (constant) equation of state parameter for the dark energy, $N_\mathrm{eff}$ is the number of relativistic species at recombination, $A_\mathrm{lens}$ is the phenomenological rescaling of the lensing power that smears the CMB power~\cite{Calabrese:2008rt}, $Y_\mathrm{He}$ is the primordial Helium abundance, $r$ is the tensor to scalar ratio. \textit{Upper section}: constraints are from the full grid of results from the Planck collaboration (see text for details). BAO data are from 6dFGS, SDSS MGS, BOSS LOWZ DR11 and BOSS CMASS DR11 (see~\cite{PlanckXIII} for details). \textit{Lower section}: Forecasted constraints are from~\cite{COREPar}. BAO refers to simulated data for DESI and Euclid surveys. The fiducial model adopted for the analysis is the following: $\sumnu=0.06\,\eV$, $\Omega_K=0$, $w=-1$, $N_\mathrm{eff}=3.046$, $Y_\mathrm{He}=0.24$, $r=0$.}\label{tab:extended}
\end{center}
\end{table*}

%%%%%%%%

\section{Cosmology and the neutrino mass hierarchy \label{sec:hierarchy}}

Cosmology is mostly sensitive to the total energy density in neutrinos, directly proportional to the sum of the neutrino masses $\sumnu\equiv m_1+m_2+m_3$. We can express $\sumnu$ in the two hierarchies as a function of the lightest eigenstate $\mlight$ (either $m_1$ or $m_3$) and of the squared mass differences $\dms$ and $\dma$: 
\begin{eqnarray}
\sumnu ^{NH}&=&\mlight+\sqrt{\mlight^2+\dms}+\sqrt{\mlight^2+|\dma|}\\
\sumnu ^{IH}&=&\mlight+\sqrt{\mlight^2+|\dma|}+\sqrt{\mlight^2+|\dma|+\dms}
\end{eqnarray}
When stating that oscillation experiments are insensitive to the absolute mass scale, one refers to the fact that the value of $\mlight$ is not accessible with oscillation data. When $\mlight = 0\,\eV$, one obtains $\sumnu^{NH}\simeq0.06\,\eV$ and $\sumnu^{IH}\simeq0.1\,\eV$. Therefore, for each hierarchy, a minimum mass scenario exists in which $\sumnu\neq0$.

It has been a long-standing issue whether or not cosmological probes are sensitive to the neutrino mass hierarchy. In principle, we expect physical effects due to the choice of the neutrino hierarchy on cosmological observables. Individual neutrino species that carry a slightly different individual mass exhibit a slightly different free-streaming scale $k_{fs}$: depending on their individual mass, neutrinos can finish suppressing the matter power at different epochs, leaving three distinct ``kinks'' in the matter power spectrum. As a consequence, the weak lensing effects on the CMB and on high redshift galaxies can be slightly affected by the choice of the hierarchy. In practice, all of these signatures are at the level of permille effects on the matter and CMB power spectra, well below the current sensitivity~\cite{Lesgourgues:2004ps}. 

Given the current sensitivity (roughly $\sumnu<0.2\,\eV$ at 95\% CL), it is then a legitimate assumption to approximate the mass spectrum as perfectly degenerate ($m_i=\sumnu/3$) when performing analysis of cosmological data. Very recently, several authors investigated the possibility that such an approximation could fail reproducing the physical behaviour of massive neutrinos when observed with the high sensitivity of future cosmological surveys~\cite{Giusarma:2016phn,Gerbino:2016sgw,COREPar,Hannestad:2016fog}. In addition, the issue of whether future survey could unravel the unknown hierarchy has been addressed by several groups~\cite{Gerbino:2016ehw,Vagnozzi,Hannestad:2016fog,Hamann:2012fe,Xu:2016ddc,Jimenez:2010ev}. We refer the reader to the relevant papers for a thorough discussion of these issues. Here, we summarise the main results: 1) the sensitivity of future experiments will not be enough to clearly separate the effects of different choices of the neutrino hierarchy, \emph{for a given value of $\sumnu$}; therefore the fully-degenerate approximation is still a viable way to model the neutrino mass spectrum in the context of cosmological analysis; 2) the possibility to clearly identify the neutrino hierarchy  with future cosmological probes is related 
to the capability of measuring $\sumnu < 0.1\eV$ at high statistical significance, in order to exclude the IH scenario. It is clear that the possibility
to do this strongly depends on the true value of $\sumnu$: the closer it is to $\sumnu^{NH,\mathrm{min}}=0.06\,\eV$, the larger will be the statistical significance by which we can exclude IH. This is true independently of whether we approach the issue from a frequentist or Bayesian perspective. In the latter case, however, since a detection of the hierarchy would driven by volume effects, this posits the question of what is the correct prior choice for $\sumnu$. The issue is extensively discussed in~\cite{Gerbino:2016ehw,Simpson:2017qvj,Schwetz:2017fey,Caldwell:2017mqu,Long:2017dru,Hannestad:2017ypp}.

\section{Complementarity with laboratory searches \label{sec:lab}}
Cosmological observables are ideal probes of the neutrino absolute mass scale, though they are not the only probes available. In fact, laboratory avenues such as kinematic measurements in $\beta$-decay experiments (see e.g.~\cite{Drexlin:2013lha}) and neutrino-less double-$\beta$ decay ($\bb$) searches (see e.g.~\cite{Cremonesi:2013vla,DellOro:2016tmg}) provide complementary pieces of information to those carried by cosmology.

Kinematic measurements are carried on with $\beta$-decay experiments mostly involving tritium $^{3}\mathrm{H}$. The shape of the decay spectrum close to the end point is sensitive to the (electron) neutrino mass and can be parametrized in terms of constraints on the electron neutrino effective mass\footnote{It has to be noticed that the observable which $\beta$-decay experiments are sensitive to is $m_\beta^2$, rather than $m_\beta$. Nevertheless, it is useful to quote constraints in terms of $m_\beta$ to facilitate the comparison with results from other probes.} 
defined in Eq. (\ref{eq:mb}). The current best limits on $m_\beta$ come from the Troitzk and Mainz experiments, with $m_\beta<2.05\,\eV$~\cite{Aseev:2011dq} and $m_\beta<2.3\,\eV$~\cite{Kraus:2004zw} at 95\% CL. The new generation $^{3}\mathrm{H}$ $\beta$-decay experiment KATRIN (Karlsruhe Tritium Neutrino\footnote{ \url{https://www.katrin.kit.edu}}) is expected either to reach a sensitivity of $m_\beta<0.2\,\eV$ at 90\% CL, an order of magnitude improvement with respect to current sensitivities, or to detect the neutrino mass if it is higher than $m_\beta=0.35\,\eV$. Note that a detection of non-zero neutrino mass in KATRIN would imply $\sumnu \gtrsim 1\,\eV$, and would then be in tension with the cosmological constraints obtained in the framework of the $\lcdm$ model. This could point to the necessity of revising the standard cosmological model, although it should be noted that none of the simple one-parameter extensions reported in Tab.~\ref{tab:extended} could accommodate for such a value.

Future improvements in kinematic measurements involve technological challenges, since KATRIN reaches the experimental limitations imposed to an experiment with spectrometers. Future prospects are represented by the possibility of calorimetric measurements of $^{136}\mathrm{Ho}$ (HOLMES experiment~\cite{Giachero:2016xnn}) and measurements of the $^{3}\mathrm{H}$ decay spectrum via relativistic shift in the cyclotron frequency of the electrons emitted in the decay (Project8 experiment\footnote{\url{http://www.project8.org/index.html}}~\cite{Esfahani:2017dmu}). Although the bounds coming from $\beta$-decay experiments are very loose compared to bounds from cosmology, nevertheless they are appealing for the reason that they represent model-independent constraints on the neutrino mass scale, only relying on kinematic measurements.  

$\bb$ decay is a rare process that is allowed only if neutrinos are Majorana particles. A detection of $\bb$ events thus would solve the issue related to the nature of neutrinos, whether they are Dirac or Majorana particles. Searches for $\bb$ directly probe the number of $\bb$ events, which is related to the half life of the isotope involved in the decay $T_{1/2}$. The latter can be translated in limits on the Majorana mass $\mbb$ (defined in Eq. \ref{eq:mbb})  once a nuclear model has been specified. In practice, a bound on $T_{1/2}$ is reflected in a range of bounds on $\mbb$, due to the large uncertainties associated with the exact modelling of the nuclear matrix elements. Additional complications are due to model dependencies: when translating bounds on $T_{1/2}$ to bounds on $\mbb$, a mechanism responsible for the $\bb$ decay has to be specified. This is usually the exchange of light Majorana neutrinos, though alternative mechanisms could be responsible for the lepton number violation that not necessarily allow a direct connection between $T_{1/2}$ and $\mbb$. Finally, it can be shown that in the case of NH, disruptive interference between mixing parameters could prevent a detection of $\bb$ events, regardless of the neutrino nature and the lepton-number violation mechanism. 

We report here some of the more recent limits on $\mbb$ from $\bb$ searches. Constraints are reported as a range of 90\% CL upper limits, due to the uncertainty on the nuclear matrix elements. We also specify the isotope used in each experiment. The current bounds are $\mbb<0.120-0.270\,\eV$ from Gerda Phase-II ($\mathrm{^{76}Ge}$) \cite{Agostini:2017dxu,PandolaTAUP17}, $\mbb<0.061-0.165\,\eV$ from KamLAND-Zen~\cite{KamLAND-Zen:2016pfg} ($^{136}\mathrm{Xe}$), $\mbb < 0.147 - 0.398 \,\eV$ from EXO-200 ($^{136}\mathrm{Xe}$)~\cite{Albert:2017owj}, $\mbb < 0.140 - 0.400\,\eV$ from CUORE ($^{130}\mathrm{Te}$)~\cite{Alduino:2017ehq}.  
%The current sensitivity for experiments using different isotopes is: $\mbb<0.15-0.33\,\eV$ at 90\% CL for Gerda Phase-II~\cite{Agostini:2017iyd} ($\mathrm{^{76}Ge}$), $\mbb<0.061-0.165\,\eV$ at 90\% CL for KamLAND-Zen~\cite{KamLAND-Zen:2016pfg} ($^{136}\mathrm{Xe}$). 
The next generation $\bb$ experiments, such as LEGEND, SuperNEMO, CUPID, SNO+, KamLAND2-Zen, nEXO%\footnote{\url{https://www-project.slac.stanford.edu/exo/about.html}}
, NEXT, PANDAX-III, aims to cover the entire region of IH, reaching a $3\sigma$ discovery sensitivity for $\mbb$ of $20\,\mathrm{meV}$ or better, roughly an order of magnitude improvement with respect to the current limits (see Ref.~\cite{Agostini:2017jim} for a more detailed discussion and for a full list of references).  

As outlined above, laboratory searches and cosmology are sensitive to different combinations of neutrino mixing parameters and individual masses. Therefore, it makes sense to compare their performances in terms of constraints on the neutrino mass scale. It is also beneficial to combine these different probes of the mass scale, in order to overcome the limitations of each single probe and increase the overall sensitivity to the neutrino masses~\cite{Caldwell:2017mqu,Capozzi:2017ipn,Gerbino:2015ixa}. This is possible because, once the elements of the mixing matrix are known, specifying one of three mass parameters among $(m_\beta,\,\mbb,\,\sumnu)$, together with the solar and atmospheric mass splittings, uniquely determines the other two. Oscillation experiments measure precisely the values of the mixing angles and of the squared mass differences, with an ambiguity on the sign of $\Delta m^2_{31}$, so that these parameters can be simply fixed to their best-fit values, given the larger uncertainties on the absolute mass parameters.
The value of the Dirac phase, on the other hand, is known with lesser precision, and the Majorana phases, relevant for the interpretation of $\bb$ searches, are not probed at all by oscillation experiments. However this ignorance can be folded into the analysis using standard statistical techniques. Finally, the relation between the mass parameters also depends on the mass hierarchy. This can be taken into account either by performing different analysis for NH and IH, or by marginalizing over the hierarchy itself (see e.g. Ref. \cite{Gerbino:2016ehw}).

\begin{figure}
\begin{center}
\includegraphics[width=0.9\textwidth]{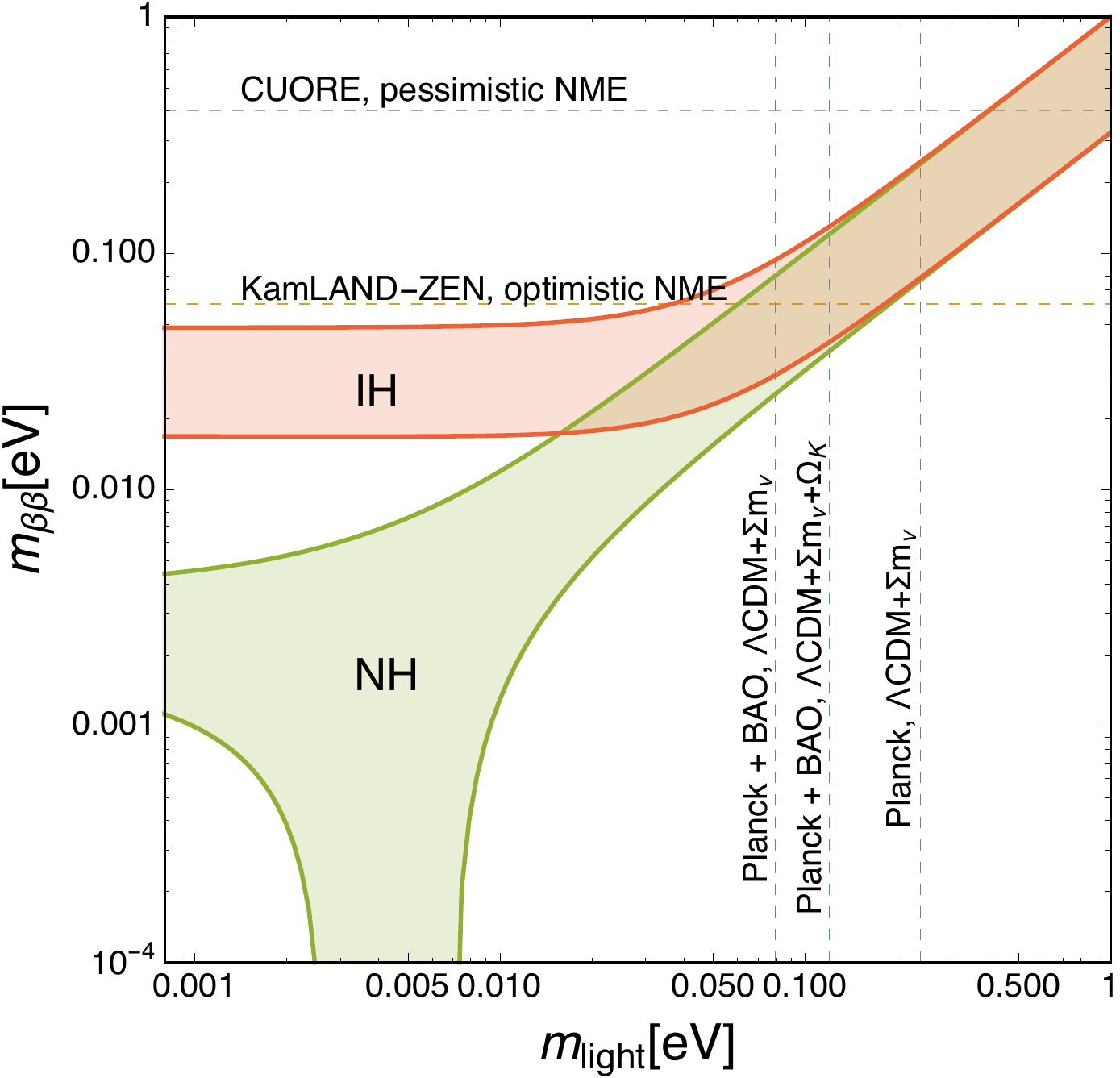}
\end{center}
\caption{Majorana mass $\mbb$ of the electron neutrino as a function of the mass $m_\mathrm{light}$ of the lightest neutrino eigenstate, for normal (green) or inverted (red) hierarchy. The filled regions correspond to the uncertainty related to the CP-violating phases. The horizontal dashed lines show 95\% current upper limits from $\bb$ searches. In particular, we show the tightest and loosest limits among those reported in the text, namely the most stringent from KamLAND-Zen (labeled ``KamLAND-Zen, optimistic NME''), and the less stringent from CUORE (labeled ``CUORE, pessimistic NME''). NME refers to uncertainty related to the nuclear matrix elements. We also show vertical dashed lines corresponding to 95\% upper limits on $\sumnu$ from cosmological observations, translated to upper limits on $m_\mathrm{light}$ using the information from oscillation experiments. In particular we show different model and dataset combinations, from right to left: PlanckTT+lowP in the $\lcdm+\sumnu$ model, PlanckTT+lowP+BAO in the $\lcdm+\sumnu+\Omega_K$ model, PlanckTT+lowP+BAO in the $\lcdm+\sumnu$ model. The vertical lines shown in the plot assume normal hierarchy, but the difference with the case of inverted hierarchy is very small on the scale of the plot.
\label{fig:mbb}}
\end{figure}

Combining the different probes of the absolute mass scale, with the support of oscillation results, leads to some interesting considerations. First of all, basically all of the information on the absolute mass scale comes from cosmology and $\bb$ searches. This confirms the naive expectation 
that can be made by comparing the sensitivity of the different probes. However, we recall again that the robust limits on $m_\beta$ from kinematic experiments represent an invaluable test for the consistency of the more model dependent constraints coming from cosmology and $\bb$ decay experiments. At the moment, cosmology still provides most of the information on the neutrino masses, although the sensitivity of $\bb$ experiments is rapidly approaching that of cosmological observations. A summary of the current limits is reported in Fig.~3 of Ref. \cite{Gerbino:2016ehw}. 
To better illustrate the complementarity of cosmology and $\bb$ searches, we show in Fig.~\ref{fig:mbb} how they constrain, together with oscillation experiments, the allowed space in the $(\mbb,\, m_\mathrm{light})$ plane. In more detail, we show the region in that plane that is singled out by oscillation experiments, for normal and inverted hierarchy. The width of the allowed regions traces  the uncertainties on the CP-violating phases. We show current upper 95\% bounds on $\mbb$ from $\bb$ searches as horizontal lines, and current 95\% bounds on $m_\mathrm{light}$ from cosmology as vertical lines. These are translated from the bounds on $\sumnu$ using information from oscillation experiments and assuming normal hierarchy. Assuming inverted hierarchy would however make a barely noticeable difference on the scale of the plot. It can be seen that in general cosmological observations are more constraining than $\bb$ searches.

\begin{figure}
\begin{center}
\includegraphics[width=0.9\textwidth]{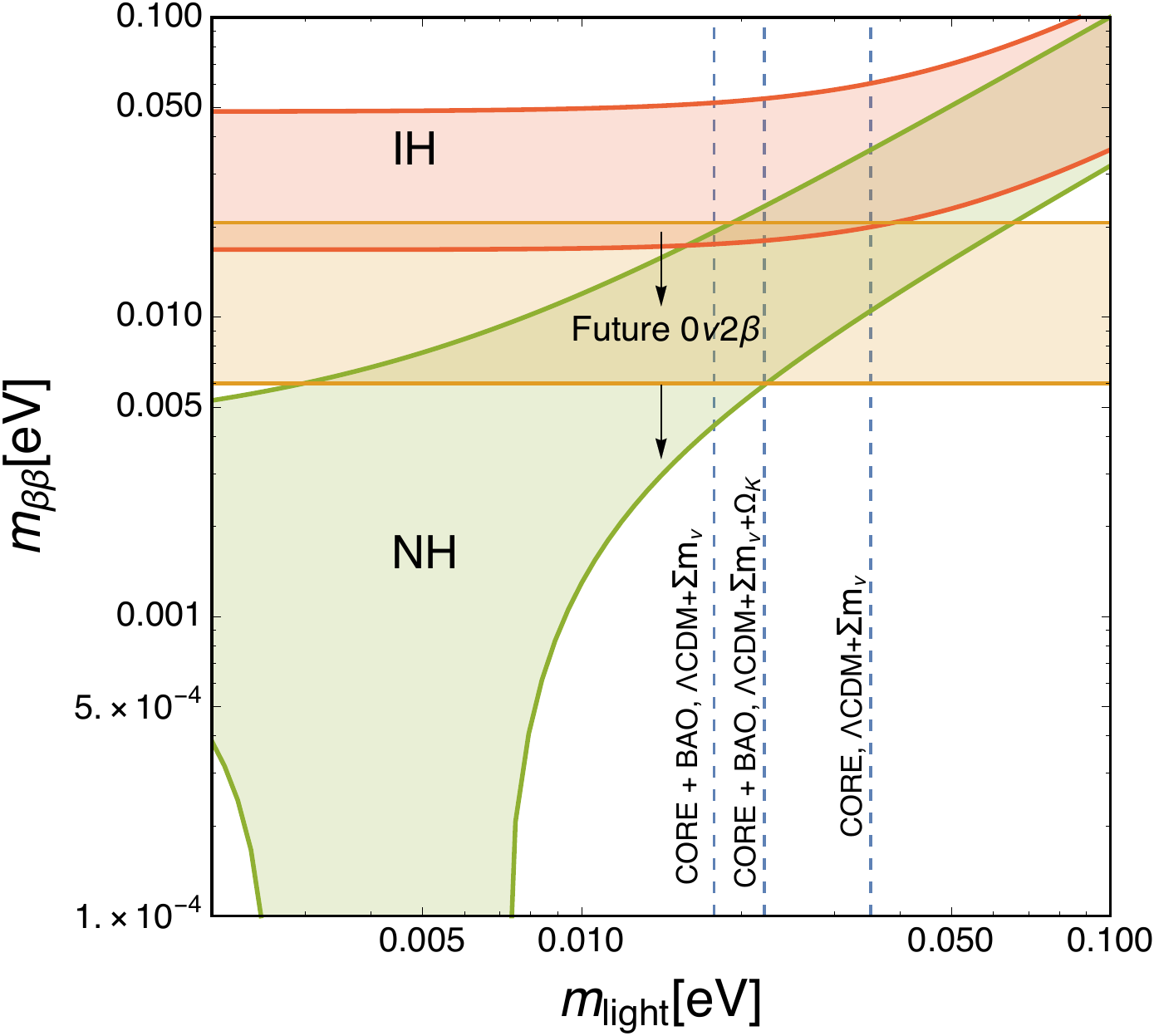}
\end{center}
\caption{The same as Fig~\ref{fig:mbb}, but for future cosmological observations and $\bb$ experiments. Note that in this figure we show $95\%$ upper limits for both $\mbb$ and $m_\mathrm{light}$, assuming that the true values of both quantities are much smaller that the corresponding experimental sensitivities. The horizontal yellow band labeled ``Future $\bb$'' is the union of the regions that contain the 95\% upper limits for LEGEND 1K, CUPID and nEXO, assuming 5 years of live time.
The vertical dashed lines correspond to $95\%$ upper limits on $\sumnu$. From right to left: CORE TT, TE, EE, PP in the $\lcdm+\sumnu$ model, CORE TT, TE, EE, PP + the DESI and EUCLID BAO in the $\lcdm+\sumnu+\Omega_K$ model,  CORE TT, TE, EE, PP + the DESI and EUCLID BAO in the $\lcdm+\sumnu$ model. The vertical lines shown in the plot assume normal hierarchy.
\label{fig:mbb_future}}
\end{figure}

In the future, however, one can expect that the constraining power of these two probes will be roughly equivalent. This can be seen in Fig.~\ref{fig:mbb_future} where, similarly to Fig~\ref{fig:mbb}, we show the allowed space in the $(\mbb,\, m_\mathrm{light})$ plane for
future cosmological and $\bb$ probes. As shown in Ref. \cite{Gerbino:2016ehw}, the constraining power of $\bb$ searches for $\sumnu$ would also depend crucially on the possibility of reducing the uncertainty on the nuclear matrix elements for the $\bb$ isotopes. In fact, provided that neutrinos are Majorana particles and that the leading mechanism responsible for the decay is a mass mechanism, the combination of cosmological probes and $\bb$ measurements could not only lead to a detection of the mass scale, but could also solve the hierarchy dilemma and provide useful information about (at least one of) the Majorana phases~\cite{Gerbino:2015ixa,Minakata:2014jba,Dodelson:2014tga}.

%. Cosmology is still leading the field, although the sensitivity of $\bb$ experiments is approaching that from cosmology. On the other hand, $\beta$-decay searches are not competitive yet with the other searches. This would be true in the future as well, when cosmological constraints will be as competitive as $\bb$ limits. 

%%%%%%%%

\section{Constraints on $\Neff$ }\label{sec:beyond}

Until now, we have focused on the capability of cosmological observations to constrain neutrino masses.
However, as noted in the introduction, cosmology is also a powerful probe of other neutrino properties.
The main example is without any doubt the \emph{effective number of neutrino families} (also called
effective number of relativistic degrees of freedom) $\Neff$, defined in equation Eq. (\ref{eq:Neff}).
As it is clear from its definition, $\Neff$ is simply a measure of the total cosmological density during the radiation-dominated era.
More precisely, it represents the density in relativistic species, other than photons, normalized to the energy density
of a massless neutrino that decouples well before electron-positron annihilation (that, we remember, is not actually the case).
As explained in Sec.~\ref{sec:cosmonu}, the standard framework, in which photons and active neutrinos are the only relativistic degrees of freedom present, 
and neutrino interactions follow the SM of particle physics, predicts $\Neff = 3.046$ after electron-positron annihilation \cite{Dolgov:2002wy,Mangano:2005cc,deSalas:2016ztq}.

Given its meaning, it is clear that a deviation from the expected value of $\Neff$ can hint to a broad class of effects - in fact,
all those effects that change the density of light species in the early Universe. Those effects 
are not necessarily related to neutrino physics, as the definition of $\Neff$ in terms of the number of relativistic degrees of freedom suggests.
For example, the existence of a Goldstone boson that decouples well before the QCD phase transition would appear as an increased number of degrees of freedom, with $\Delta\Neff \equiv \Neff - 3.046 = 0.027$ \cite{Baumann:2016wac}. Speaking however about changes in $\Neff$ that are somehow related to neutrino physics, 
the most notable example is probably the existence of one (or more) additional, sterile light eigenstate, produced through some mechanism in the early Universe. In such a situation, one would have $\Neff > 3.046$, as well as an additional contribution to $\sumnu$. Note that a light sterile neutrino 
would not necessarily contribute with $\Delta\Neff = 1$, as it does not share the same temperature as the active neutrinos. 

In this section we will focus on cosmological constraints on sterile neutrinos. However, for completeness, we mention a few other examples of scenarios in which $\Delta\Neff$ can possibly be different from zero. One is the presence of primordial lepton asymmetries, related to the presence of a non-vanishing chemical potential in the neutrino distribution function, Eq. (\ref{eq:FD}). Constraints on the allowed amount of lepton asymmetry, obtained taking into account the effect of neutrino oscillations, have been reported in Ref.~\cite{Castorina:2012md} using CMB and BBN data. Another possibility is the so-called low-reheating scenario~\cite{deSalas:2015glj,Kawasaki:2000en,Hannestad:2004px}, in which the latest reheating episode of the Universe happens just before BBN, at temperatures of the order of a few MeV, so that neutrinos do not have time to thermalize completely. In this case, one has $\Delta\Neff \le 0$. Finally, non-standard interactions between neutrino and electrons can modify the time of neutrino decoupling \cite{Mangano:2006ar}, so that 
the entropy transfer from $e^+e^-$ annihilation is larger than in the standard picture and $\Neff$ is larger. We note that the effects related to these 
new scenarios are often more complicated that just a change in $\Neff$: for example, both in the case of lepton asymmetries and low reheating,
the neutrino distribution function is changed in a non trivial way, affecting also the other moments of the distribution (like the number density, the average velocity, etc). Finally, to mention a possibility that is not related to changes in $\Neff$, cosmology can also probe the free-streaming nature of neutrinos, for example by looking for the effects of non-standard interactions among neutrinos~\cite{Forastieri:2015paa,Lancaster:2017ksf,Oldengott:2017fhy,Archidiacono:2016kkh}, or between neutrinos and dark matter~\cite{Wilkinson:2014ksa,Mangano:2006mp,Serra:2009uu}.

Let us briefly recall how $\Neff$ is constrained by cosmological observations \cite{neffdamping}. Increasing $\Neff$ will make the Universe expand faster (larger $H$) during the radiation-dominated era, and thus be younger at any given redshift. Then the comoving sound horizon at recombination will be smaller, going like $1/H$, while the angular diameter distance to recombination stays constant, because $H$ is unchanged after equality, so that $\theta_s$ is smaller.  Also, for fixed matter content, this will make the radiation-dominated era last longer. Recalling our discussion in Sec.~\ref{sec:obscmb}, the effect on the CMB spectrum is that the first peak is enhanced due to the larger early ISW, and all the peaks are moved to the right. However, as we have already learned, these effects can be canceled by acting on other parameters. There is however a more subtle and peculiar of effect of $\Neff$, that is related to the scale of Silk damping. The damping scale roughly scales as $1/\sqrt{H}$, i.e., as $\sqrt{t}$, as expected for a random walk process. Then the ratio between the angle subtended by the sound horizon and that subtended by the damping length scales like $H^{-1}/H^{-1/2} = H^{-1/2}$. Since $\theta_s$ is fixed by the position 
of the first peak, this means that increasing $\Neff$, the damping length is projected on larger angular scales, or, equivalently, that damping at a given scale
is larger. In conclusion the net effect is to lower the damping tail of the CMB spectrum. This effect is difficult to mimic with other parameters, at least in the standard framework. The damping length also depends on the density of baryons, so in principle one could think of changing this to compensate for the effect of $\Neff$; however, the baryon density is very well determined by the ratio of the heights of the first and second peak, so that it is in practice fixed. One possibility, in extended models, is to vary the fraction of primordial helium. Since the mean free path of photons depends on the number of free electrons, and helium recombines slightly before hydrogen, changing the helium to hydrogen ratio alters the Silk scale. However, this requires the assumption of nonstandard BBN, since, in the framework of standard BBN, 
the helium fraction is fixed by $\omega_b$ and $\Neff$ themselves, so it is not a free parameter.

We first review constraints on $\Neff$ in a simple one-parameter extension of $\Lambda$CDM, in which $\Neff$ is left free to vary, and
the mass of active neutrinos is kept fixed to the minimum value allowed by oscillations. This case can be considered as the most agnostic, in some sense,
in which one does not make any hypothesis on the new physics that is changing $\Neff$ (and thus on any other effects this new physics might produce).
Moreover, one can think of these as limits for a very light (massless) sterile neutrino. Finally, constraining $\Neff$ is a robustness check for the standard
$\Lambda$CDM mode. In fact, measuring $\Neff=3.046$ within the experimental uncertainty can be see as a great success of the standard cosmological model. It can be regarded as an indirect detection of the C$\nu$B, or, at least, of some component who has the same density, within errors, as we would expect for the three active neutrinos\footnote{The fact that, when probed, there is no hint for deviations from the free-streaming behaviour should strengthen
our belief that we are really observing the C$\nu$B.}. From PlanckTT+lowP, one gets $\Neff=3.13 \pm 0.32$; adding BAO gives $\Neff =3.15 \pm 0.23$~\cite{PlanckXIII}.
Both measurements, with a precision of $\sim 10\%$, are in excellent agreement with the standard prediction. Moreover, according to these results, $\Delta\Neff = 1$ is excluded at least at the $3\sigma$ level. Using also information about the full shape of the matter power spectrum, the BOSS collaboration finds $\Neff=3.03\pm 0.18$~\cite{BOSSDR12}. We note that adding information from direct measurements of the Hubble constant results in larger values of $\Neff$ ($\Neff=3.41\pm0.22$ from Planck TT,TE,EE+lowP+lensing+BAO+JLA+$H_0$, see~\cite{Riess}); this is due to the tension with the value of $H_0$ that is inferred from the CMB, that is alleviated in models with larger $\Neff$.
The next generation of cosmological experiments will improve these constraints by roughly an order of magnitude, getting close to the theoretical threshold of $\Delta\Neff=0.027$ discussed at the beginning of this section, corresponding to a Goldstone boson decoupling before the QCD phase transition. Moreover, it will be possible to confirm the effects of non-instantaneous decoupling, since future sensitivities will allow
to distinguish, at the $1$-$\sigma$ level, between $\Neff = 3 $ and $\Neff=3.046$.
The combination of CORE TT,TE,EE,PP will put an upper bound at 68\% CL of $\Delta\Neff<0.040$ on the presence of extra massless ($m\ll0.01\,\eV$) species\footnote{This constraint has been obtained in the context of a $\lcdm+\sumnu$ cosmology, with $\sumnu^{\mathrm{fid}}=0.06\,\eV$.}~\cite{COREPar} in addition to the three active neutrino families. The CORE collaboration puts limits also on the scenario in which the three active neutrinos have a fixed temperature, but their energy density is rescaled as $(\Neff/3.046)^{3/4}$. This scenario can account for an enhanced neutrino density (if $\Neff>3.046$) and reduced neutrino density (if $\Neff<3.046$ as for example in the case of low-reheating scenarios). In this case, CORE TT,TE,EE,PP yields $\Neff=3.045\pm0.041$.
Forecasts from S4 show that, in order to get closer to the threshold of $\Delta\Neff=0.027$, a sensitivity of $1\uka$ and $f_\mathrm{sky}>50\%$ are needed for a $1'$ beam size~\cite{S4}. Efficient de-lensing will help improve the limits on $\Neff$: delensed spectra will have sharper acoustic peaks, allowing to constrain $\Neff$ not only through the impact on the Silk scale, but also through the phase shift in the acoustic peaks~\cite{Baumann:2015rya}. Finally, having access to a larger sky fraction -- and therefore to a larger number of modes observed -- will be beneficial for constraints on $\Neff$~\cite{S4}. We conclude this summary about future limits by noticing that the inclusion of LSS data, such as BAO measurements from DESI and Euclid, provides only little improvements with respect to CMB only constraints (e.g., from CORE TT,TE,EE,PP+DESI BAO+Euclid BAO, $\Delta\Neff<0.038$ at 68\% CL for extra massless species and $\Neff=3.046\pm0.039$ for three neutrinos with rescaled energy density~\cite{COREPar}). For a summary of current and future limits on $\Neff$, we refer to Tab.\ref{tab:neff}.

\begin{table*}
\begin{center}
\begin{threeparttable}
\begin{tabular}{l|c|c}
Dataset	&$Bounds$	&Reference\\
\hline
\hline
Planck TT+lowP	&$\Neff=3.13\pm0.32$	&\cite{PlanckXIII}\\
Planck TT+lowP+BAO	&$\Neff=3.15\pm0.23$	&\cite{PlanckXIII}\\
Planck TT+lowP+BAO+FS	&$\Neff=3.03\pm0.18$	&\cite{BOSSDR12}\\
\hline
\hline
CORE TT,TE,EE,PP\,\tnotex{tnote:massless}	&$\Delta\Neff<0.040$	&\cite{COREPar}\\
CORE TT,TE,EE,PP\,\tnotex{tnote:massive}	&$\Neff=3.045\pm0.041$	&\cite{COREPar}\\
S4 TT,TE,EE,PP\,\tnotex{tnote:delens} 	&$\sigma(\Neff)=0.027$	&\cite{S4}\\
CORE TT,TE,EE,PP+DESI BAO+Euclid BAO\,\tnotex{tnote:massless}	&$\Delta\Neff<0.038$	&\cite{COREPar}\\
CORE TT,TE,EE,PP+DESI BAO+Euclid BAO\,\tnotex{tnote:massive}	&$\Neff=3.046\pm0.039$	&\cite{COREPar}\\
\hline
\end{tabular}
\begin{tablenotes}
\item\label{tnote:massless}The constrain applies to the scenario of extra light relics in addition to the three massive neutrino families, i.e. $\Neff\ge3.046$.
\item\label{tnote:massive}The constrain applies to the scenario of three massive neutrinos with energy density rescaled by $\Neff$, i.e. $\Neff$ can be either lower or greater than 3.046.
\item\label{tnote:delens}The combination includes delensed CMB spectra and a gaussian prior on the optical depth $\tau=0.06\pm0.01$.
\end{tablenotes}
\end{threeparttable}
\caption{Constraints on $\Neff$ from different combinations of cosmological data. \textit{Upper part:} current constraints on $\Neff$. BAO in row no. 2 are from 6dFGS~\cite{6dfgs}, SDSS MGS~\cite{mgs}, BOSS LOWZ DR11 and BOSS CMASS DR11~\cite{dr11} (see~\cite{PlanckXIII} for details). BAO and FS (full shape measurements) in row no. 3 are from BOSS DR12~\cite{BOSSDR12}. \textit{Lower section:} forecasts for future cosmological surveys. Unless otherwise stated, the sensitivity on $\Neff$ is forecasted assuming a standard cosmological model with $\Neff=3.046$ and also marginalising over $\sumnu$. DESI and Euclid BAO refer to the simulated BAO datasets based on expected experimental performances~\cite{DESI,EuclidStudyRep} (see~\cite{COREPar} for details).}\label{tab:neff}
\end{center}
\end{table*}

Let us now come to the case of a massive sterile neutrino. A sterile neutrino would contribute both to $\Neff$ and to $\omega_\nu$.
Its effect on the cosmological observables will thus be related to changes in these two quantities, as explained through this review.
In fact, in principle, we should specify the full form of the distribution function of the sterile neutrino, and its effects could not be
fully parameterized through $\Neff$ and to $\omega_\nu$. Fortunately, one has that, when the distribution function
is proportional to a Fermi-Dirac distribution, all the effects on the perturbation evolution of a light fermion can be mapped 
into two parameters  \cite{Colombi:1995ze}: its energy density in the relativistic limit (and thus its contribution to $\Neff$) and its energy density in the nonrelativistic limit
(and thus its density parameter, let us denote it with $\omega_s$ to distinguish it from the active neutrinos). This covers 
several physically interesting cases, namely those of a sterile neutrino that either (i) has a thermal distribution with arbitrary temperature $T_s$, or (ii) is distributed
proportionally to the active neutrinos, but with a suppression factor $\chi_s$  (this  corresponds to the 
Dodelson-Widrow (DW) prediction for the non-resonant production scenario \cite{Dodelson:1993je}; see also Ref. \cite{Merle:2015vzu}).
Defining an effective mass $m_s^\mathrm{eff}$ by mimicking Eq. \ref{eq:omeganu}, i.e.:
\begin{equation}
m^\mathrm{eff}_s \equiv 93.14\,\omega_s  \, \eV\, ,
\end{equation}
the actual mass $m_s$
of the sterile is related to the effective parameters by:
\begin{align}
&m_s = (T_s/T_\nu)^{-3}m^{\mathrm{eff}}_s = \Delta\Neff^{-3/4}m^{\mathrm{eff}}_s  \qquad \mathrm{(thermal)} \\
&m_s = \chi_s^{-1} m^{\mathrm{eff}}_s = \Delta\Neff^{-1} m^{\mathrm{eff}}_s \qquad \mathrm{(DW)}.
\end{align}

Planck data are consistent with no sterile neutrinos: the 95\% allowed region in parameter space is $\Neff < 3.7$, $m_s^{\mathrm{eff}} < 0.52\ \eV$
from PlanckTT + lowP + lensing + BAO. However, it should be noted that they do not exclude a sterile neutrino, provided it contribution to the total energy
density is small enough. We recall that a light sterile neutrino has been proposed as an explanation of the anomalies
observed in short-baseline (SBL) experiments (see e.g. Ref. \cite{Gariazzo:2015rra} and references therein). However, 
a sterile neutrino with the mass ($m_s \simeq 1$~eV) and coupling required to explain reactor anomalies
would rapidly thermalize in the early Universe (see e.g. Refs. \cite{Mirizzi:2013gnd,Hannestad:2015tea}) 
and lead to $\Delta\Neff = 1$, strongly at variance with cosmological constraints 
(excluded at more than 99\% confidence considering the above combination of Planck and BAO data). 
We conclude this section by quoting the forecasts for future cosmological probes. In the context of a $\lcdm+\sumnu$ model with $\sumnu^\mathrm{fid}=0.06\,\eV$ and $m_s^\mathrm{fid}=0\,\eV$, the combination of CORE TT,TE,EE,PP with BAO measurements from DESI and Euclid will provide $\Delta\Neff<0.054$ and $m_s<0.035\,\eV$~\cite{COREPar}.

%%%%%%%%

\section{Summary \label{sec:concl}}

The absolute scale of neutrino masses is one of the main open questions in physics to date. Measuring the neutrino mass
could shed light on the mechanism of mass generation, possibly
related to new physics at a high energy scale. From the experimental point of view, neutrino masses can be probed
in the laboratory, with $\beta$- and double $\beta$-decay experiments, and with cosmological observations. In fact,
cosmology is at the moment the most sensitive probe of neutrino masses. Upper limits from cosmology on the sum of neutrino masses
are possibly based on combinations of different observables. Results from the CMB alone can be regarded as very robust: 
these are of the order of $\sumnu <0.7\,\eV$ (95\%~CL). The addition of geometrical measurements, like those provided by BAO - also very robust -
bring down this limit to $\sumnu < 0.2\,\eV$ (95\%~CL). More aggressive analyses can get the bound very close to the minimum value allowed by oscillation experiments in the case
of inverted hierarchy, but are based on observations where control of systematics is more difficult and thus should be taken with caution.
It should also be borne in mind that cosmological inferences of neutrino masses are somehow model dependent. In extended cosmological models, especially those involving non-vanishing spatial curvature or dark energy, the constraints on $\sumnu$ are degraded, even though
they still remain very competitive with those obtained from laboratory experiments.
Combination of future CMB and LSS experiments could reach, if systematics are kept under control, a sensitivity of 15 meV in the first half of the next decade, allowing
a $4\sigma$ detection of neutrino masses if the hierarchy is normal and the lightest eigenstate is massless. In that case, it will also 
be possible to exclude the inverted hierarchy scenario with a high statistical significance. 

Present data are also compatible with the standard description
of the neutrino sector, based on the standard model of particle physics. CMB measurements constrain the number of relativistic species at recombination to be $N_\mathrm{eff}=3.13\pm0.32$ at 68\% CL. The inclusion of LSS data further tightens the constraints to $N_\mathrm{eff}=3.03\pm0.18$ at 68\% CL. These results exclude the presence of an additional thermalized species at more than $3\sigma$ level. Cosmological data are also consistent with no sterile neutrinos. Thus no new physics in the neutrino sector is presently required to interpret cosmological data. The standard picture will be tested more thoroughly by future experiments, that will allow to probe to an unprecedented level the physics of neutrino decoupling. An example would be the possibility to constrain non-standard neutrino-electron interactions. Future cosmological probes will also possibly reach the sensitivity necessary to detect, at the $1$-$\sigma$ level, the increase in the number of degrees of freedom due to a Goldstone boson that decouples well before the QCD phase transition. 

%%%%%%%%

\vspace{1cm}
\paragraph{Acknowledgments}
M.G. acknowledges support by the Vetenskapsr\aa det (Swedish Research Council) through contract No. 638-2013-8993 and the Oskar Klein Centre for Cosmoparticle Physics.

\end{document}